\documentclass[pre,twocolumn,amssymb]{revtex4-1}
\usepackage{mathtools}
\usepackage{amsmath}
\usepackage{float}
\usepackage{graphicx,array}
\usepackage{color}
\usepackage{epsfig}
\usepackage{latexsym}
\usepackage{bm}
\usepackage[colorlinks=true, allcolors=blue]{hyperref}
\usepackage{color}
\usepackage[normalem]{ulem}

\begin{document}
\newcommand\newfonts{\fontsize{4}{16}\selectfont}
\newcommand\smallfonts{\fontsize{5.75}{16}\selectfont}
\newcommand\modifyfonts{\fontsize{5.75}{16}\selectfont}
\newcommand{\rar}{\rightarrow}
\newcommand{\lar}{\leftarrow}
\newcommand{\rlh}{\rightleftharpoons}
\newcommand{\eref}[1]{Eq.~(\ref{#1})}%
\newcommand{\Eref}[1]{Equation~(\ref{#1})}%
\newcommand{\fref}[1]{Fig.~\ref{#1}} %
\newcommand{\Fref}[1]{Figure~\ref{#1}}%
\newcommand{\sref}[1]{Sec.~\ref{#1}}%
\newcommand{\Sref}[1]{Section~\ref{#1}}%
\newcommand{\aref}[1]{Appendix~\ref{#1}}%

\renewcommand{\ni}{{\noindent}}
\newcommand{\dprime}{{\prime\prime}}
\newcommand{\be}{\begin{equation}}
\newcommand{\ee}{\end{equation}}
\newcommand{\bea}{\begin{eqnarray}}
\newcommand{\eea}{\end{eqnarray}}
\newcommand{\nn}{\nonumber}
\newcommand{\bk}{{\bf k}}
\newcommand{\bQ}{{\bf Q}}
\newcommand{\q}{{\bf q}}
\newcommand{\s}{{\bf s}}
\newcommand{\bN}{{\bf \nabla}}
\newcommand{\bA}{{\bf A}}
\newcommand{\bE}{{\bf E}}
\newcommand{\bj}{{\bf j}}
\newcommand{\bJ}{{\bf J}}
\newcommand{\bs}{{\bf v}_s}
\newcommand{\bn}{{\bf v}_n}
\newcommand{\bv}{{\bf v}}
\newcommand{\la}{\left\langle}
\newcommand{\ra}{\right\rangle}
\newcommand{\dg}{\dagger}
\newcommand{\br}{{\bf{r}}}
\newcommand{\brp}{{\bf{r}^\prime}}
\newcommand{\bq}{{\bf{q}}}
\newcommand{\hx}{\hat{\bf x}}
\newcommand{\hy}{\hat{\bf y}}
\newcommand{\bS}{{\bf S}}
\newcommand{\cU}{{\cal U}}
\newcommand{\cD}{D}
\newcommand{\bR}{{\bf R}}
\newcommand{\pll}{\parallel}
\newcommand{\sumr}{\sum_{\vr}}
\newcommand{\cP}{{\cal P}}
\newcommand{\cQ}{{\cal Q}}
\newcommand{\cS}{{\cal S}}
\newcommand{\ua}{\uparrow}
\newcommand{\da}{\downarrow}
\newcommand{\red}{\textcolor {red}}
\newcommand{\black}{\textcolor {black}}
\newcommand{\1}{{\oldstylenums{1}}}
\newcommand{\2}{{\oldstylenums{2}}}
\newcommand{\mDelta}{\varepsilon}
\newcommand{\m}{\tilde m}
\def\lsim {\protect \raisebox{-0.75ex}[-1.5ex]{$\;\stackrel{<}{\sim}\;$}}
\def\gsim {\protect \raisebox{-0.75ex}[-1.5ex]{$\;\stackrel{>}{\sim}\;$}}
\def\lsimeq {\protect \raisebox{-0.75ex}[-1.5ex]{$\;\stackrel{<}{\simeq}\;$}}
\def\gsimeq {\protect \raisebox{-0.75ex}[-1.5ex]{$\;\stackrel{>}{\simeq}\;$}}

\title{\large{Time-dependent properties of run-and-tumble particles: Density relaxation }}

\author{Tanmoy Chakraborty}
%\email{tanmoy.chakraborty@bose.res.in}  
\author{Punyabrata Pradhan}
\affiliation{Department of Physics of Complex Systems, S. N. Bose National Centre for Basic Sciences, Block-JD, Sector-III, Salt Lake, Kolkata 700106, India}

\begin{abstract}

  We characterize  collective diffusion of hardcore run-and-tumble particles (RTPs) by explicitly calculating the bulk-diffusion coefficient $D(\rho, \gamma)$ for arbitrary density $\rho$ and tumbling rate $\gamma$, in systems on a $d$ dimensional periodic lattice. We study two minimal models of RTPs: Model I is the standard version of hardcore RTPs introduced in [Phys. Rev. E \textbf{89}, 012706 (2014)], whereas model II is a long-ranged lattice gas (LLG) with hardcore exclusion - an analytically tractable variant of model I.
  We calculate the bulk-diffusion coefficient analytically for model II and  numerically for model I through an efficient Monte Carlo algorithm; notably, both models have qualitatively similar features.
  In the strong-persistence limit $\gamma \rightarrow 0$ (i.e., dimensionless ratio $r_0 \gamma/v \to 0$), with $v$ and $r_0$ being the self-propulsion speed and particle diameter, respectively, the fascinating interplay between persistence and interaction is quantified in terms of two length scales: (i) Persistence length $l_p=v/\gamma$ and (ii) a ``mean free path'', being a measure of the average empty-stretch or gap size in the hopping direction. We find that the bulk-diffusion coefficient varies as a power law in a wide range of density: $D \propto \rho^{-\alpha}$, with exponent $\alpha$ gradually crossing over from $\alpha=2$ at high densities to $\alpha=0$ at low densities. As a result, the density relaxation is governed by a nonlinear diffusion equation with anomalous spatiotemporal scaling.
  In the thermodynamic limit, we show that the bulk-diffusion coefficient - for $\rho, \gamma \to 0$ with $\rho/\gamma$ fixed - has a scaling form $D(\rho, \gamma) = D^{(0)} {\cal F}(\rho a v/ \gamma)$, where $a \sim r_0^{d-1}$ is particle cross-section and $D^{(0)}$ is proportional to the diffusion coefficient of noninteracting particles; the scaling function ${\cal F}(\psi)$ is calculated analytically for model II (LLG) and numerically for model I. Our arguments are independent of dimensions and microscopic details.

\end{abstract}

\maketitle

\section{Introduction}

Transport characteristics of self-propelled particles (SPPs), also called active matter, never cease to surprise \cite{Ramaswamy2013}; they are often anomalous, implying the presence of nontrivial correlations and being the rule rather than the exception.
Numerous studies have been conducted to examine the nature of transport in SPPs - spanning from living to non-living materials (e.g., bacterial colonies and active Janus particles, respectively) - through a variety of experiments \cite{Libchaber_2000, granick_pnas_2009, goldstein_prl_2009,Kudroli_2010, granick_nature, ariel2015swarming, Clement_PRL_2015, Cavagna_2017, Sokolov_2018, Metzler_2018, Takahiro_2020}, simulations \cite{Graham_PRL_2008, Suriyanarayanan_2019, Lowen_2020}, and theories \cite{Hatwalne_PRL_2004, Kardar_2019, Brady_PRE_2020, majumdar_epl_2020}. For instance, in a recent experiment, the expansion of localized Janus swimmers on a two-dimensional substrate was found to exhibit an early-time ballistic growth \cite{takatori2016}, \black{with the width of the density perturbation growing with time $t$ as $\sim t^{1/z}$, characterized by the dynamic exponent $z=1$ (ultimately crossing over to normal diffusion with $z=2$)}. In a related simulation study of single-file transport of interacting SPPs, the space-time scaling of the two-point density correlations in the early-time regimes was shown to display superdiffusion with $z \approx 1.67$ \cite{dolai2020}. Presently it is unclear precisely what causes the anomalous growth, the resultant exponent, and whether or not one should expect a universal behavior. Despite recent progress \cite{Bar_2012, Blythe_PRL_2016, Tailleur_PRL_2018, pagonabarraga_2019, Blythe_2020,Chate_2020, Kafri_2021}, a good theoretical understanding of the time-dependent properties of SPPs, by taking into consideration {\it many-body correlations}, is missing.

Various models of SPPs have been investigated in the literature over the last decade, with studies of their dynamical aspects generally streamlined into characterizing self-diffusion coefficient \cite{Peruani_PRL_2013, Levis_berthier2014, Voigtmann_2017_PRE, Benichou_2018, Vanderzande_2019, Anupam_2021, Debets_PRL_2021, Benichou_2022, Tailleur_2022, Kurzthaler_PRL_2018, Brady_PRE_2020, Irani_PRL_2022} and other quantities such as effective self-propulsion velocity, mechanical pressure, and chemical potential, among others \cite{Wijland_2019_PRL, Brady_PRL2014, Solon_2015_PRL, Brady_2023_PNAS, Speck_2021_PRE, Subhadip_PRE_2016}.
In this paper, we consider a many-particle description of a subclass of SPPs - the hardcore run-and-tumble particles (RTPs), and  provide a microscopic theory for large-scale (hydrodynamic) density relaxations in the systems.
Hardcore RTPs are possibly the simplest examples of interacting active-matter systems, which have a single conserved quantity - number of particles. Also, despite persistent motion of particles on short time scales, the steady-state current in the system on a  periodic domain is exactly zero. Indeed, models of RTPs constitute a paradigm for theoretically examining time-dependent properties of a typical active-matter system and their studies, one believes, could lead to simple explanation for several fascinating aspects observed in more realistic settings.
In order to understand particle transport in interacting RTPs, one must distinguish between the two density-dependent transport coefficients: The self-diffusion coefficient for characterizing diffusion of individual (tagged) particles and the collective- or bulk-diffusion coefficient for spread of a density perturbation. While the former determines fluctuations, the latter determines relaxation in the system. In fact, mean-square cumulative displacement of all particles can be attributed to current fluctuations \cite{Tanmoy-condmat2023}. On the other hand, the bulk-diffusion coefficient $D(\rho)$, which depends on the local density $\rho(X)$ in general, is related to density relaxation, via Fick's law for particle current,
\begin{equation}
  \label{local_current}
J_D = -D(\rho) \frac{\partial \rho}{\partial X},
\end{equation}
generated due to density gradient $\partial\rho/\partial X$.

Understanding  bulk diffusion in many-body systems is of fundamental interest in statistical physics; indeed, it is a crucial first step in formulating fluctuating hydrodynamics for diffusive systems \cite{Derrida_2004, Bertini_2001}. Despite significant progress made in the past \cite{Landim_1998, Spohn_2012}, obtaining the bulk-diffusion coefficient exactly as a function of density and other parameters  has been a challenge for such systems, particularly the {\it conventional} models of SPPs, which typically have nontrivial spatio-temporal correlations. Perhaps not surprisingly, exact studies  are often limited to systems with a product-measure steady state \cite{Krapivsky_2014,  Carlson_1993, Tailleur_PRL_2018}, and with one \cite{Malakar} or two particles \cite{Blythe_PRL_2016, Blythe_2017, Das_2020}.  Of course, there are phenomenological theories \cite{Fily_2012, Baskaran_2013}, however, their validity has not been rigorously proven.

In the past,  there has been some success in developing a theoretical framework to characterize bulk diffusion, and diffusive instability, in the context of active Brownian particles (ABPs) \cite{Speck_2013_EPL}. Recently, progress has been made in characterizing bulk diffusion in a specific class of RTPs, where the tumbling rate and particle velocity  are system-size dependent and vanishingly small; as a result, the system can be exactly described by a {\it mean-field} hydrodynamics \cite{Tailleur_PRL_2018}. On the contrary, for the conventional RTPs having {\it finite} tumbling rate and velocity, the system possesses nonzero spatial correlations (which can be even long-ranged for small tumbling rates); in that case, the hydrodynamic description of Ref. \cite{Tailleur_PRL_2018} is not applicable any more. 
In this scenario, a rigorous study of the bulk-diffusion coefficient in minimal lattice models of conventional  SPPs are desirable.

In this paper, by using microscopic  calculations, we characterize collective diffusion in hardcore run-and-tumble particles (RTPs) on a $d$ dimensional periodic lattice for {\it arbitrary} density $\rho$ and tumbling rate $\gamma$, i.e., inverse of the persistence time $\tau_p$ (proportional to the persistence length $l_p = v \tau_p$, with $v$ being self-propulsion speed).
We specifically consider two minimal models: Model I is the standard version of hardcore RTPs introduced in Ref. \cite{Soto_2014}, whereas model II is a long-ranged lattice gas (LLG) - an analytically tractable variant of model I. We consider periodic boundary condition, where total number of particles $N$ remains conserved, with (global) density being $\rho=N/L^d$ and $L$ being (linear) size of the system.
In these two model systems, we perform a detailed study of large-scale relaxations of local density $\rho({\bf X},t)$, which is defined at position ${\bf X}$ and time $t$ and relax from an initial density profile $\rho({\bf X}, t=0)=\rho_{in}({\bf x}={\bf X}/L)$. We take large system-size limit $L \to \infty$, with $\rho_{in}({\bf x})$ fixed, and typically consider long-wave-length fluctuations around background density $\rho_0$ with $\rho_{in}({\bf x}) = \rho_0 + {\rm const.} \sin (2 \pi x)$; that is, we simply take an intital sinusoidal density profile $\rho({\bf X},0)=\rho_0 + {\rm const.} \sin (q X)$ with wave number $q=2 \pi/L  \to 0$ as $L \to \infty$.
We numerically demonstrate the diffusive scaling limit concerning the particle transport, which is governed by the Fick's law as in Eq. \eqref{local_current}. More specifically, for any finite tumbling rate and an initial coarse-grained profile $\rho_{in}({\bf X}/L)$, the time-dependent density profile is shown to possess a diffusive scaling $\rho({\bf X},t) \equiv \rho({\bf x}= {\bf X}/L,\tau=t/L^2)$, with $L \gg 1$. In that case, the time evolution of the coarse-grained density field $\rho({\bf x},\tau)$ is governed by a nonlinear diffusion equation
$$\partial_{\tau} \rho({\bf x},\tau) = \nabla [ D(\rho,\gamma) \nabla \rho({\bf x},\tau)],$$ where $D(\rho,\gamma)$ is the bulk-diffusion coefficient, depending on both local density $\rho$ and tumbling rate $\gamma =\tau_p^{-1}$. 
We substantiate the above assertion by calculating the bulk-diffusion coefficients, numerically for model I through an efficient Monte Carlo algorithm and analytically for model II, for arbitrary density and tumbling rate.
We also consider an interesting scaling regime with density and tumbling rate being small, i.e., $\rho \to 0$ and $\gamma \to 0$ with the ratio $\rho/\gamma$ fixed. We show that, in the thermodynamic limit and regardless of dimensions and microscopic details, there exists a scaling function ${\cal F}(\psi)$ governing large-scale relaxation through a precise scaling law for the bulk-diffusion coefficient, 
\be
D(\rho, \gamma) = D^{(0)} {\cal F}\left( \frac{\rho a v}{\gamma} \right),
\label{D_scaling}
\ee
where $v$, $a \sim r_0^{d-1}$ and $r_0$ are the self-propulsion speed, cross-section and diameter of particles, respectively (we throughout put $v=1$ and $r_0=1$). We analytically calculate ${\cal F}(\psi)$ for model II and verify the scaling of Eq. \eqref{D_scaling} through direct simulations of both the models. The scaling variable $\psi=\rho a v/\gamma$ quantifies the interplay between persistence and interaction, with the prefactor $D^{(0)}$ is proportional to the bulk-diffusion coefficient in the noninteracting limit. 
Although persistent random walks are long known  in the literature \cite{Schnitzer93, WEISS2002_PRW}, the scaling function ${\cal F}(\psi)$ as given in Eq. \eqref{D_scaling}, to the best of our knowledge, has not yet been reported.

Our central idea is that, in the strong-persistence limit with persistence length being much larger than the particle diameter (i.e., $\l_p \gg r_0$), collective particle transport is governed by the two length scales - persistence length and a ``mean free path'', the latter being a measure of the average empty-stretch size, or {\it gap}, in the {\it direction of particle hopping}. Indeed, one can argue by using a simple scaling theory that the bulk-diffusion coefficient has a highly nonlinear - a power law - density dependence. Interestingly, this power-law scaling of the bulk-diffusion coefficient accounts for the  {\it early-time} (roughly, at the onset of diffusion) anomalous growth previously observed in Ref. \cite{takatori2016}. We demonstrate that the early-time relaxation of initially localized density profiles are {\it non-Gaussian}, where their spatiotemporal scaling form, to a very good approximation, is determined within our theory. Our results suggest that the related growth exponent is {\it not} universal, but rather depends on specific parameter regimes.

We organize the paper as following. We begin by describing two models of hardcore RTPs in Sec.~\ref{sec:models} and present a scaling theory in Sec.~\ref{sec:scaling_argument}. Next, we obtain a large-scale hydrodynamic description concerning the time-evolution of a suitably coarse-grained density profile in Sec.~\ref{sec:hydrodynamics} by calculating the density- and tumbling-rate-dependent bulk-diffusion coefficient $D(\rho,\gamma)$ in both models I and II. In Sec.~\ref{sec:D_result}, we study the explicit parameter dependence of the bulk-diffusion coefficient $D(\rho,\gamma)$ and show the existence of a scaling law for both the models in one and two dimensions.  In Sec.~\ref{sec-density_relaxation}, we also verify the diffusive hydrodynamic time evolution by studying density relaxation through direct simulations of the models. In Sec.~\ref{sec:emergent_hydrodynamics}, we verify, in the limit of strong persistence, the existence of an emergent hydrodynamics in both the models. We also study relaxation of a $delta$ initial density perturbation, relaxing on an infinite domain, to demonstrate an early-time anomalous (ballistic) transport. Finally, in Sec.~\ref{sec:summary}, we summarise our findings with some concluding remarks.

\section{Model description}\label{sec:models}

\begin{figure}
           \centering
            \includegraphics[width=0.4\textwidth]{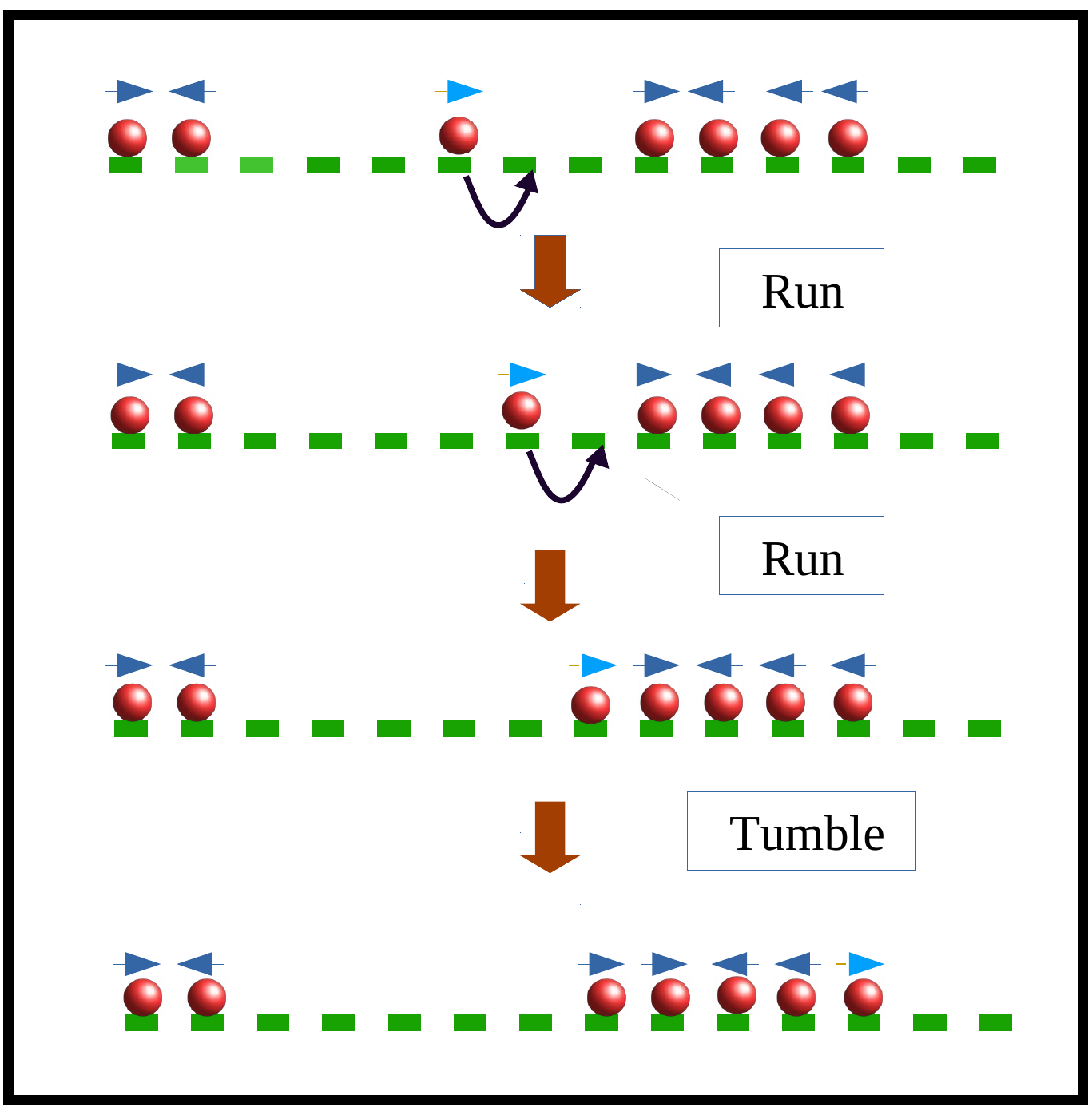}\hfill
            \includegraphics[width=0.4\textwidth]{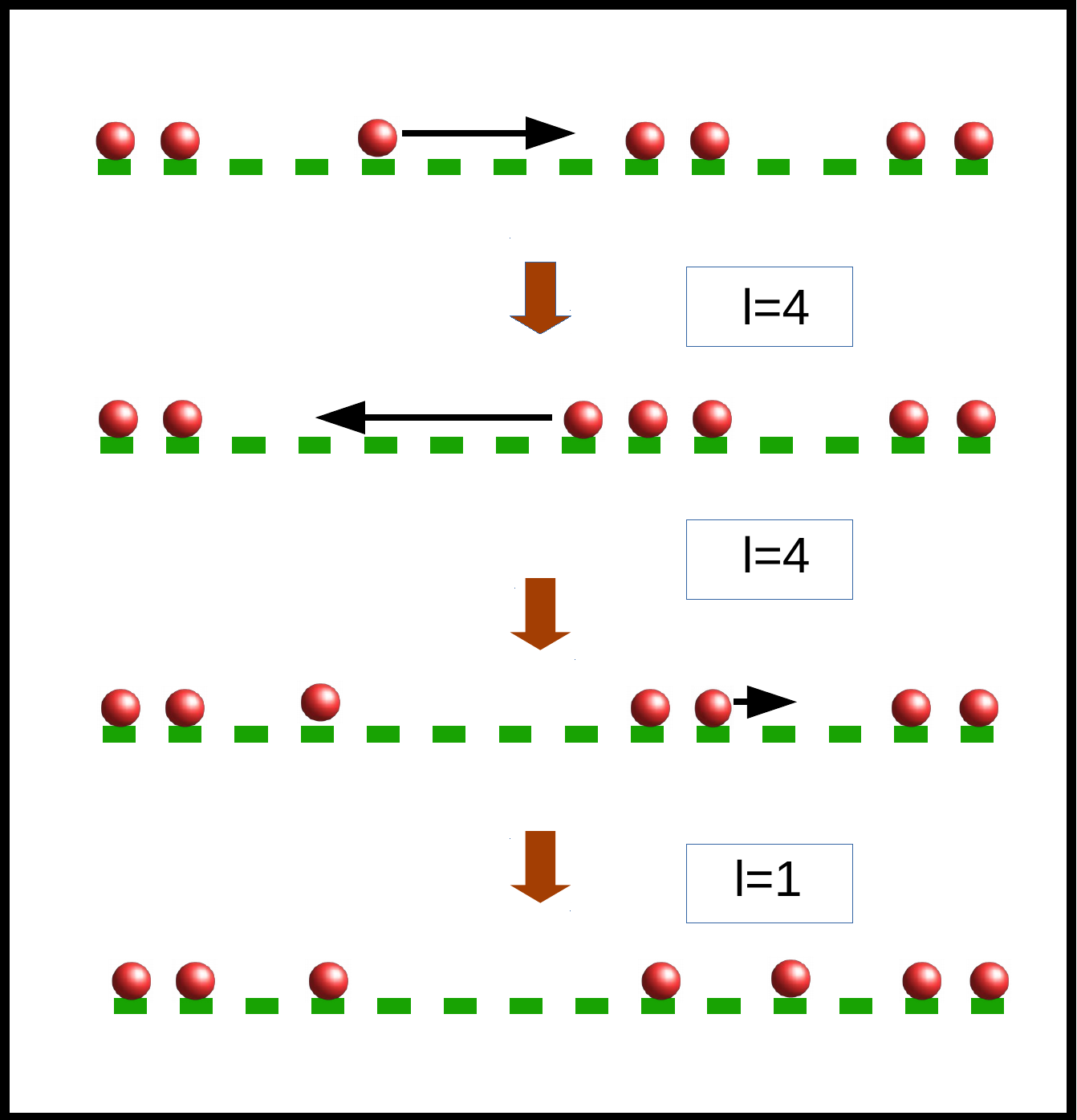}
            \caption{\textit{Schematic diagram of model I and II.--} Top panel: We illustrate microscopic dynamics of model I (i.e., run and tumble), which is composed of standard hardcore RTPs (red circles) on a one-dimensional lattice. Particles hop to their nearest neighbor site along the direction of their spins, indicated by the arrows above them. Bottom panel: We demonstrate the dynamics of hardcore particles (red circle) in model II on a one dimensional lattice. Particles hop symmetrically, to right or left, by length $l$ drawn from an exponential distribution $\phi(l) \propto e^{-l/l_p}$ with $l \in [0, 1, 2, \dots ]$.}
             \label{fig:model_description}
  \end{figure}

We consider two minimal models of interacting RTPs, model I and II, consisting of $N$ particles on a $d$ dimensional periodic hypercubic lattice with length $L$ and volume $V=L^d$ with constant global density $\rho=N/V$. While model I is the standard hardcore RTPs, model II is a ``long-range" variant of model I that is amenable to analytical calculations. In both models, particles interact via {\it excluded-volume repulsion}, i.e., a site can be occupied by at most one particle and particle crossing is not allowed. We introduce the occupancy variable $\eta_{\textbf{X}}$ at a site $\textbf{X}$, where $\eta_{\textbf{X}}=1$ or $0$, depending on whether a site  is occupied or vacant, respectively.

\subsection*{Model I: Standard hardcore RTPs}

We study the paradigmatic model of hardcore RTPs introduced in Ref.~\cite{Soto_2014}. In this model, particles are associated with a spin variable $\textbf{s}$, randomly oriented along any of the $2d$ directions. We consider continuous-time stochastic dynamics, which are specified below.
\begin{itemize}
\item[A.] {\it Run:} With rate $v$, a particle hops by unit lattice spacing along its spin direction, provided the destination site is vacant.
\item[B.] {\it Tumble:} With rate $\gamma$ - the inverse of persistence time $\tau_p$, the spin of a particle flips by choosing a new direction at random out of $2d -1$ other available directions (i.e., excluding the initial direction) with equal probability.
\end{itemize}
We illustrate the schematic representation of the above model in Fig.~\ref{fig:model_description}(a). Clearly, between two successive tumblings which happen on a timescale $\tau_p=1/\gamma$, RTPs move ballistically with constant speed $v$ along the empty stretches in the direction of their spins. The typical distance traveled by a single particle on a timescale $\tau_p$ is known as {\it persistence length}  $l_p=v \tau_p =v/\gamma$. 
Although the model is computationally quite efficient, it is difficult to characterize the model analytically due to the extra spin variables. To overcome the difficulty, we also consider an analytically tractable idealized variant of the above model and refer to the variant as model II - a long-ranged lattice gas (LLG).

\subsection*{Model II: Hardcore long-ranged lattice gas (LLG)}

We consider a long-ranged lattice gas (LLG) of $N$ hardcore particles on a $d$-dimensional periodic hypercubic lattice of length $L$, as depicted in Fig. \ref{fig:model_description}(b). With unit rate, a particle attempts to hop, symmetrically along any one of the $2d$ directions, by hop-length \textcolor{black}{$l\in[0, 1, 2, \dots]$} drawn from a distribution $\phi(l)$. Provided that the stretch of consecutive vacancies, called the {\it gap}, in the hopping direction is at least of length $l$ (i.e., if gap size $g \ge l$), the attempted particle hop is successful; otherwise, due to the hard-core constraint, the particle traverses the entire stretch and sits adjacent to its nearest occupied site in that direction. Although the hop-length distribution $\phi(l)$ can in principle be arbitrary, we choose, for simplicity, a single parameter family of an exponential distribution,
\begin{eqnarray}\label{hop-length}
\phi(l)=B e^{-l/l_p},
\end{eqnarray}
where $l_p=v/\gamma$ is called persistence length (as compared to that in model I) and $B=1-e^{-1/l_p}$ is the normalization constant.

\begin{figure*}
           \centering
            \includegraphics[width=0.25\textwidth]{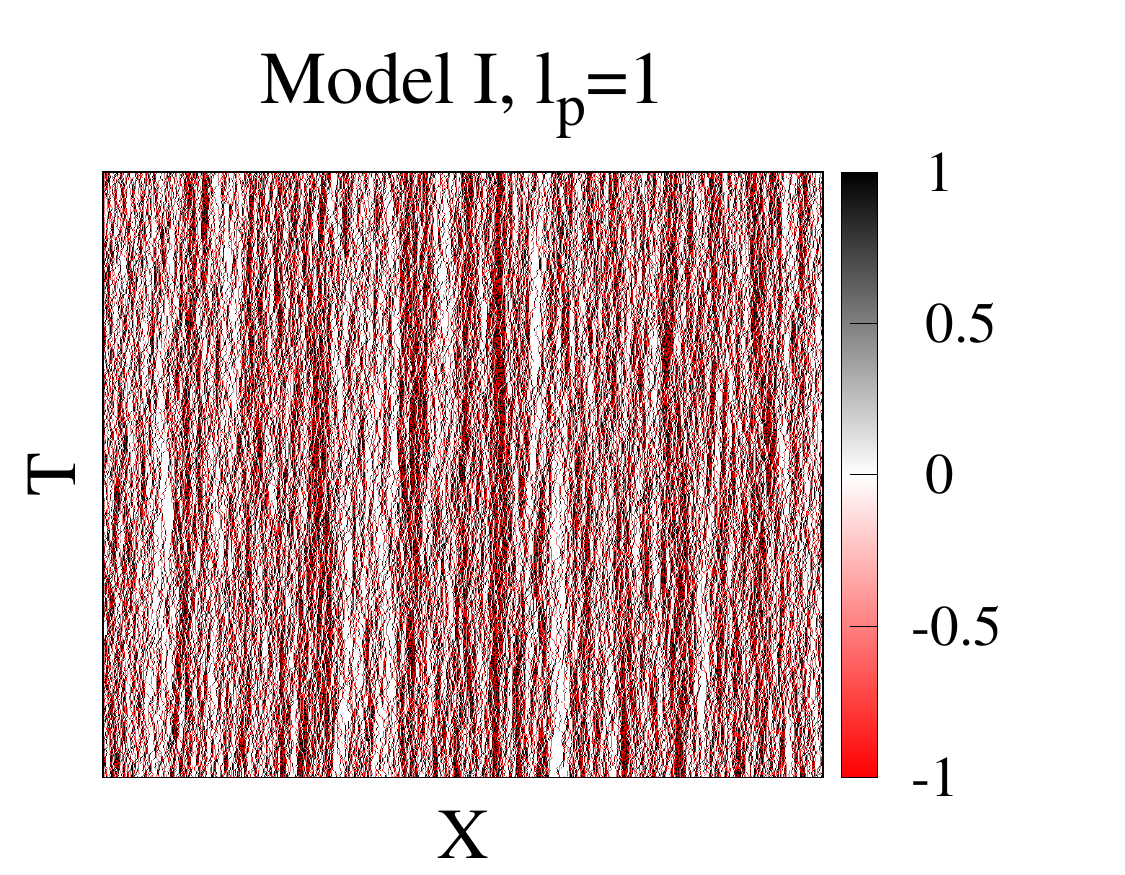}\hfill
            \includegraphics[width=0.25\textwidth]{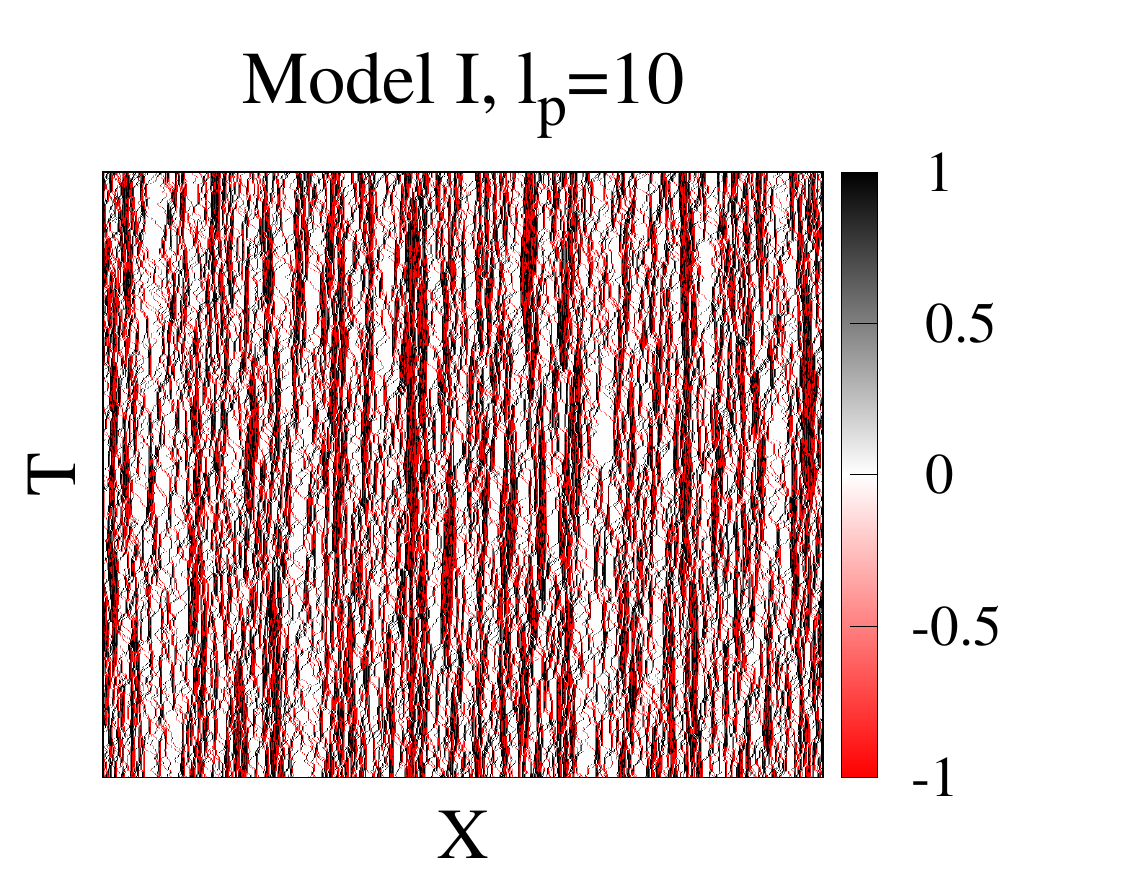}\hfill
            \includegraphics[width=0.25\textwidth]{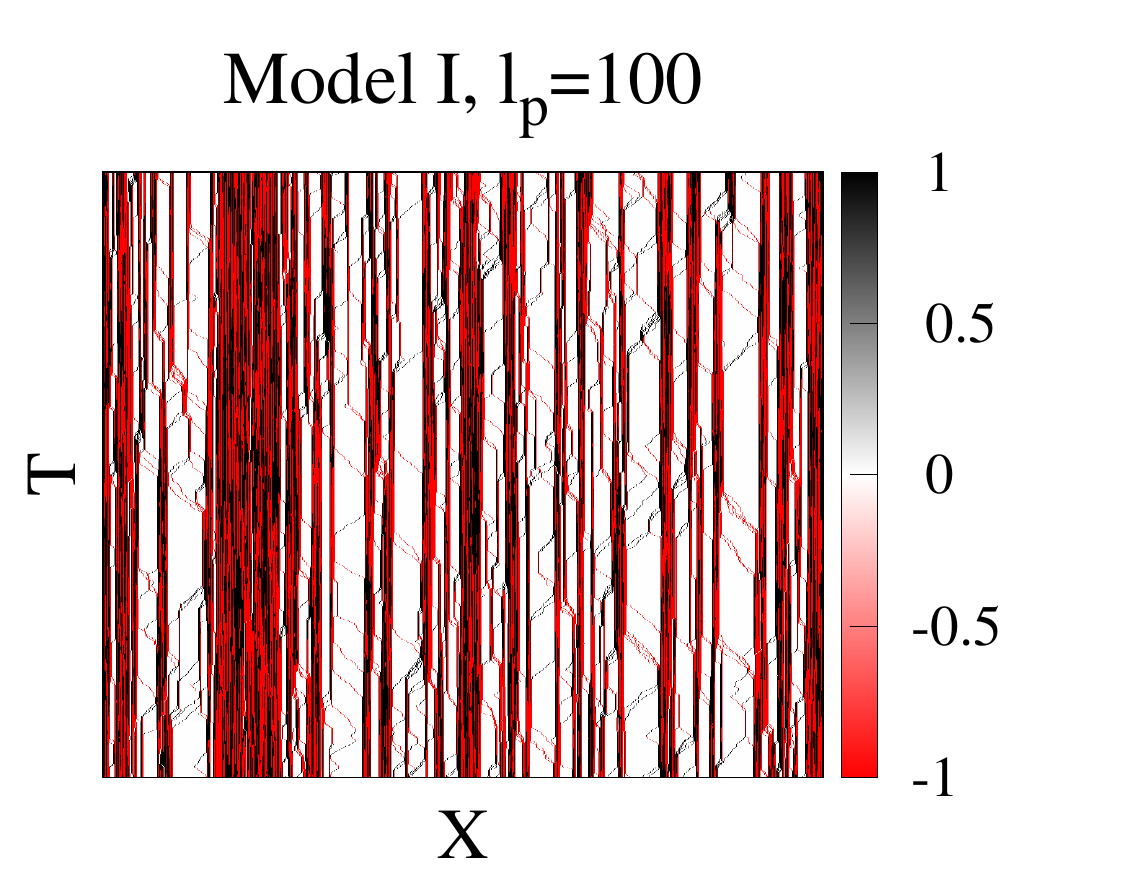}\hfill
            \includegraphics[width=0.25\textwidth]{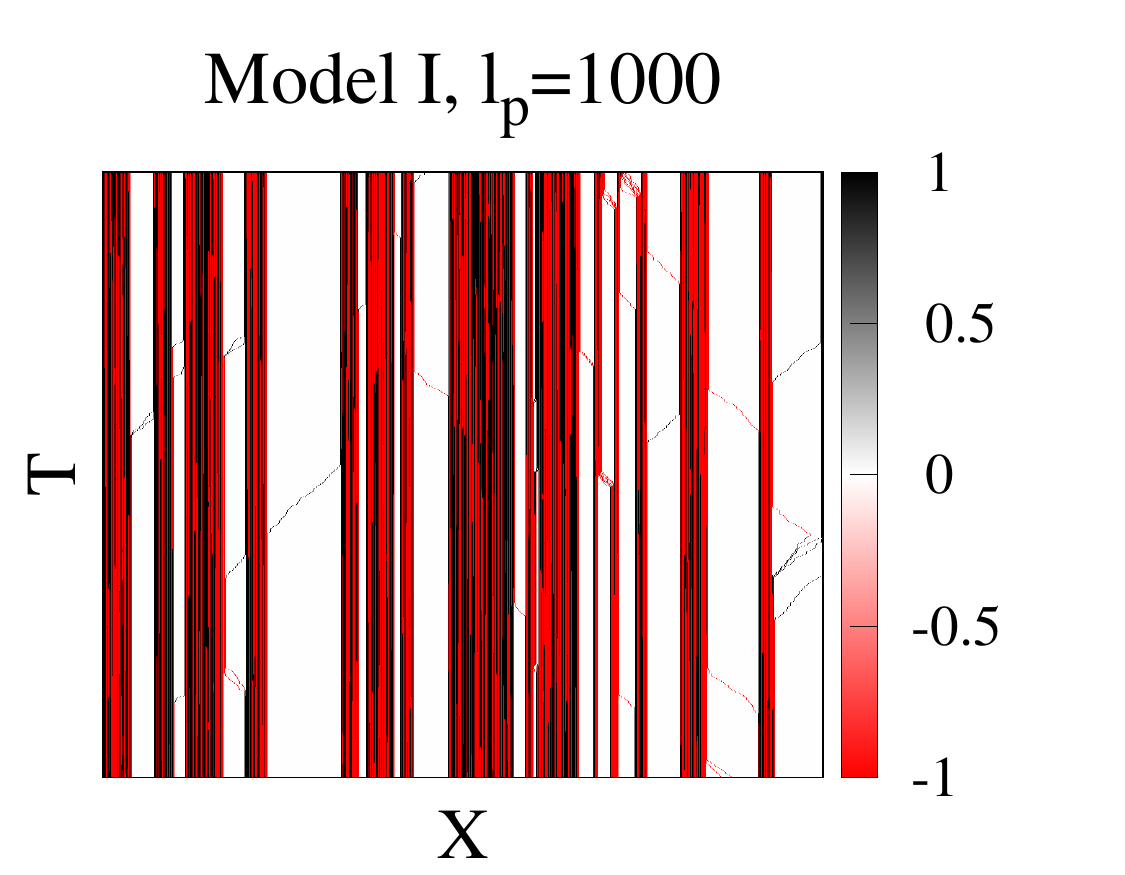}\vfill
             \includegraphics[width=0.25\textwidth]{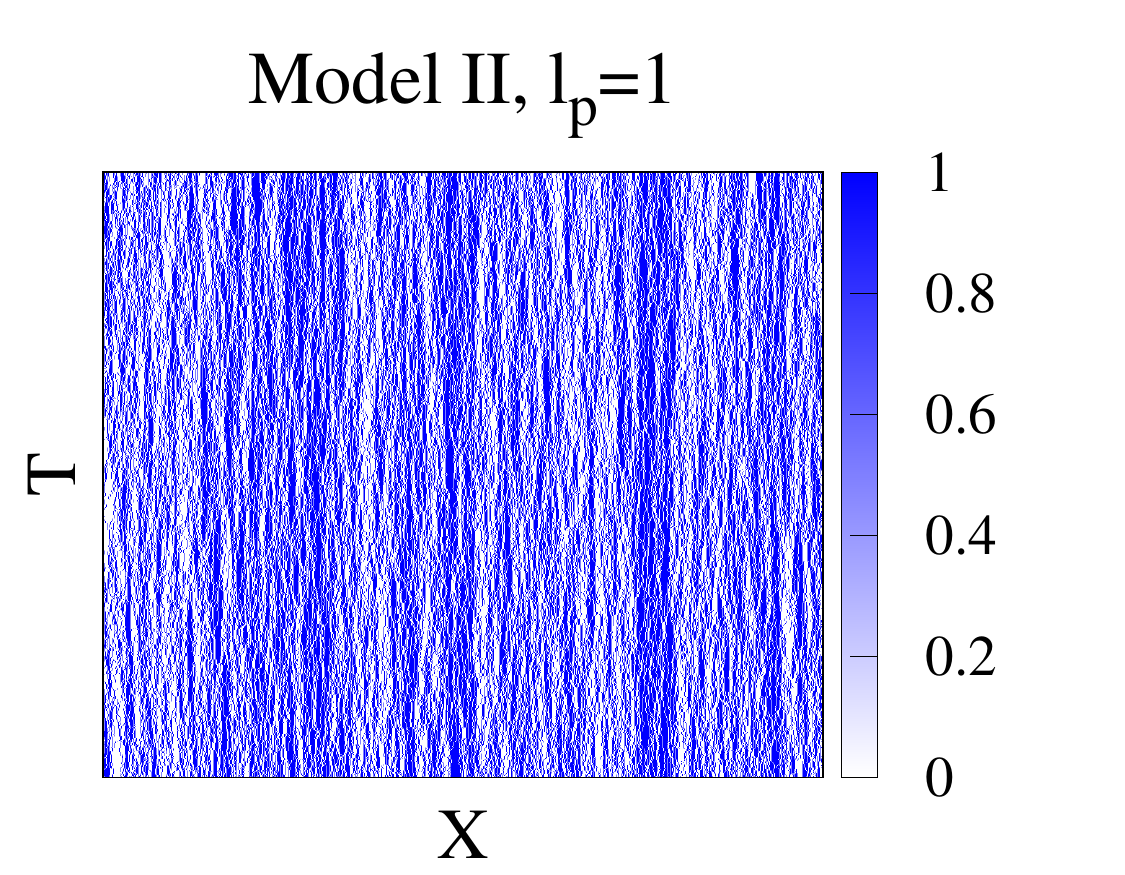}\hfill
            \includegraphics[width=0.25\textwidth]{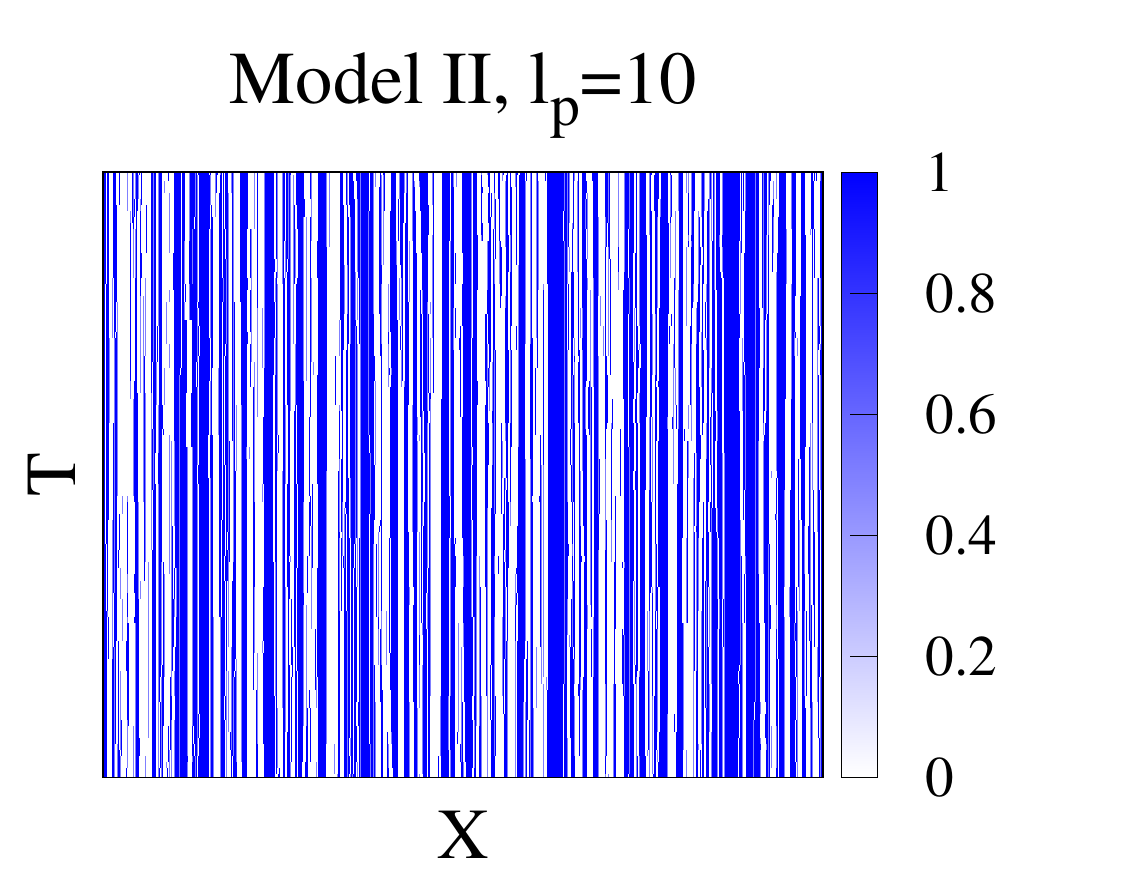}\hfill
            \includegraphics[width=0.25\textwidth]{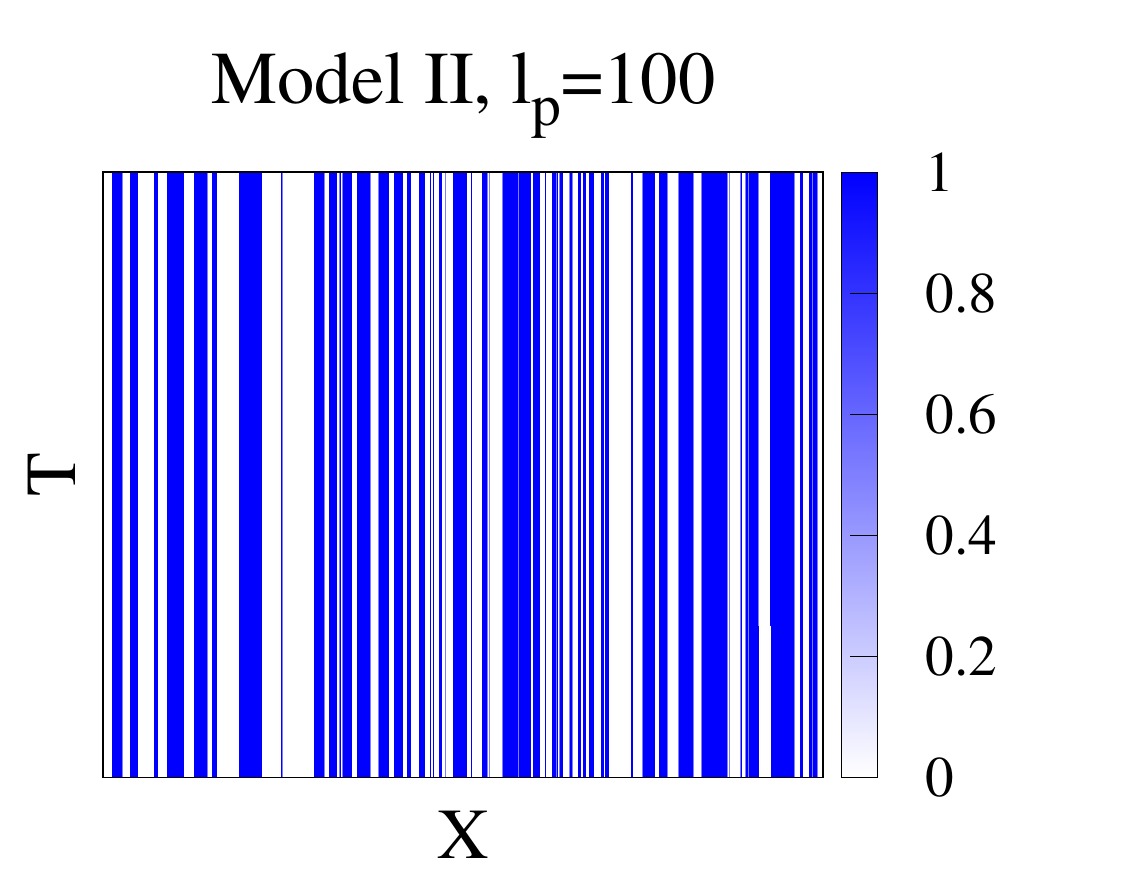}\hfill
            \includegraphics[width=0.25\textwidth]{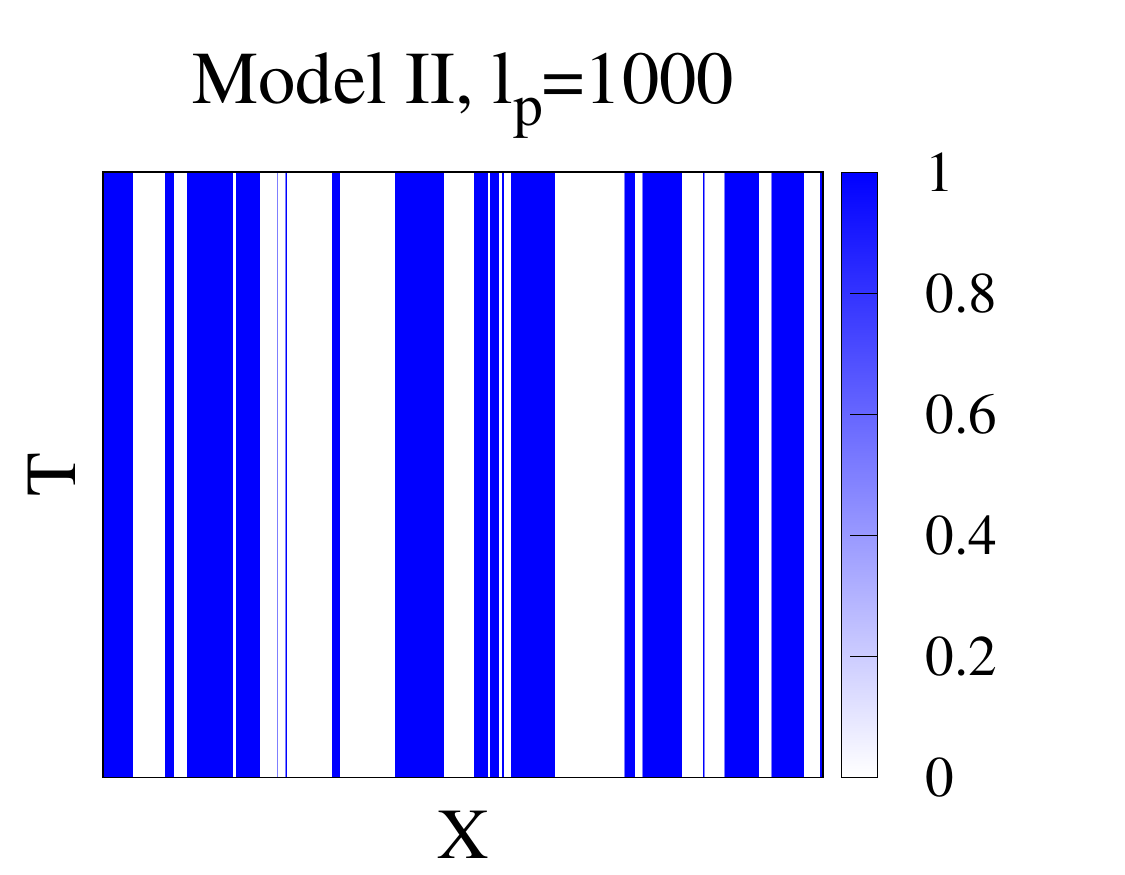}
            \caption{{\it Steady-state space-time trajectories of RTPs.--} We plot the steady-state space-time trajectories of run-and-tumble particles in models I (top panel) and II (bottom panel) at the same density $\rho=0.5$ and persistence lengths $l_p= \gamma^{-1} =1$, $10$, $100$, and $1000$. The black and red colors in model I represent particles with spin $s=1$ (right mover) and $s=-1$ (left mover), respectively, while white represents the vacancy or hole. In model II, however, blue denotes particles and white denotes vacancies or holes in the system. }
            \label{fig:space_time_trajectory}
  \end{figure*}

Clearly, the long-ranged hopping in model II mimics the ballistic ``run'' of standard RTPs, having exponentially distributed random ``run''-lengths with a typical persistence length $l_p$, and is expected to capture the collective behavior of model I (standard RTPs) on the persistence time scale $\tau_p$. Despite its simplicity, model II does not have a product-measure steady-state and generates nontrivial spatial correlations, resulting in clustering and large density fluctuations \cite{Tanmoy_PRE_2020}.
 In this particular context, we examine the space-time trajectories of models I and II as illustrated in Fig.~\ref{fig:space_time_trajectory}, by considering various persistence lengths $l_p$ (as shown in the figure). Here we perform the previously mentioned time scaling: the time axis of model II is multiplied by the persistence time $\tau_p=l_p/v$ (with $v=1$). The observed qualitative resemblances between these two models indicate that despite their distinct microscopic dynamics, models I and II are indeed connected through simple time-rescaling and share remarkably similar features (however, there is no exact one-to-one mapping between the two models). In the following sections, we calculate the bulk-diffusion coefficient and provide a detailed quantitative comparison between models I and II.

\section{Relevant length scales and a scaling theory}
\label{sec:scaling_argument}

Let us begin by discussing the relevant length scales in the strongpersistence regime in the context of model I (it is not difficult to extend the arguments to other variants of RTPs). We do not consider thermal diffusion, being irrelevant in the strong-persistence regime. Due to persistence, particles move ballistically with a characteristic speed $v$ (``run'') and then randomly change direction (``tumble'') with rate $\gamma = 1/\tau_p=v/l_p$ where $\tau_p$ and $l_p$ are the persistence time and length, respectively. Clearly, $l_p$ is a microscopic length scale induced by the persistence effect in the system. However, the persistent motion of individual particles is cut off by the presence of other particles along the direction of motion due to excluded-volume, or hardcore, interactions. Consequently, there is another microscopic length-scale induced by interaction - the mean {\it gap} $\langle g \rangle$ (equivalently, a ``mean free path'') between two consecutive particles {\it along the direction of motion}. The physics is essentially determined by how far, on average, the next particle is in the direction of a moving one; it makes no difference where all the neighboring particles are (as characterized by the inter-particle spacing). Indeed, $l_p$ and $\langle g \rangle$ are the only relevant length scales in the system, carrying the signatures of persistence and interaction, respectively. Therefore, from dimensional ground, any quantities are expected to be constructed by combining the two length scales $l_p$ and $\langle g \rangle$. This implies that the bulk-diffusion coefficient can be cast in a form $D=D^{(0)} {\cal F}\left( l_p/\langle g \rangle \right)$, where scaling function ${\cal F}(\psi)$ depends on a {\it single} dimensionless  scaling variable $\psi=l_p/\langle g \rangle=v/\langle g \rangle \gamma$ with the prefactor $D^{(0)} \sim \gamma l_p^{2}=v^2/\gamma$ being proportional to the effective diffusion coefficient of noninteracting RTPs. Furthermore, in the low-density limit $\rho \rightarrow 0$, as the average number of particles in any row (consisting of $L$ sites) is  equal to $L/\langle g \rangle$ and therefore $(L/\langle g \rangle) \times (L^{d-1}/a) = N$, we obtain an exact relation, 
\be 
\langle g \rangle = \frac{1}{\rho a},
\label{avg_g}
\ee 
where $a \sim r_0^{d-1}$ and $r_0$ are particle cross-section and diameter, respectively. The above relation immediately leads to Eq. (\ref{D_scaling}) where the scaling variable is given by $\psi=\rho a v/\gamma$ (now onwards, we put $r_0=1$, $a=1$ and $v=1$ on a hypercubic lattice of unit spacing).
The asymptotic form of ${\cal F}(\psi)$ is determined as follows. 
In {\it noninteracting} limit $\rho \ll 1$ and $\langle g \rangle \gg l_p$, we have $D \sim l_p^{2}/\tau_p \sim 1/\gamma$, implying ${\cal F}(\psi)=\rm{const.}$ as $\psi \rightarrow 0$. However, in {\it strongly interacting} limit $l_p \gg \langle g \rangle$, $D$ is proportional to $\gamma$ as microscopic events occur on a time scale $\tau_p = \gamma^{-1}$, implying ${\cal F}(\psi) \sim 1/\psi^2$ as $\psi \rightarrow \infty$. This explains the power-law dependence of $D$ on $\rho$. The above arguments, although plausible, must be validated, which we do next in both models I and II by obtaining their hydrodynamic descriptions.

\section{Hydrodynamics}\label{sec:hydrodynamics}

Hydrodynamics deals with large-scale spatiotemporal properties of slow variable(s). As particle number is conserved an we are interested in finite-$\gamma$ regime, particle density $\rho(\textbf{X}, t)=\langle \eta_{\textbf{X}}(t)\rangle$ is the only slow variable in the system. \textcolor{black}{Though, at small time scales $\mathcal{O}(\tau_p)$, the motion of a typical RTP is expected to be ballistic (given that the density is small), on time scales much longer than $\tau_p$ the RTP would undergo multiple tumblings, thus making the collective motion effectively diffusive. In that case, large-scale density relaxation should be governed by diffusive processes only} and, by using Eq.~\eqref{local_current} and continuity equation, we can immediately write the time-evolution of local density as
 \begin{eqnarray}\label{diffusion_equation_micro}
 \frac{\partial \rho(\textbf{X},t)}{\partial t} = \frac{\partial}{\partial \textbf{X}}.\left[D(\rho,\gamma) \frac{\partial \rho(\textbf{X},t)}{\partial \textbf{X}}\right].
\end{eqnarray}   
Note that, due to an explicit density dependence of the bulk-diffusion coefficient, the density relaxation is in fact governed by a {\it nonlinear} diffusion equation and has interesting consequences. Clearly, Eq.~\eqref{diffusion_equation_micro} is invariant under the scale transformation $\textbf{X} \rightarrow \lambda \textbf{X}$ and $t \rightarrow \lambda^{2} t$. Therefore, on large spatio-temporal scales, i.e., typically space $\sim {\cal O}(L)$ and time $\sim {\cal O}(L^{2})$, the time-evolution equation as in Eq.~\eqref{diffusion_equation_micro} is reduced to the following diffusion equation for the coarse-grained density variable $\rho(\textbf{x},\tau)$,
\begin{eqnarray}\label{diffusion_equation_macro}
\frac{\partial \rho(\textbf{x},\tau)}{\partial \tau}=\nabla. \left[  D(\rho,\gamma) \nabla \rho(\textbf{x},\tau) \right], 
\end{eqnarray}
where we transform the density field as $\rho(\textbf{X}, t) \equiv \rho\left(\textbf{x}={\textbf{X}}/{L},\tau={t}/{L^{2}}\right)$, which is now a function of the coarse-grained space $\textbf{x}=\textbf{X}/L$ and time $\tau = t/L^{2}$ variables.
In the subsequent section, we verify the above hydrodynamic description by explicitly calculating the density- and tumbling-rate-dependent bulk-diffusion coefficient $D(\rho,\gamma)$ for both models I and II. To calculate $D(\rho, \gamma)$, we develop a microscopic theory for model II (LLG) and an efficient Monte Carlo algorithm for model I.

\subsection{Theory for model II (LLG)}

In this section, we describe the details of our analytical calculation scheme, which, for simplicity, are performed in one dimension, unless mentioned otherwise; generalization to higher dimensions is straightforward and the final results are given for arbitrary dimension $d$. 

An advantage of dealing with model II  is that the system possesses a \textit{gradient property} \cite{Krapivsky_2014}, which allows for the local instantaneous current to be expressed as the gradient of a local observable; thus the bulk-diffusion coefficient can be exactly determined in terms of the steady-state gap distribution $P(g)$. We begin our calculation by introducing the cumulative, or time-integrated, bond-current $Q_{X}(t)$, which measures the net particle current across  bond $[X, X+1]$  in a time interval $t$. Notably, the stochastic variable $Q_{X}(t)$ is a quantity, which can be measured in simulations quite efficiently. However, this particular quantity is related the instantaneous current $J_X(t)$ in the following manner, 
\begin{eqnarray}\label{inst_current_defn}
J_X(t)=\lim_{\Delta t \rightarrow 0}\frac{\Delta Q_{X}}{\Delta t},
\end{eqnarray}  
where $\Delta Q_{X}(t)=\int_{t}^{t+\Delta t} J_X(t) dt$ is the time-integrated current in the time interval $\Delta t$. 
In this section, we characterize the behavior of average bond current to calculate the bulk-diffusion coefficient for model II.

Note that, whenever a particle makes a long-range hop of length $l$ and reach its destination site, it travels the entire stretch of vacant sites of that length. As a result, for a rightward (leftward) hop, current across {\it all} bonds in that stretch increases (decreases) by unity. Accordingly, the continuous-time evolution for the time-integrated current $Q_X(t)$ in an infinitesimal time interval $[t, t+dt]$ can be written as
\begin{eqnarray} 
 Q_X(t+dt) = 
\left\{
\begin{array}{ll}
\vspace{0.15 cm}
 Q_X(t)  + 1,            ~~~  & {\rm prob.}~~~ \mathcal{P}^{R}_{X}(t) dt , \\
\vspace{0.15 cm}
 Q_X(t) - 1,            ~~~  & {\rm prob.}~~~ \mathcal{P}^{L}_{X}(t) dt, \\
 \vspace{0.15 cm}
 Q_X(t),                ~~~  & {\rm prob.}~~~~  1 - (\mathcal{P}^{R}_{X} + \mathcal{P}^{L}_{X}) dt, \\
\end{array}
\right.
\label{Q_update_eq}
\end{eqnarray}
where $\mathcal{P}^{R}_{X} dt$ and $\mathcal{P}^{L}_{X} dt$ are the probabilities of the corresponding hopping events; the explicit identification of the stochastic variables $\mathcal{P}^{R}_{X}$ and $\mathcal{P}^{L}_{X}$ are discussed below.
By using the above microscopic update rules, average instantaneous current $\left \langle J_{X}(t) \right \rangle$ can be immediately written as
\begin{eqnarray}\label{time-derivative-int-current_1}
\left \langle J_{X}(t) \right \rangle  &=& \frac{d \left \langle Q_{X}(t) \right \rangle}{dt} = \left \langle \mathcal{P}^{R}_{X}(t) \right\rangle - \left \langle \mathcal{P}^{L}_{X}(t) \right\rangle.
\end{eqnarray}
We now define  the following stochastic variables, which will be required in the subsequent calculations,
  \begin{eqnarray}
 \mathcal{U}_{X+l}^{(l)} &\equiv& \overline{\eta}_{X+1} \overline{\eta}_{X+2} \dots \overline{\eta}_{X+l} , \\ 
 \mathcal{V}_{X+l+1}^{(l+2)} &\equiv& \eta_{X}\overline{\eta}_{X+1} \overline{\eta}_{X+2} \dots \overline{\eta}_{X+l}\eta_{X+l+1} , 
 \end{eqnarray}
 where $\bar{\eta}_X=(1-\eta_X)$, $\mathcal{U}^{(l)}$ is an indicator function of $l$ consecutive sites being vacant and $\mathcal{V}^{(l+2)}$ is that of a vacancy cluster of size $l$. Note that, in Eq.~\eqref{Q_update_eq}, rightward hopping events increases $Q_{X}(t)$ by $1$, and the corresponding probability term is given by $\mathcal{P}^{R}_{X}(t) dt$. Therefore, one can identify the stochastic variable $\mathcal{P}^{R}_{X}(t)$ by simply considering the rightward hopping and accordingly computing the probability of increment in current $Q_X(t)$. Now, depending on hop-length $l$ and gap size $g$, we must consider the following two possibilities.

 \textit{Case I. $l > g$.--}  In this case, a particle crosses the entire empty stretch (a cluster of vacancies) of size $g$ during an infinitesimal time $dt$, and consequently the current across all $g$ bonds is increased by unity. Therefore, for the increment in $Q_{X}(t)$, caused by the above hopping event, the vacancy cluster must contain the bond $(X, X+1)$ itself. However, we can keep a specific bond $(X, X+1)$ inside a hole cluster of size $g$ by translating the entire cluster in $g$ different ways, each of which corresponds to a unit increment of $Q_{X}(t)$. Therefore the probability of increment in current by unit amount  is given by $\mathcal{P}^{>,R}_{X} dt$, with
\begin{eqnarray}
\mathcal{P}^{>,R}_{X}(g) &\equiv&  \frac{1}{2}\sum_{k=1}^{g}  \eta_{X+k-g}(t) \overline{\eta}_{X+k-g+1} \dots \overline{\eta}_{X+k} \eta_{i+k+1},  \nonumber \\ &=& \frac{1}{2} \sum_{k=1}^{g} \mathcal{V}_{X+k+1}^{(g+2)}.
\end{eqnarray}

\textit{Case II. $l \leq g$.--} In this case, during time $dt$, the particle hops rightward by $l$ units, thus increasing current by unity  across all the $l$ bonds. Consequently, $Q_{X}(t)$ increases by unity if the bond $(X, X+1)$ is a part of these $l$ bonds. Similar to case I, the current $Q_{X}(t)$  increases by unity in $l$ possible ways and the corresponding probability of the increment is given by $\mathcal{P}^{\leq,R}_{X}(t) dt$, where
   \begin{eqnarray}
    \mathcal{P}^{\leq,R}_{X}(l) &\equiv& \frac{1}{2} \sum_{k=1}^{l}  \eta_{X+k-l}\overline{\eta}_{X+k-l+1} \overline{\eta}_{X+k-l+2} \dots \overline{\eta}_{X+k},  \nonumber
    \\ &=& \frac{1}{2} \sum_{k=1}^{l} \left(\mathcal{U}_{X+k}^{(l)} - \mathcal{U}_{X+k}^{(l+1)} \right).
\end{eqnarray}
  Therefore, by combining cases I and II, the total probability that $Q_{X}(t)$ increases by unity during time interval $dt$, can be written as $\mathcal{P}^{R}_{X} dt$, where
\begin{eqnarray}
  \label{pr}
  \mathcal{P}^{R}_{X} &\equiv& \sum_{l=1}^{\infty} \phi(l) \left[\mathcal{P}^{\leq,R}_{X}(l) + \sum_{g=1}^{l-1} \mathcal{P}^{>,R}_{X}(g) \right],  \nonumber
  \\
  &=& \frac{1}{2}\sum_{l=1}^{\infty} \phi(l) \left[\sum_{k=1}^{l} \left(\mathcal{U}_{X+k}^{(l)} - \mathcal{U}_{X+k}^{(l+1)}\right) + \sum_{g=1}^{l-1}\sum_{k=1}^{g}\mathcal{V}_{X+k+1}^{(g+2)}\right]. \nonumber \\
\end{eqnarray}
Similarly the total probability of current $Q_{X}(t)$ decreasing by unity, during time interval $dt$, due to leftward hopping can be written as $\mathcal{P}^{L}_{X} dt$, where
\begin{eqnarray}\label{pl}
\mathcal{P}^{L}_{X} \equiv \frac{1}{2}\sum_{l=1}^{\infty} \phi(l) \left[\sum_{k=1}^{l} \left(\mathcal{U}_{X+k-1}^{(l)} - \mathcal{U}_{X+k}^{(l+1)}\right) + \sum_{g=1}^{l-1}\sum_{k=1}^{g}\mathcal{V}_{X+k}^{(g+2)}\right]. \nonumber \\
\end{eqnarray}
By substituting $\mathcal{P}^{R}_{X}(t)$ and $\mathcal{P}^{L}_{X}(t)$ in Eq.~\eqref{time-derivative-int-current_1}, the average instantaneous current can be immediately obtained as
\begin{eqnarray}
  \label{time-derivative-int-current_2}
\left \langle J_{X}(t) \right \rangle &=& \frac{1}{2}\sum_{l=1}^{\infty}\phi(l)\left[\sum_{g=1}^{l-1}\Big( \langle \mathcal{V}_{X+g+1}^{(g+2)} \rangle - \langle \mathcal{V}_{X+1}^{(g+2)} \rangle \right) \nonumber \\ && \hspace{2.75 cm} + \left( \langle \mathcal{U}_{X+l}^{(l)} \rangle -\langle \mathcal{U}_X^{(l)}\rangle \Big)\right].
\end{eqnarray}
Note that, in the above equation, $\langle J_{X}(t) \rangle$ is written as a (generalized) gradient of the observables $\langle \mathcal{V}^{(g+2)}\rangle$ and $\langle\mathcal{U}^{(l)} \rangle $. Now, on a long time scale, we assume that the system will reach a \textit{local equilibrum-like state}; as a result, the time-dependent ``fast'' observables $\langle \mathcal{V}_{X}^{(g+2)}\rangle(t)$ and $\langle\mathcal{U}_{X}^{(l)}\rangle(t)$ are effectively a function of only  the local density  $\rho(X, t)= \langle \eta_{X}(t) \rangle$, which varies slowly in time. To this end, by substituting $\langle \mathcal{V}_{X}^{(g+2)} \rangle(t) \equiv \langle \mathcal{V}^{(g+2)}\rangle [\rho(X, t)]$ and $\langle \mathcal{U}_{X}^{(l)}\rangle(t) \equiv \langle \mathcal{U}^{(l)}\rangle[\rho(X,t)]$, and then by performing Taylor series expansion in Eq.~\eqref{time-derivative-int-current_2}, we obtain
\begin{equation}\label{time-derivative-int-current_3}
\left \langle J_{X}(t) \right \rangle =\frac{1}{2}\sum_{l=1}^{\infty}\phi(l)\frac{\partial}{\partial \rho}\left[\sum_{g=1}^{l-1}g \langle \mathcal{V}^{(g+2)} \rangle + l \langle \mathcal{U}^{(l)} \rangle \right]\frac{\partial \rho}{\partial X},
\end{equation}
Now, by using the expression of $\langle J_{X}(t) \rangle$, the time-evolution equation of local-density $\rho(X,t)$ can be written as  continuity equation,
\begin{eqnarray}\label{rho_evolution_LLG}
\frac{\partial \rho}{\partial t} &=& \left \langle J_{X-1}(t) \right \rangle - \left \langle J_{X}(t) \right \rangle \\ &\equiv& \frac{\partial}{\partial X} \left[ D_{II}(\rho,\gamma) \frac{\partial \rho}{\partial X}  \right],
\end{eqnarray}
where $D_{II}(\rho,\gamma)$ is the desired bulk-diffusion coefficient for model II in one dimension, and is given by
\begin{eqnarray}
D_{II}(\rho,\gamma) = -\frac{1}{2}\sum_{l=1}^{\infty}\phi(l)\frac{\partial}{\partial \rho}\left[\sum_{g=1}^{l-1}g \langle \mathcal{V}^{(g+2)} \rangle + l \langle \mathcal{U}^{(l)} \rangle \right]. \nonumber \\ 
\end{eqnarray}
However, in arbitrary (finite) dimensions $d$, the above expression can be written as
\begin{eqnarray}
D_{II}(\rho,\gamma) = -\frac{1}{2d}\sum_{l=1}^{\infty}\phi(l)\frac{\partial}{\partial \rho}\left[\sum_{g=1}^{l-1}g \langle \mathcal{V}^{(g+2)} \rangle + l \langle \mathcal{U}^{(l)} \rangle \right]. \nonumber \\
\end{eqnarray}
Note that, the above equation expresses $D_{II}(\rho,\gamma)$ in terms of the correlation functions $\langle \mathcal{V}^{(g+2)} \rangle$ and $\langle \mathcal{U}^{(l)} \rangle$.
Now, identifying the correlators as a function of gap distribution function $P(g)$,
\begin{eqnarray}
\langle \mathcal{V}^{(g+2)}(\rho, \gamma) \rangle &=&\rho P(g), \\
\langle \mathcal{U}^{(l)}(\rho, \gamma) \rangle &=& \rho \sum_{g=l-1}^{\infty} (g-l+1) P(g),
\end{eqnarray}
we can explicitly write $D_{II}(\rho,\gamma)$ in terms of $P(g)$,
\begin{widetext}
\begin{equation}\label{bulk-diffusivity-LLG}
 D_{II}(\rho,\gamma)=-\frac{1}{2d}\frac{\partial}{\partial \rho}\left[\rho \sum_{l=1}^{\infty} \phi(l) \left(\sum_{g=1}^{l-1} g P(g) + l\sum_{g=l-1}^{\infty} (g-l+1) P(g)  \right)  \right].
 \end{equation}
\end{widetext}
Of course, explicit calculation of $P(g)$ is difficult in general. However, $P(g)$ for large $g$ is expected to be an exponential, {$P(g) \simeq N_* e^{-g/g_*},$} where $N_*(\rho,\gamma)$ and  $g_*(\rho,\gamma)$ are the proportionality constant and typical gap size, respectively; $N_*$ is determined from the normalization condition $\sum_gP(g)=1$ and Eq. (\ref{avg_g}). Now, using the above form of gap distribution $P(g)$ in Eq.~\eqref{bulk-diffusivity-LLG}, we obtain the  expression of the bulk-diffusion coefficient,
\begin{eqnarray}\label{bulk-diffusivity-LLG_most_general}
D_{II}(\rho,\gamma) = -\frac{B}{2d} \frac{\partial}{\partial \rho} \left[(1-\rho)\Bigg\{\frac{1}{e^{1/l_p}-1} +  \frac{1}{(e^{1/\xi}-1)^{2}} \Bigg\}\right],\nonumber \\
\end{eqnarray}
where we determine the following length scale,
\begin{eqnarray}
\xi=\frac{1}{l_p^{-1}+g_{*}^{-1}}.
\end{eqnarray}
Notably, the above form of $D_{II}(\rho,\gamma)$ is valid for arbitrary $\rho$ and $\gamma$. In the subsequent analysis, we consider the following special cases.

\subsubsection*{Case I: $\rho$ arbitrary and \textcolor{black}{$\gamma$ large}, i.e., $l_p \ll \langle g \rangle $}

In the small-persistence limit, density correlations vanish in the thermodynamic limit. As a result, the steady-state distribution is a product measure and $P(g) \sim (1-\rho)^{g} \simeq e^{-g/g_*}$, where $g_*=-1/\log (1-\rho)$.  By using $g_*$ in Eq.~\eqref{bulk-diffusivity-LLG_most_general}, and by setting the condition $\gamma \gg 1$ and $\rho$ finite, we obtain
\begin{eqnarray}\label{D_ssep_limit}
D_{II}(\rho,\gamma) \simeq  \frac{e^{-\gamma}}{2d} =e^{-\gamma} D_{SSEP}.
\end{eqnarray}
\textcolor{black}{Note that, the above expression is independent of density and, upon an appropriate scaling of time ($t \rightarrow t'=te^{\gamma}$), reduces to the bulk-diffusion coefficient $D_{SSEP}=1/2d$ for the symmetric simple exclusion process (SSEP); see Appendix B for details.} The appearance of the exponential prefactor $e^{-\gamma}$ in the above equation can be explained as following. In this case, the distribution $\phi(l)$ has the most weight for $l=0$, while all other hop-lengths, i.e., $l >0$, have an exponentially smaller probability $1-\phi(0)=e^{-\gamma}$. Furthermore, among these nonzero hop-lengths, clearly the nearest-neighbor or unit-distance hop with $l=1$ dominates, and contributions from larger $l$ are negligible in the relaxation process. As a result, in the limit $\gamma \gg 1$ or, equivalently, $l_p \ll 1$, particles effectively perform SSEP-like dynamics with an exponentially small rate, $e^{-\gamma}$, immediately explaining the prefactor in Eq.~\eqref{D_ssep_limit}.

\subsubsection*{Case II: High density $\rho \rightarrow 1$ and small tumbling rate $\gamma \rightarrow 0$}

In this regime of strong persistence and large density, persistence length $l_p$ is much larger than the mean gap-size $\langle g \rangle$. Although it is easy to calculate $g_*$ numerically, its analytical form is known only  in one dimension \cite{subhadip_PRE_2021} and it has been calculated to be $g_*=\sqrt{\langle g \rangle/\gamma}$ with $\langle g \rangle = 1/\rho -1$ - the average gap for arbitrary density $\rho$ ($\langle g \rangle$ can be calculated by using arguments in Sec.~\ref{sec:scaling_argument}. By substituting this expression of $g_*$ and the condition $g_* \ll l_p \rightarrow \infty$ in Eq.~\eqref{bulk-diffusivity-LLG_most_general}, we obtain the bulk-diffusion coefficient,
\begin{eqnarray}
D_{II}(\rho,\gamma) \simeq \frac{1}{2\rho^{2}};
\end{eqnarray}
see Appendix B for details.
It is interesting to note that, in this parameter regime, $D_{II}(\rho,\gamma)$ is independent of $\gamma$, due to an effective rescaling of time in model II, and it has a power-law behavior as a function of density $\rho$.

\subsubsection*{Case III: $\rho \rightarrow 0$ and $\gamma \rightarrow 0$}

In this  case, both the length scales $l_p$ and $\langle g \rangle$ are diverging, i.e., $l_p \sim \langle g \rangle \rightarrow \infty$, and are competing with each other. Since, they are the only relevant length scales, one may immediately construct a dimensionless quantity $\psi=l_p/\langle g \rangle$, which must be the parameter to characterize any physical quantities in the system (see Sec~\ref{sec:scaling_argument}). Then the typical gap size should have a scaling form, 
\begin{eqnarray}\label{g_*}
g_*\simeq \frac{1}{\rho}\mathcal{G}(\psi),
\end{eqnarray}
with $\psi=\rho v/\gamma=l_p \rho$; these assertions are verified in simulations, see Appendix A for details. The prefactor $1/\rho$ is fixed from the fact that in the limit $\psi \rightarrow 0$, the system reduces to a noninteracting one and therefore ${\cal G}(0)=1$. Now, using the above form of $g_*$ in Eq.~\eqref{bulk-diffusivity-LLG_most_general} and putting $B=1/l_p$, we obtain
 \begin{equation}\label{bulk_diff_LLG_final}
 D_{II}(\rho,\gamma)=\frac{ e^{1/\xi}}{l_p d(e^{1/\xi}-1)^{3}}\frac{1}{\mathcal{G}(\psi)}\left(1 - \psi \frac{\mathcal{G}'(\psi)}{\mathcal{G}(\psi)} \right),
 \end{equation}
where $\xi$ is given by
\begin{eqnarray}
\xi=\frac{l_p}{1+\psi/\mathcal{G}(\psi)}.
\end{eqnarray}
Furthermore, using the fact that $\xi \gg 1$ for large $l_p$, we perform an asymptotic analysis of Eq.~\eqref{bulk_diff_LLG_final}, leading to $D_{II}(\rho,\gamma)$ which satisfies the following scaling law,
\begin{eqnarray}\label{D_scaling_relation}
D_{II}(\rho,\gamma)/D^{(0)}_{II} \equiv \mathcal{F}_{II}(\psi),
\end{eqnarray}
\textcolor{black}{where $D^{(0)}_{II}= \langle l^{2} \rangle/2d = l_p^{2}/d$ is the diffusion coefficient calculated using the central limit theorem in the noninteracting limit (i.e., the single-particle case),} and $\mathcal{F}_{II}(\psi)$ is the desired scaling function, which is given by
 \begin{equation}\label{D_scaling_function_G(psi)}
 \mathcal{F}_{II}(\psi)=  \frac{\mathcal{G}^{2}(\psi)}{(\mathcal{G}(\psi)+\psi)^{3}}\left(1 - \psi \frac{\mathcal{G}^{\prime}(\psi)}{\mathcal{G}(\psi)} \right)
 \end{equation} 
 see Appendix B for details.

For the purpose of illustration, we consider below model II in one dimension. In the limit of $\psi$ large, as shown in \cite{subhadip_PRE_2021}, the typical gap size in in one dimension is given by $g_* = \sqrt{\psi}/\rho$. Thus, by combining the two limiting behaviors for small and large $\psi$, we can simply write $\mathcal{G}(\psi) \simeq (1+\psi)^{1/2}$, which is then substituted in Eq.~(\ref{D_scaling_function_G(psi)}) to explicitly obtain the scaling function,
 \begin{equation}\label{D_scaling_function}
  \mathcal{F}_{II}(\psi)=\frac{(2+\psi)}{2(\psi + \sqrt{1+\psi})^{3}}.
 \end{equation}
  Note that, for $\psi \ll 1$,  ${\cal F}_{II}(\psi)=1$ and consequently $D_{II} \simeq l_p^2$, which diverges when the persistence length $l_p \rightarrow \infty$; for $\psi \gg 1$, ${\cal F}_{II}(\psi) \simeq 1/2\psi^2$ and $D_{II} \simeq 1/2\rho^2$. Interestingly, expanding $\mathcal{F}_{II}(\psi)$ around $\psi=0$, we obtain
  \begin{eqnarray}
   \mathcal{F}_{II}(\psi) \simeq 1 - 4 \psi - \mathcal{O}(\psi^{2}).
  \end{eqnarray}
  Now, putting back all the dimensional factors explicitly in the expression of the bulk-diffusion coefficient in Eq.~\eqref{D_scaling_relation}, we get
  \begin{eqnarray}
  D_{II}(\rho,\gamma) \simeq D^{(0)}_{II}\left( 1 - \frac{\rho}{\rho_{*}} \right),
  \end{eqnarray}
where we find a characteristic density $\rho_{*}=1/4al_p$ and $a$ is the particle cross-section; although the expansion is derived in one dimension, the above form of bulk-diffusion coefficient should be valid in any dimension with $\rho_{*} \sim 1/al_p$. Now, if one assumes that the above low-density approximation $(\rho \ll \rho_{*})$ is valid also in the high-density regime, it would imply that there would be a diffusive instability in the system for $\rho > \rho_{*}$, where the bulk-diffusion coefficient would become negative. However, this is actually not the case as, in the higher density regime $(\rho \gg \rho_{*})$, the bulk-diffusion coefficient gradually crosses over to a power-law form,  $D_{II} \sim \rho^{-\alpha}$ with $\alpha>0$, [see  Eq.\eqref{power_law_D}] and never becomes zero in the range $0 < \rho < 1$.

\subsection{Numerical scheme for model I}
\label{sec:bulk-diff_RTPs}

\begin{figure}
           \centering
            \includegraphics[width=0.5\textwidth]{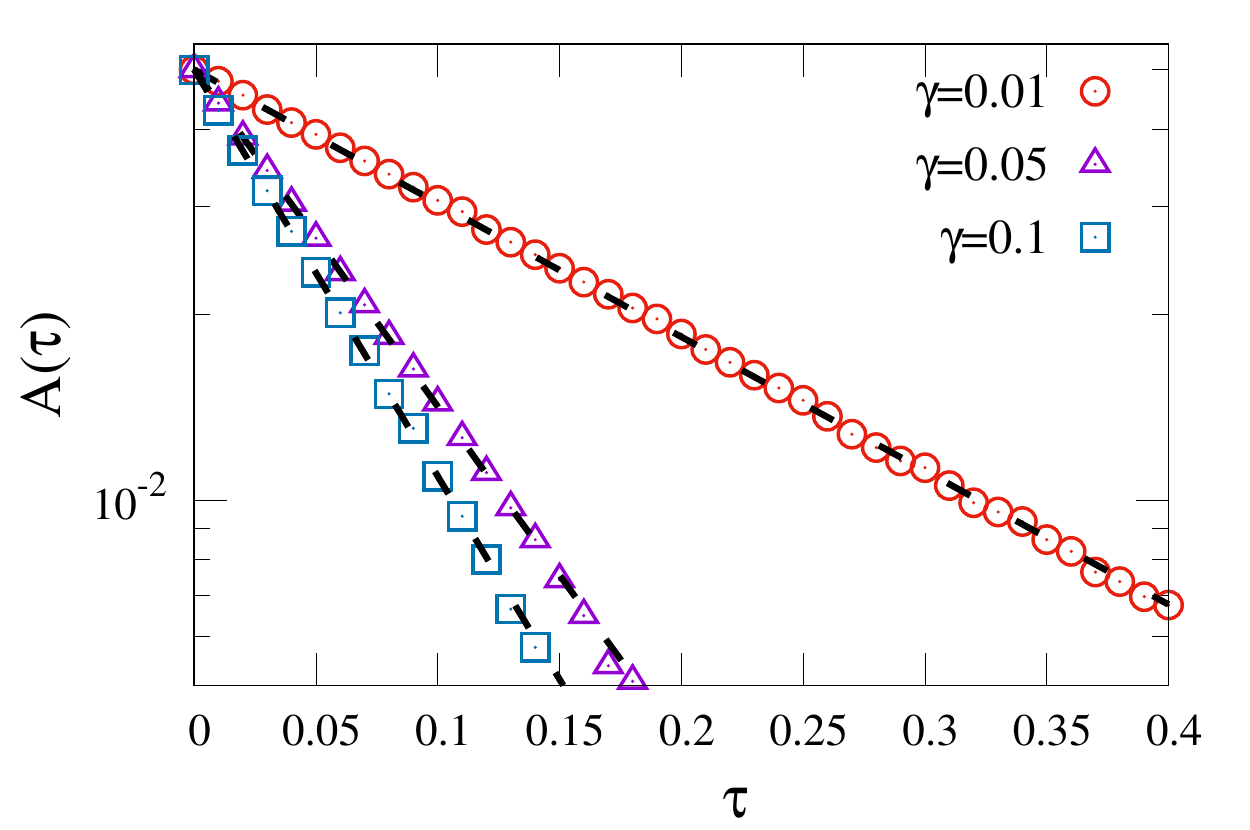}
          \caption{\textit{Verification of Eq.~\eqref{A(t)_vs_t} and determination of relaxation rates.--} We plot the perturbation amplitude $A(\tau)$, as a function of the hydrodynamic time $\tau$, for tumbling rates $\gamma= 0.01$ (red circle), $0.05$ (magenta triangle) and $0.1$ (blue square). The dotted black lines are the best fitted exponential functions with relaxation rates $\Gamma = 5.03$, $12.82$ and $15.24$ for $\gamma = 0.01$, $0.05$ and $0.1$, respectively. For both these panels, we
have fixed $\rho_0= 0.5$ and $A(0) = 0.05$.}
\label{fig:amp_decay}
  \end{figure}

According to our scaling argument, for model I too, the scaled bulk-diffusivity $D_{I}/D^{(0)}_{I} = {\cal F}_{I}(\rho v/\gamma)$ should be expressed in terms of a scaling function, which we check next. As an analytic calculation of $D_{I}(\rho, \gamma)$ for arbitrary densities and tumbling rates is not possible at this stage, we resort to an efficient numerical technique described below \cite{katz1984nonequilibrium}.
We discuss the numerical method in the context of a one dimensional system and the method can be immediately generalized to higher dimensional ones. We study relaxation of a long-wave-length - typically sinusoidal - initial density perturbation $\rho(X,0) = \rho_{in}(X/L)$ with wave number $2\pi/L$,
\begin{eqnarray}
\label{sinusoidal_initial_condition}
\rho(X,0) = \rho_0 + A(0) \sin \left(\frac{2 \pi X}{L} \right),
\end{eqnarray}
i.e., $\rho_{in}(x) = \rho_0 + A(0) \sin (2 \pi x)$, where $X= x L$ is the lattice position and $\rho_0$ is the global density, around which the perturbation is applied. For higher dimensions ($d>1$), perturbation is applied in one spatial direction, while the density profile is kept uniform in all other direction. On long-time scales $t \sim {\cal O} (L^{2}) \gg 1/\gamma$, the memory of short-time ballistic motion is washed out after innumerable tumbling events and the system is expected to be diffusive. As a result,  on the coarse-grained (macroscopic) space $x=X/L$ and time $\tau=t/L^{2}$ scales, the density profile $\rho(x, \tau)$ must evolve according to the nonlinear diffusion equation as given in Eq.~\eqref{diffusion_equation_macro}. In the limit of weak perturbation $A(0)\ll \rho_0 $, we can simply take $D_{I}[\rho(x, \tau),\gamma] \simeq D_{I}(\rho_0,\gamma)$, and thus Eq.~\eqref{diffusion_equation_macro} can be written as
\begin{equation}\label{linear_diff_eqn_RTPs}
\frac{\partial \rho(x,\tau)}{\partial \tau}=D_{I}(\rho_0,\gamma) \frac{\partial^{2} \rho(x,\tau)}{\partial x^{2}}.
\end{equation}
The above diffusive scaling limit will be verified shortly in direct Monte Carlo simulations of model I. Now, by using Eq.~\eqref{linear_diff_eqn_RTPs}, the density perturbation at later time can be straighforwardly calculated to have the following form: $\delta \rho(x,\tau) =\rho(x,\tau) - \rho_0 = A(\tau) \sin (2 \pi x)$, with the amplitude 
\begin{eqnarray}\label{A(t)_vs_t}
A(\tau) = A(0) e^{-\Gamma \tau},
\end{eqnarray}
where $\Gamma(\rho_0,\gamma)$ is the density and tumbling rate dependent  relaxation rate and is given by
\begin{eqnarray}\label{relaxation_rate}
\Gamma(\rho_0,\gamma)=4 \pi^{2} D_{I}(\rho_0,\gamma).
\end{eqnarray}
In simulations, we calculate the average excess number of particles $\Delta(\tau)$ in the first half of the system, which is directly related with the amplitude $A(\tau)$, where we have
$\Delta(\tau) = L^{d}\int_{0}^{1/2}\delta \rho(x,\tau) dx=L^{d}\pi A(\tau)$. Upon obtaining $\Delta(\tau)$ [hence $A(\tau)$] at various $\tau$, we measure the relaxation rate and calculate the bulk-diffusion coefficient from the relation in Eq.~\eqref{relaxation_rate}. To ensure diffusive relaxation, one should first take the thermodynamic limit  ($L, N \rightarrow \infty$ with $\rho_0=N/L$ fixed) and then vary the tumbling rate $\gamma$ and analyze the case of small $\gamma \ll 1$. We validate Eq.~\eqref{A(t)_vs_t} in one dimension by plotting the numerically obtained $A(\tau)$ as a function of $\tau$ in Fig.~\ref{fig:amp_decay} for various tumbling rates $\gamma=0.01$, $0.05$, and $0.1$ at density $\rho_0=0.5$ and system size $L=1000$. Indeed, we find $A(\tau)$ to be a simple exponential as predicted by Eq.~\eqref{A(t)_vs_t}. Fitting the simulation data with Eq.~\eqref{A(t)_vs_t} yields the corresponding relaxation rates $\Gamma$, which, then by using in  Eq.~\eqref{relaxation_rate}, we finally obtain the bulk-diffusion coefficient for model I for the corresponding values of density and tumbling rate.

\subsubsection*{Numerical verification of diffusive scaling}

 \begin{figure}
           \centering
           \includegraphics[width=0.495\textwidth]{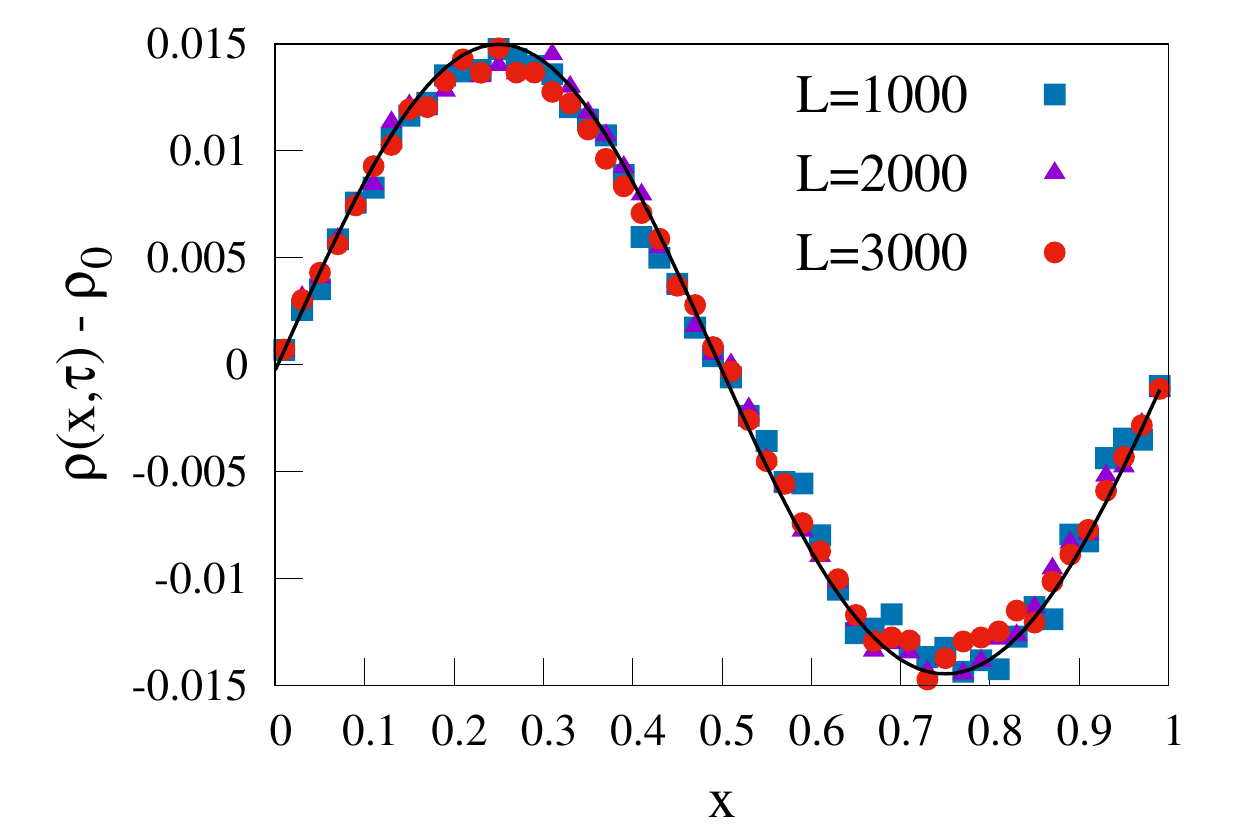}\hfill
        \includegraphics[width=0.495\textwidth]{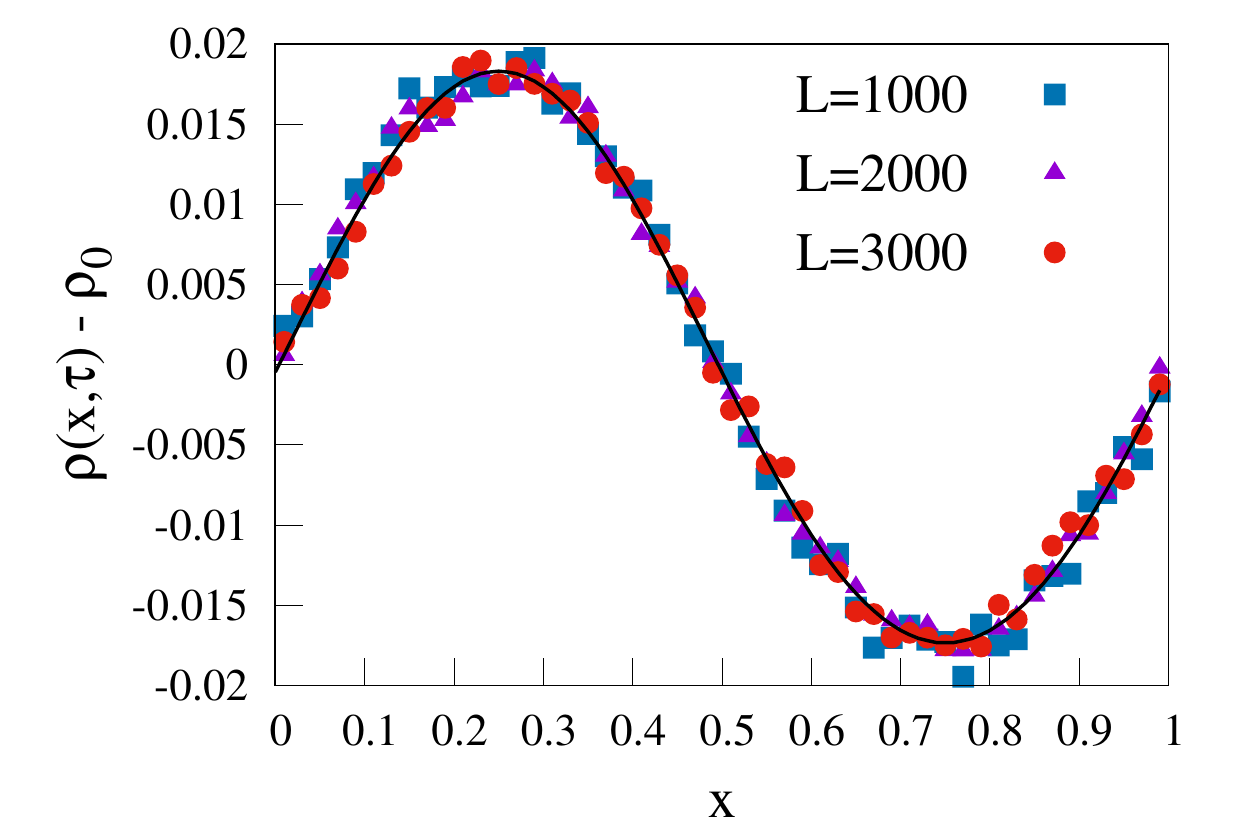}
          \caption{\textit{Verification of diffusive scaling in model I (standard hardcore RTPs).--} We plot the excess density field $\delta \rho(x,\tau) = \rho(x,\tau)-\rho_0$ for $\gamma=0.05$ (top-panel) and $0.01$ (bottom-panel), obtained from simulation, as a function of the scaled position $x=X/L$ at hydrodynamic times $\tau=0.1$ and $0.2$, respectively. We relax the system with a sinusoidal initial condition defined in Eq.~\eqref{sinusoidal_initial_condition} with $\rho_0=0.5$ and $A(0)=0.05$. Corresponding lines are obtained by numerically integrating the hydrodynamic equation Eq.~\eqref{diffusion_equation_macro} using the obtained bulk-diffusion coefficients.}
           \label{fig:diffusive_scaling_limit}
  \end{figure}

\begin{figure*}
          \centering    
     \includegraphics[width=0.249\linewidth]{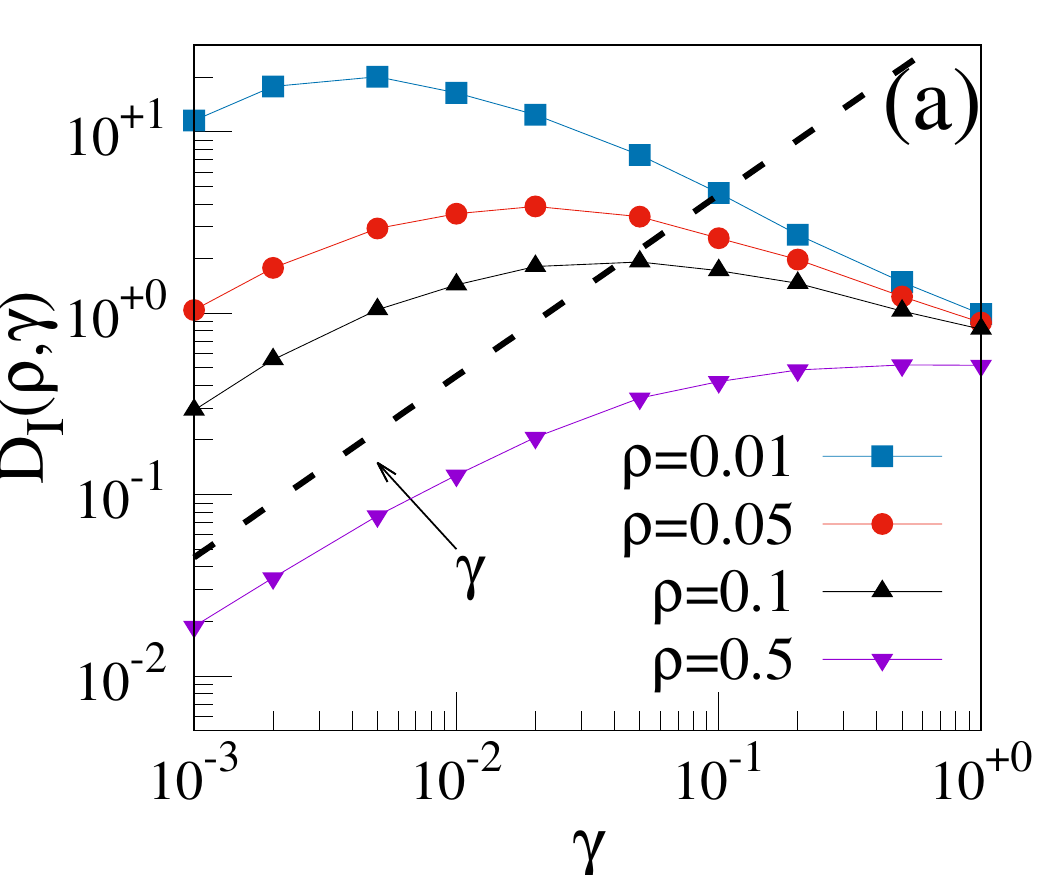}\hfill
     \includegraphics[width=0.249\linewidth]{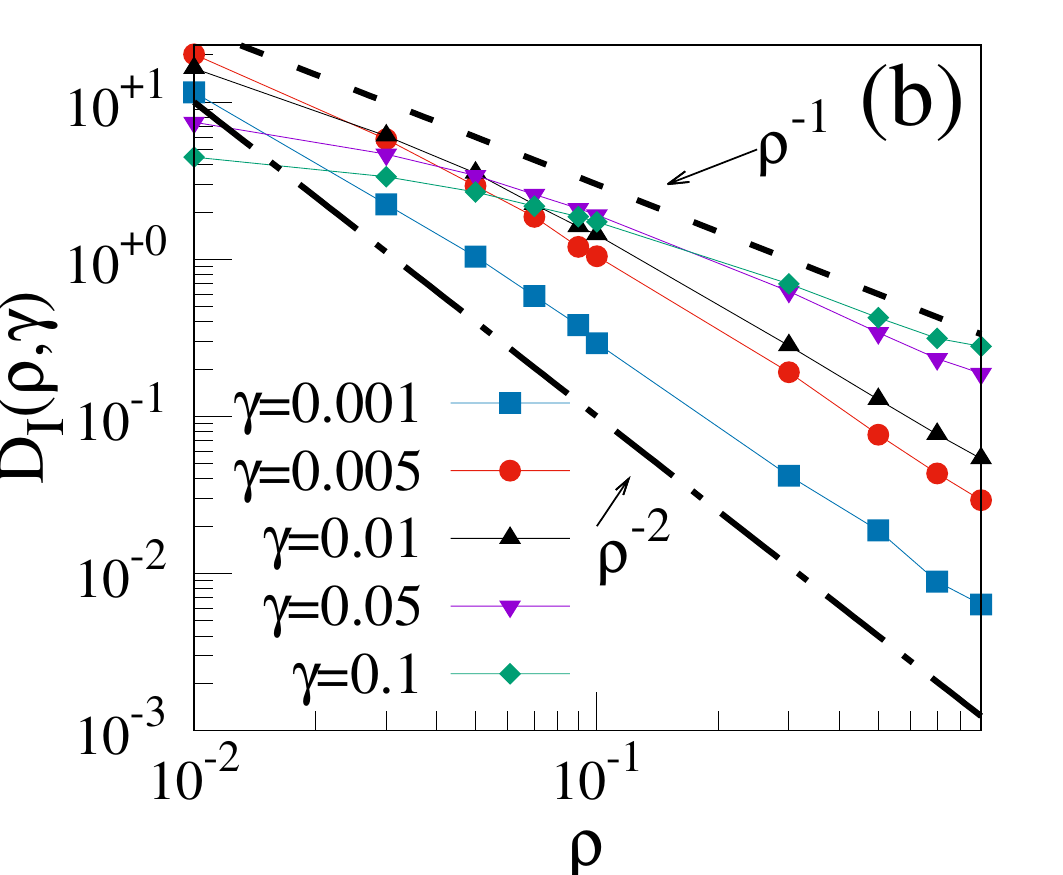}\hfill
     \includegraphics[width=0.249\linewidth]{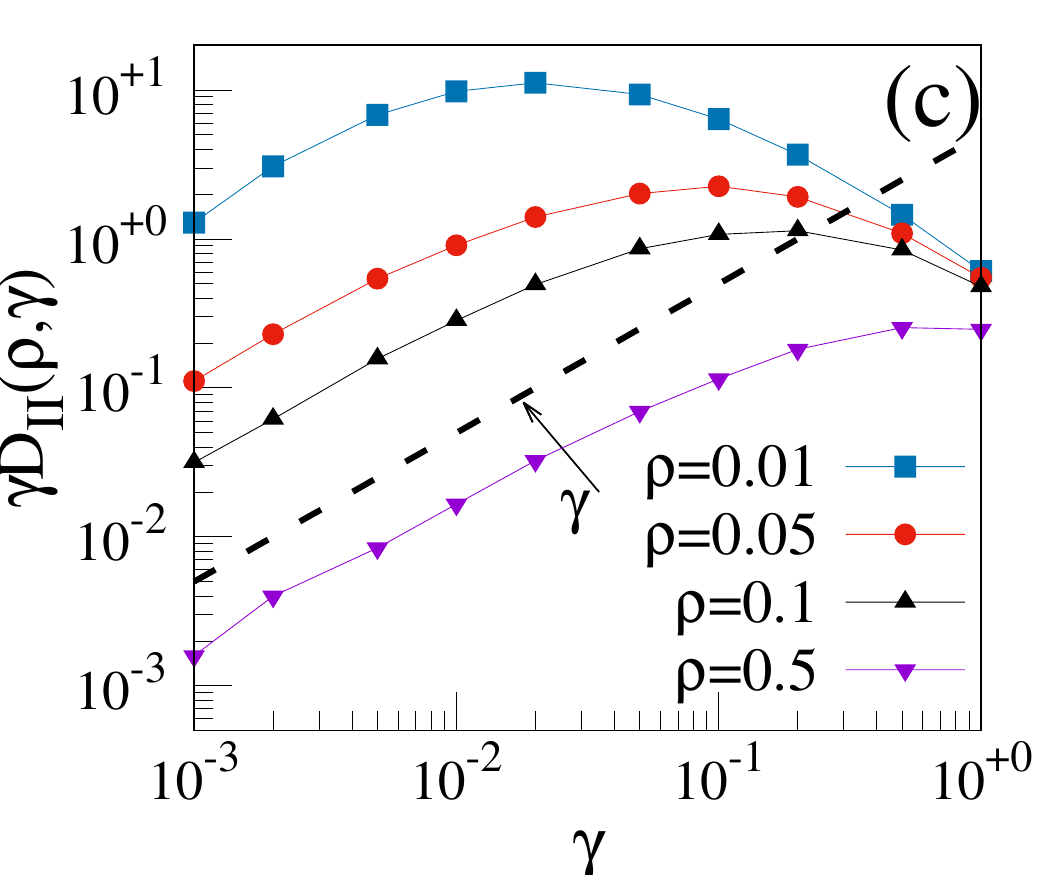}\hfill
     \includegraphics[width=0.249\linewidth]{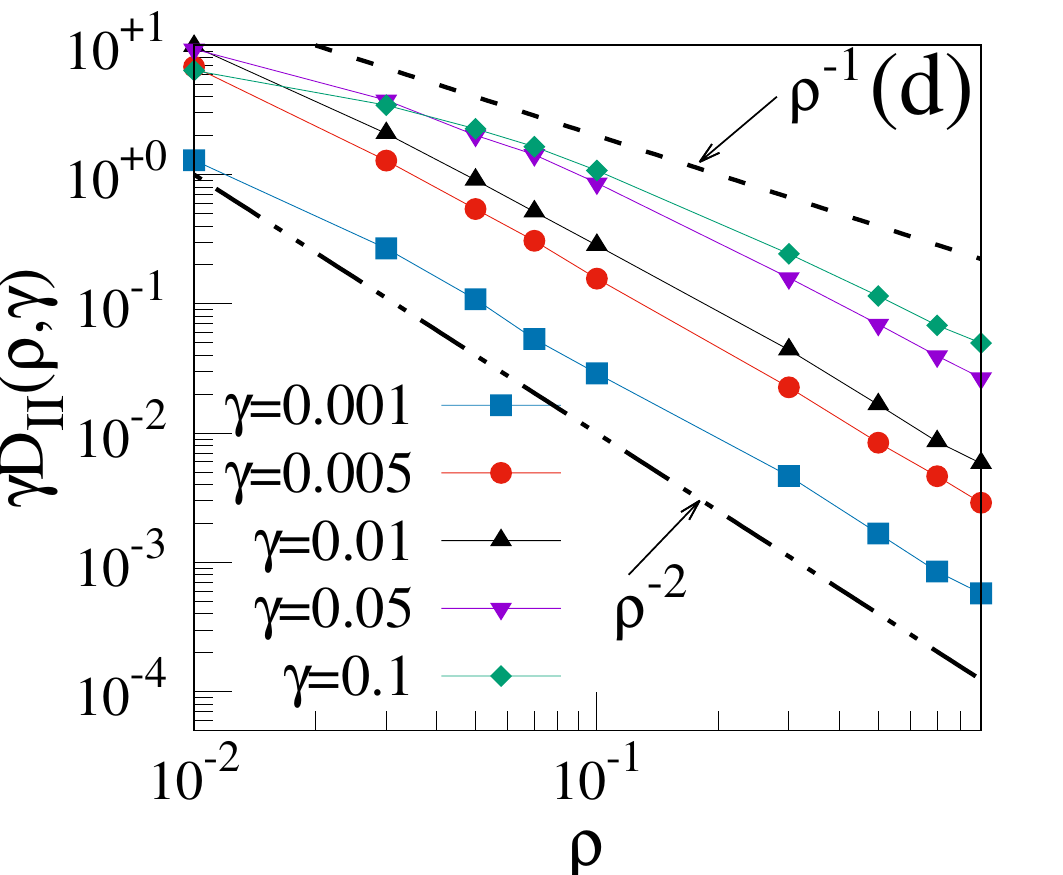}
        \caption{\textit{Bulk-diffusion coefficients in models I and II in one dimension.--} We plot $D_{I}(\rho,\gamma)$ and $\gamma D_{II}(\rho,\gamma)$, as a function of $\gamma$ at various densities $\rho=0.01$ (blue square), $0.05$ (red circle), $0.1$ (black up triangle) and $0.5$ (magenta down triangle) for model I [panel (a)] and model II [panel (c)], respectively. In panels (b) and (d), we plot $D_{I}(\rho,\gamma)$ and $\gamma D_{II}(\rho,\gamma)$, respectively as a function of $\rho$ at various $\gamma=0.001$ (blue square), $0.005$ (red circle), $0.01$ (black up triangle), $0.05$ (magenta down triangle) and 0.1 (green diamond).}
         \label{fig:bulk-diffusivity_RTP_LH_unscaled}
 \end{figure*}

One of the central assumptions in calculating $D_{I}(\rho,\gamma)$ in the preceding section is the existence of the diffusive scaling limits as expressed through Eq.~\eqref{diffusion_equation_macro}. We can immediately verify, through direct Monte Carlo simulations, the existence of such a diffusive scaling limit for model I (standard hardcore RTPs), which also leads to a direct verification of Eqs.~\eqref{diffusion_equation_macro}. To this end, we study the relaxation of coarse-grained (hydrodynamic) density field $\rho(x,\tau)$ as a function of scaled position $x=X/L$ for different system sizes $L$ and different hydrodynamic times $t \sim {\cal O}(L^2)$ such that $\tau=t/L^{2}$ remains fixed. The diffusive scaling limit would be verified if different curves corresponding to different $L$ and at the same hydrodynamic time $\tau$ collapse onto each other and, in that case, the collapsed data must be the solution of the nonlinear diffusion equation as given in Eq.~\eqref{diffusion_equation_macro}.

To implement the above numerical scheme, we simulate model I (for simplicity, data shown for one dimension only) with the initial sinusoidal density profile, as defined in Eq.~\eqref{sinusoidal_initial_condition}, with background $\rho_0=0.5$ and perturbation height $A(0)=0.05$. In Fig. \ref{fig:diffusive_scaling_limit}, we plot the numerically obtained excess density field $\delta \rho(x,\tau)= \rho(x,\tau)-\rho_0$ for tumbling rates $\gamma=0.05$ (top panel) and $0.01$ (bottom panel), as a function of the scaled position $x=X/L$, for different system sizes $L$ = $1000$ (blue square), $2000$ (magenta triangle) and $3000$ (red circle), and at hydrodynamic times $\tau=0.1$ (for $\gamma=0.05$) and $0.2$ (for $\gamma=0.01$), respectively. Throughout the analysis, we have done the averaging over $10^{4}$ realizations. In both panels of the figure, we numerically integrate Eq.~\eqref{diffusion_equation_macro} with the above-mentioned sinusoidal initial condition and the already calculated $D_{I}(\rho,\gamma)$ for the respective parameter values; the corresponding numerical solution obtained from our theory Eq.~\eqref{diffusion_equation_macro} is plotted as a black solid line. We observe that the simulation points nicely collapse onto each other and the curve for the collapsed data agrees very well with the theoretical solution, thus immediately implying the existence of a diffusive scaling limit for model I.

 \section{Simulation Results}
 
 \subsection{Bulk-diffusion coefficients in models I and II}\label{sec:D_result}

Using the methods mentioned in the previous section, we calculate the bulk-diffusion coefficient $D(\rho,\gamma)$ for both models in the parameter ranges $0.01 \leq \rho \leq 0.9$ and $0.001 \leq \gamma \leq 1$. For model II, we use the numerically obtained steady-state gap distribution $P(g)$ in Eq.~\eqref{bulk-diffusivity-LLG}, while for model I, we follow the method described in Sec.~\ref{sec:bulk-diff_RTPs}. In this section, we characterize the parameter dependence of $D(\rho,\gamma)$ in these two models.
In Fig.~\ref{fig:bulk-diffusivity_RTP_LH_unscaled}a, we plot $D(\rho,\gamma)$, for model I, as a function of the tumbling rate $\gamma$ at various densities $\rho=0.01$ (blue square), $0.05$ (red circle), $0.1$ (black up triangle )and $0.5$ (magenta down triangle). We observe thhat the bulk-diffusion coefficient $D(\rho, \gamma)$ is a nonmonotonic function of $\gamma$ and the condition for optimized diffusion occurs when $\gamma \simeq \rho$. Notably, this particular feature should be compared to the previous observations of Refs. \cite{Levis_berthier2014, Tailleur_PRL_2018, Tailleur_2022} in the context of the self-diffusion coefficient for SPPs. Indeed, this nonmonotonic behavior clearly depicts the intriguing interplay of persistence and hard-core interaction in the system. In the case of $\gamma \gg \rho$, decrease in $\gamma$ generates more persistence to the particles, leading to the increment in $D(\rho,\gamma)$. However, in the other regime, i.e., $\gamma \ll \rho$, particles are \textit{strongly interacting} and form jammed configuration. In that case, decrease in $\gamma$ value enhances the jamming condition further, thus resulting in the decrease in $D(\rho, \gamma)$. Moreover, we observe that $D(\rho,\gamma) \sim \gamma$ is proportional to tumbling rate in this regime. This linear behavior in $\gamma$ is quite expected because diffusion occurs only when one of the boundary particles from a cluster flips, and the local spin-flipping event occurs at a time scale $\tau_p=1/\gamma$, causing $D(\rho,\gamma)$ to scale as $1/\tau_p=\gamma$.
In order to characterize the density dependence, in Fig.~\ref{fig:bulk-diffusivity_RTP_LH_unscaled}b, we plot $D(\rho,\gamma)$, for model I, as a function of the density $\rho$ for various tumbling rates $\gamma=0.001$ (blue square), $0.005$ (red circle), $0.01$ (black up triangle ), $0.05$ (magenta down triangle) and $0.1$ (green diamond). Unlike the previous case, here we find that $D(\rho,\gamma)$  decreases monotonically with $\rho$ and has  a power-law tail, i.e., $D(\rho,\gamma) \sim \rho^{-\alpha}$, where the exponent $\alpha$ is parameter dependent. 
Quite interestingly, for model II, we observe $\gamma D(\rho,\gamma)$ to exhibit similar qualitative features. We show this by plotting $\gamma D(\rho,\gamma)$, for model II, as a function of $\gamma$ and $\rho$ in Figs.~\ref{fig:bulk-diffusivity_RTP_LH_unscaled}(c) and \ref{fig:bulk-diffusivity_RTP_LH_unscaled}(d), respectively, while keeping the other parameter values same as model I.

\subsubsection*{Scaling collapse of the bulk-diffusion coefficient for small density and tumbling rate}

 \begin{figure*}
           \centering    
    \hspace{1.75 cm} \includegraphics[width=0.8\linewidth]{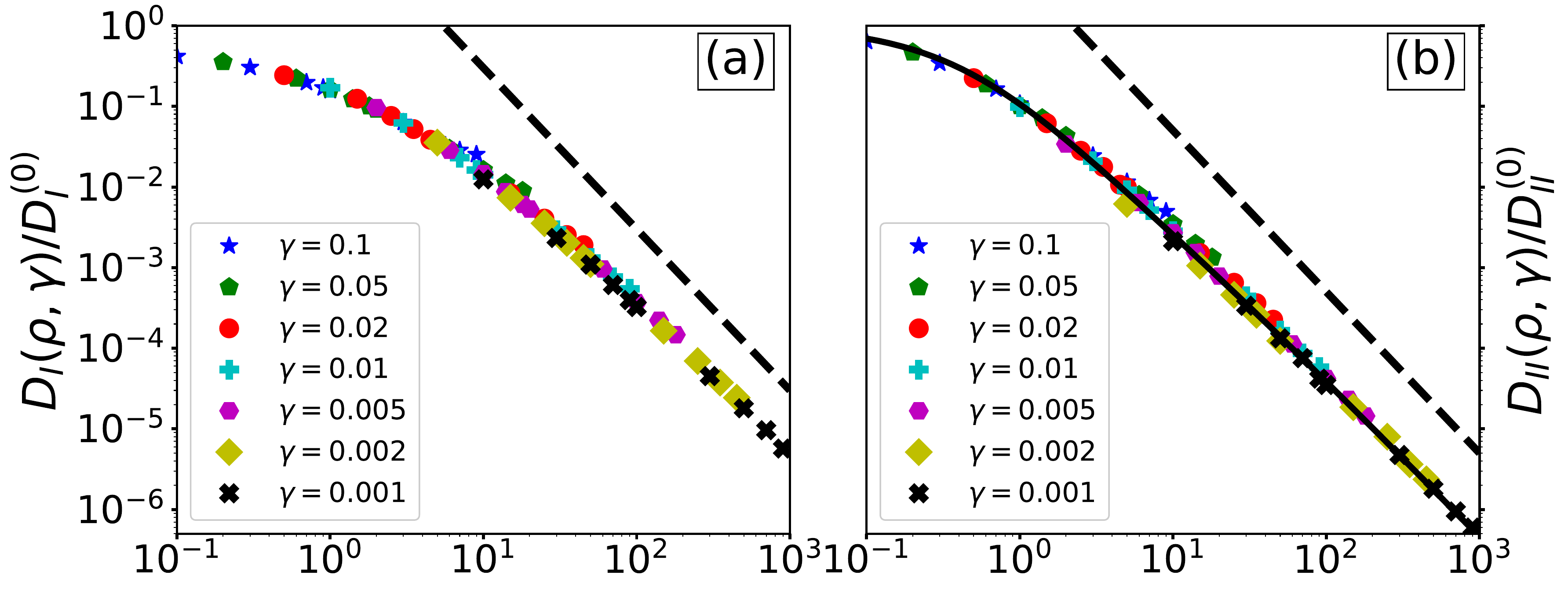} \hfill
     \vspace{-0 cm}
     \includegraphics[width=0.8\linewidth]{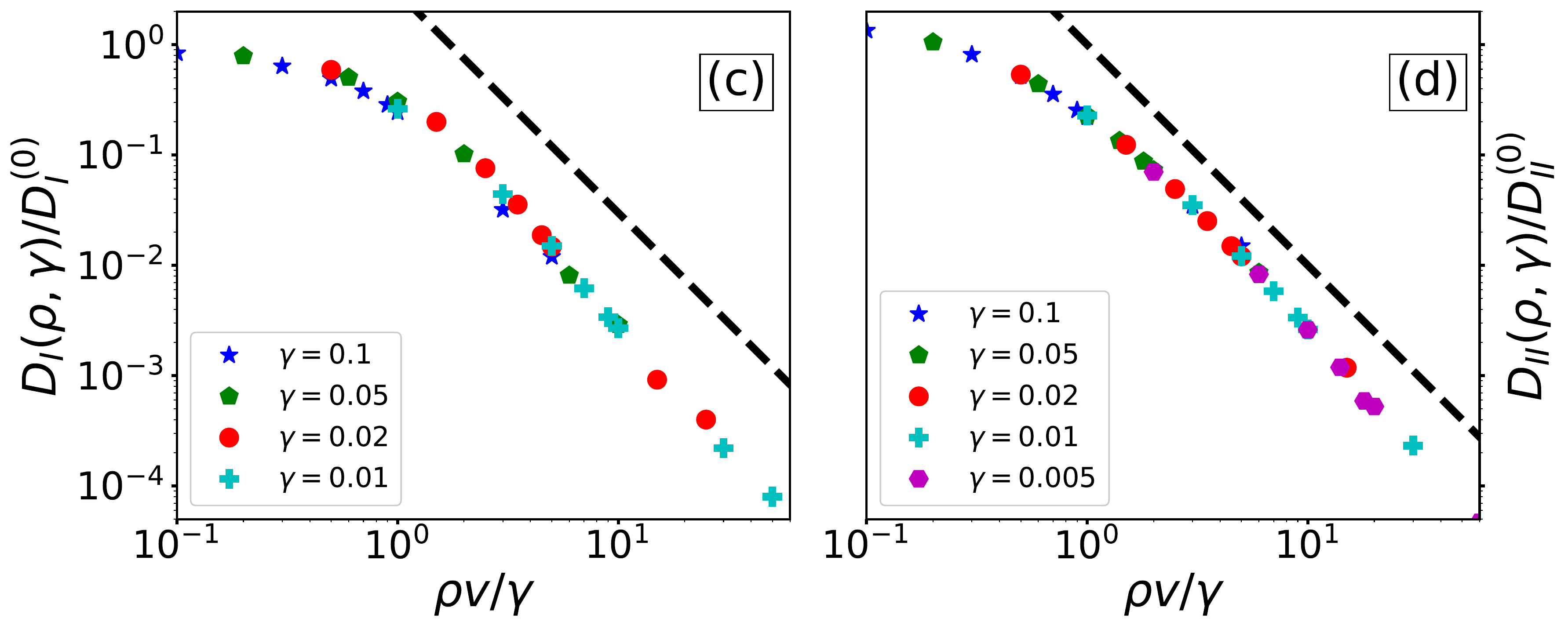}
          \caption{\textit{Scaling of the bulk-diffusion coefficient in models I and II.--} We plot scaled bulk-diffusion coefficients $D(\rho,\gamma)/D^{(0)}$  as a function of the scaled variable $\psi=\rho v/\gamma$ (with $v=1$) for model I ($(a)$ - $1$D and $(c)$ - $2$D) and model II ($(b)$ - $1$D and $(d)$ - $2$D). The solid line in panel $(b)$ represents the theory as in Eq.~\eqref{D_scaling_function}; the black dotted guiding lines represent $1/\psi^{2}$ behavior at large $\psi$.}
          \label{fig:bulk-diffusivity_RTP_LH_scaled}
 \end{figure*}
 \begin{figure*}
           \centering
     \hspace{1.5 cm} \includegraphics[width=0.4\textwidth]{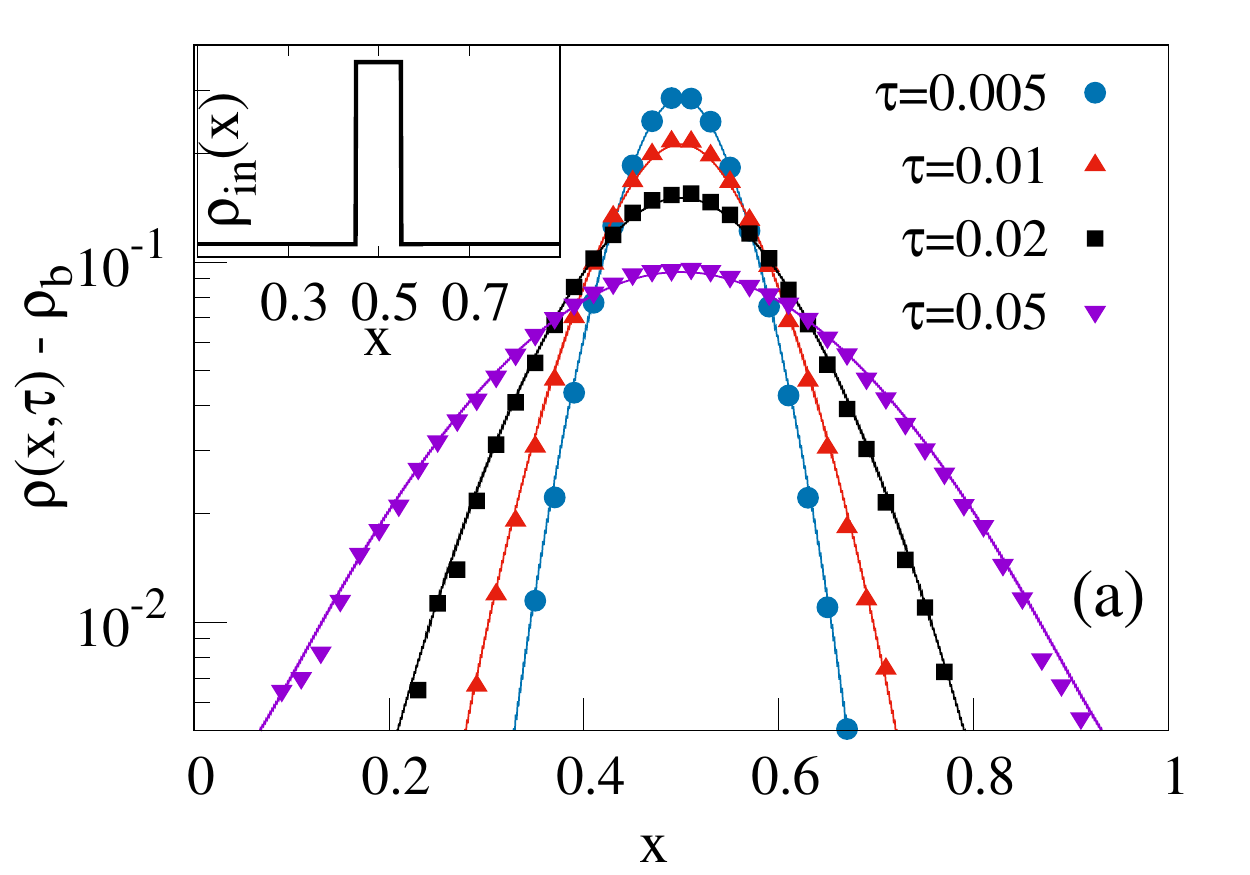}\hfill
      \includegraphics[width=0.4\textwidth]{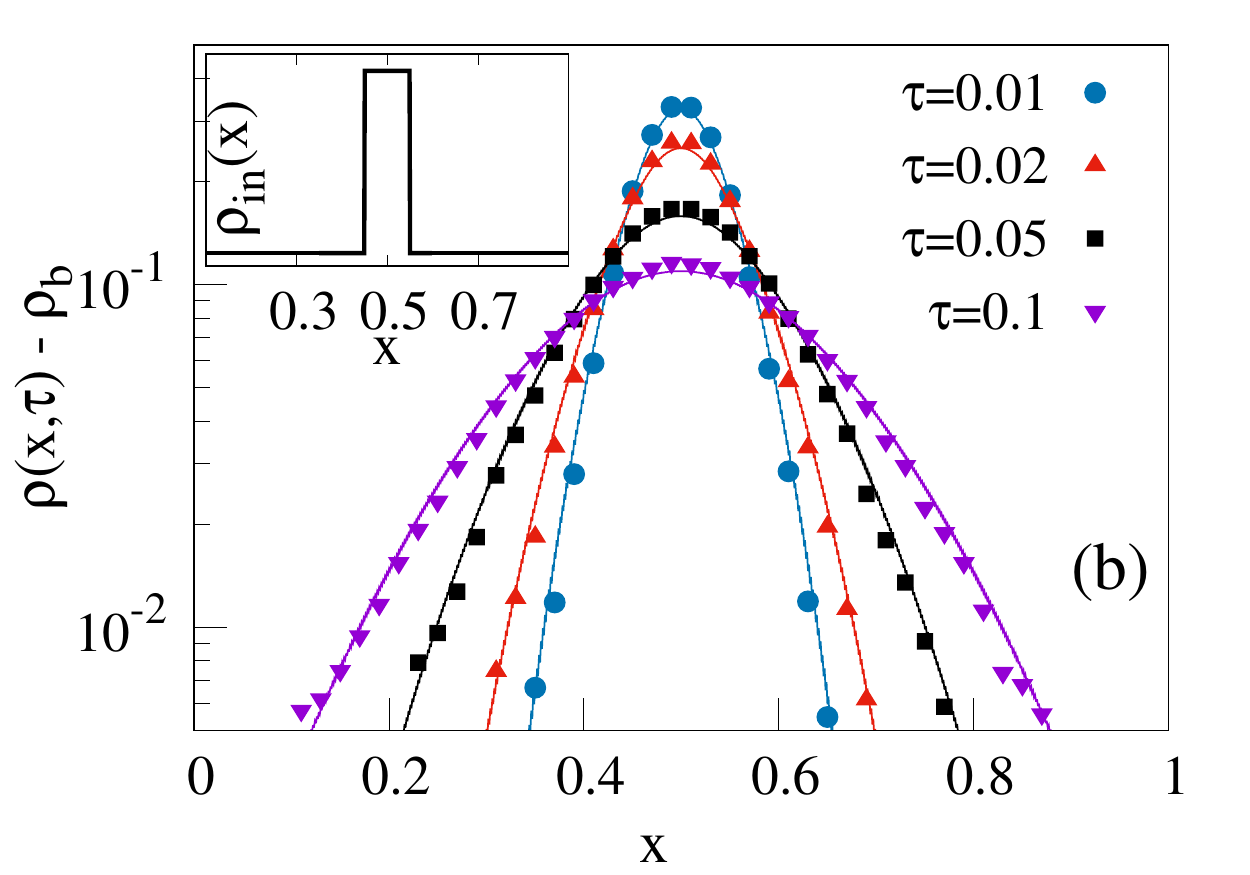} \hspace{+1.5 cm} \vfill \hspace{1.5 cm}\includegraphics[width=0.4\textwidth]
      {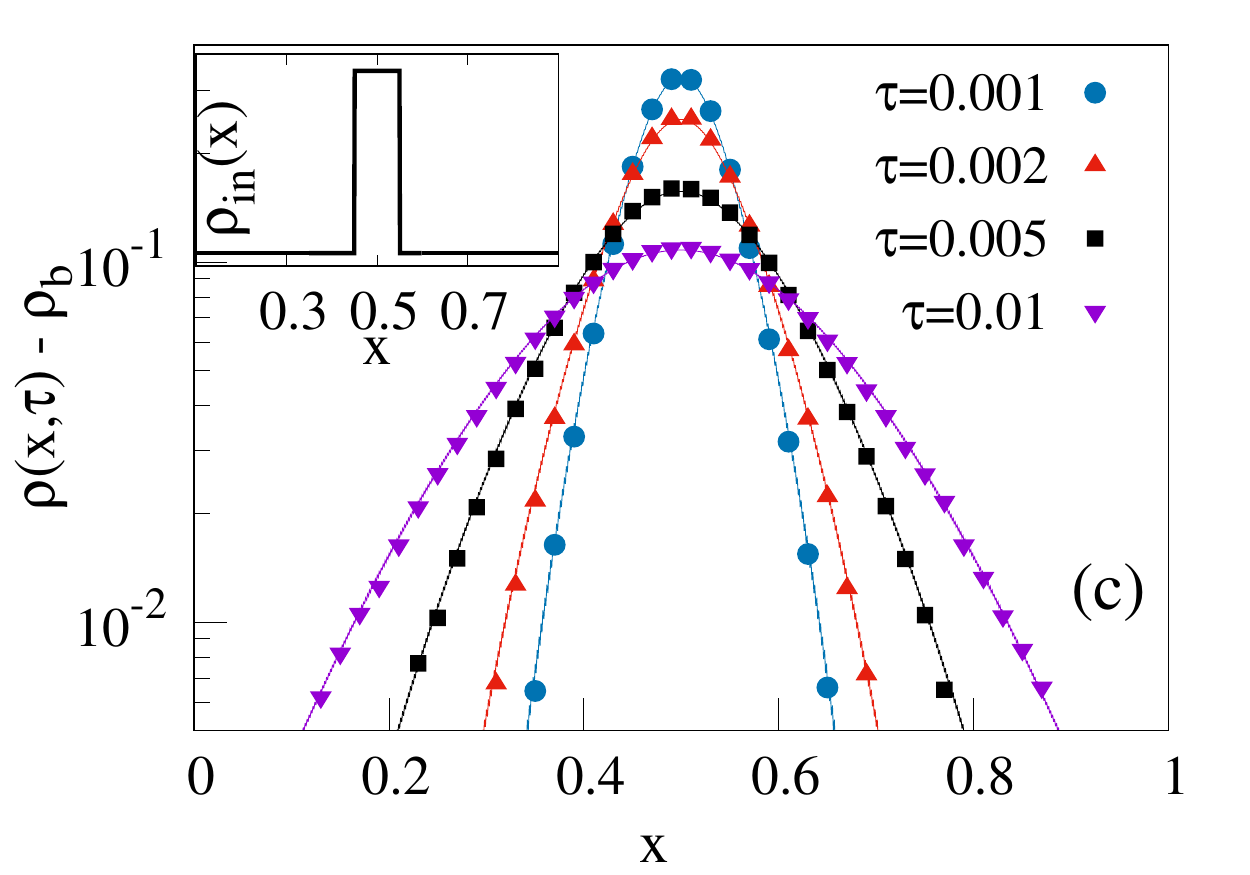} \hfill
      \includegraphics[width=0.4\textwidth]{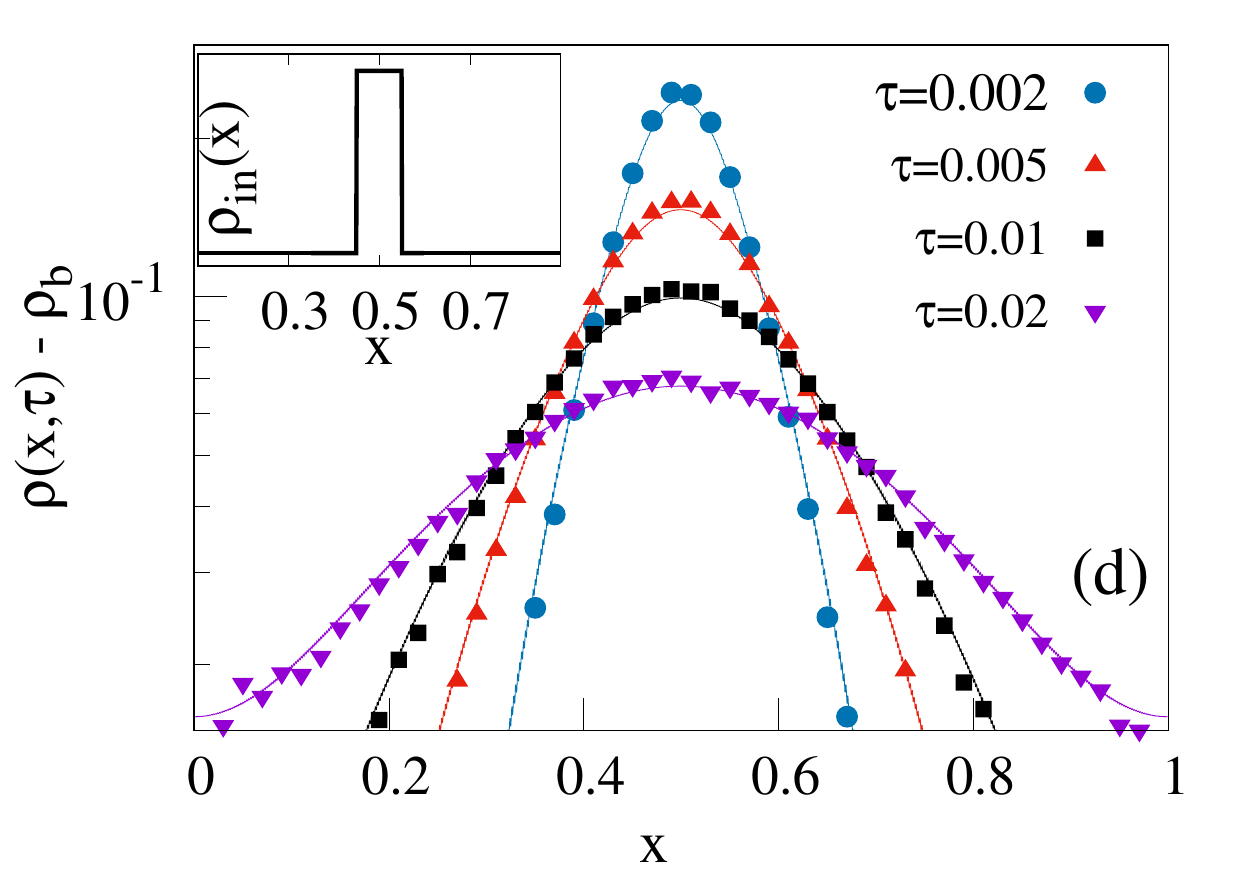} \hspace{+1.5 cm}
      \caption{\textit{Verification of nonlinear diffusion equation ~\eqref{diffusion_equation_macro} for hardcore RTPs.--} We plot the numerically obtained excess density profile $\delta \rho(x,\tau) = \rho(x,\tau)-\rho_b$ for $\gamma =0.05$ [model I (a) and model II (c)] and $0.01$ [model I (b) and model II (d)] as a function of the scaled position $x$ at various hydrodynamic time $\tau$. In all of these panels, corresponding lines are obtained by integrating the hydrodynamic equation in Eq.~\eqref{diffusion_equation_macro} using same initial condition (shown in the inset) in Eq.~\eqref{rho_initial_verification} with $\rho_b=0.5$, $\rho_1=0.4$ and $w=0.1$.}
          
          \label{fig:density_relaxation}
 \end{figure*}

We already observe that persistence and hard-core interaction compete with each other: While the former enhances diffusion, the latter diminishes it, resulting in nonmonotonic variation of the bulk-diffusion coefficient as a function of tumbling rate, see Figs.~\ref{fig:bulk-diffusivity_RTP_LH_unscaled}(a) and \ref{fig:bulk-diffusivity_RTP_LH_unscaled}(c). Notably, we have analytically characterized this competing effect for model II via a scaling law satisfied by $D(\rho,\gamma)$ as iven in Eq.~\eqref{D_scaling_relation}; also we have explicitly calculated the scaling function $\mathcal{F}(\psi)$ for $1D$ in Eq.~\eqref{D_scaling_function}. Indeed,  the scaling variable $\psi=\rho v/\gamma$ quantifies an intriguing interplay between persistence and interaction and, as in model II, one can quantify this fascinating interplay using the same scaling variable $\psi$ for model I too. We now verify these findings as in Fig.~\ref{fig:bulk-diffusivity_RTP_LH_scaled}.

In Figs.~\ref{fig:bulk-diffusivity_RTP_LH_scaled}(b) and \ref{fig:bulk-diffusivity_RTP_LH_scaled}(d), we plot the ratio $D_{II}(\rho,\gamma)/D_{II}^{(0)}$ as a function of $\psi$, for both one and two dimensional models, respectively. Indeed we observe a nice scaling collapse and see an excellent agreement with $\mathcal{F}_{II}(\psi)$ for one dimension as in Eq.~(\ref{D_scaling_function}), plotted in Fig.~\ref{fig:bulk-diffusivity_RTP_LH_scaled}b (black solid line). For the model I, in Figs.~\ref{fig:bulk-diffusivity_RTP_LH_scaled}a and \ref{fig:bulk-diffusivity_RTP_LH_scaled}c, we plot $D_{I}(\rho,\gamma)/D^{(0)}_{I} \equiv {\cal F}_{I}(\psi)$ as a function of $\psi$ with $D^{(0)}_{I}=v^{2}/\gamma d$ for both one and two dimensional cases, respectively. We observe an excellent scaling collapse in both dimensions, confirming the existence of the scaling regime of $D(\rho,\gamma)$ for the model I. Note that the difference in the microscopic time unit in models I and II is reflected through the ratio $D^{(0)}_{I}/D^{(0)}_{II}=\gamma$. Moreover, as in  model II, we find that ${\cal F}_{I}(0)= \rm{const.}$ and ${\cal F}_{I}(\psi) \simeq 1/\psi^2$ for $\psi \gg 1$. This reflects the fact that both models I and II have remarkably similar features, which are immediately evident upon time rescaling for model I (standard RTPs): $\gamma t \rightarrow t$, where time is measured in the unit of persistence time $\tau_p=\gamma^{-1}$.

 \subsection{Diffusive density relaxation }\label{sec-density_relaxation}\label{sec:density_relaxation}

In this section, we explicitly check, for simplicity only in one dimension, the long-time hydrodynamic description as in Eq.~\eqref{diffusion_equation_macro} by relaxing initial density perturbations through the already obtained bulk-diffusion coefficients $D_{I}(\rho, \gamma)$ and $D_{II}(\rho, \gamma)$ for models I and II, respectively. 
For this purpose, we take the initial density perturbation $\rho_{in}(x)=\rho(x,0)$ to be a step-function with step height $\rho_1$ and width $w$ over a uniform background density $\rho_b$, which mathematically can be expressed as,
\begin{eqnarray} 
\rho_{in}(x) = 
\left\{
\begin{array}{ll}
\rho_b + \rho_1              & {\rm for}~ \vert x-\frac{1}{2} \vert \leq \frac{w}{2} , \\
\rho_b               & {\rm otherwise}. \\
\end{array}
\right.
\label{rho_initial_verification}
\end{eqnarray}
We choose  $\rho_1=0.4$, $\rho_b=0.5$, $w=0.1$ and $L=1000$ for both the models. To verify Eq.~\eqref{diffusion_equation_macro}, we plot the excess density $\delta \rho(x,\tau)= \rho(x,\tau) - \rho_b$ in Fig.~\ref{fig:density_relaxation}, obtained from simulation (points), as a function of $x$ at various times $\tau$ for $\gamma=0.05$ [panel (a) for model I, panel (c) for model II] and $0.01$ [panel (b) for model I, panel (d) for model II]. In order to obtain the theoretical predictions (line), we perform numerical integration of Eq.~\eqref{diffusion_equation_macro} by using the above initial condition and the numerically calculated bulk-diffusion coefficients [$D_{I}(\rho, \gamma)$ for model I and $D_{II}(\rho, \gamma)$ for model II]. For both $\gamma$ values, we find the hydrodynamic theory (line) captures the simulation data (points) quite well. Such an agreement theory and simulations substantiate the correctness of calculated bulk-diffusion coefficients and thus that of the diffusive description [i.e., Eq.~\eqref{diffusion_equation_macro}] for both models I and II.

\subsection{Emergent hydrodynamics}\label{sec:emergent_hydrodynamics}

\begin{figure}
            \centering
      \includegraphics[width=0.495\textwidth]
      {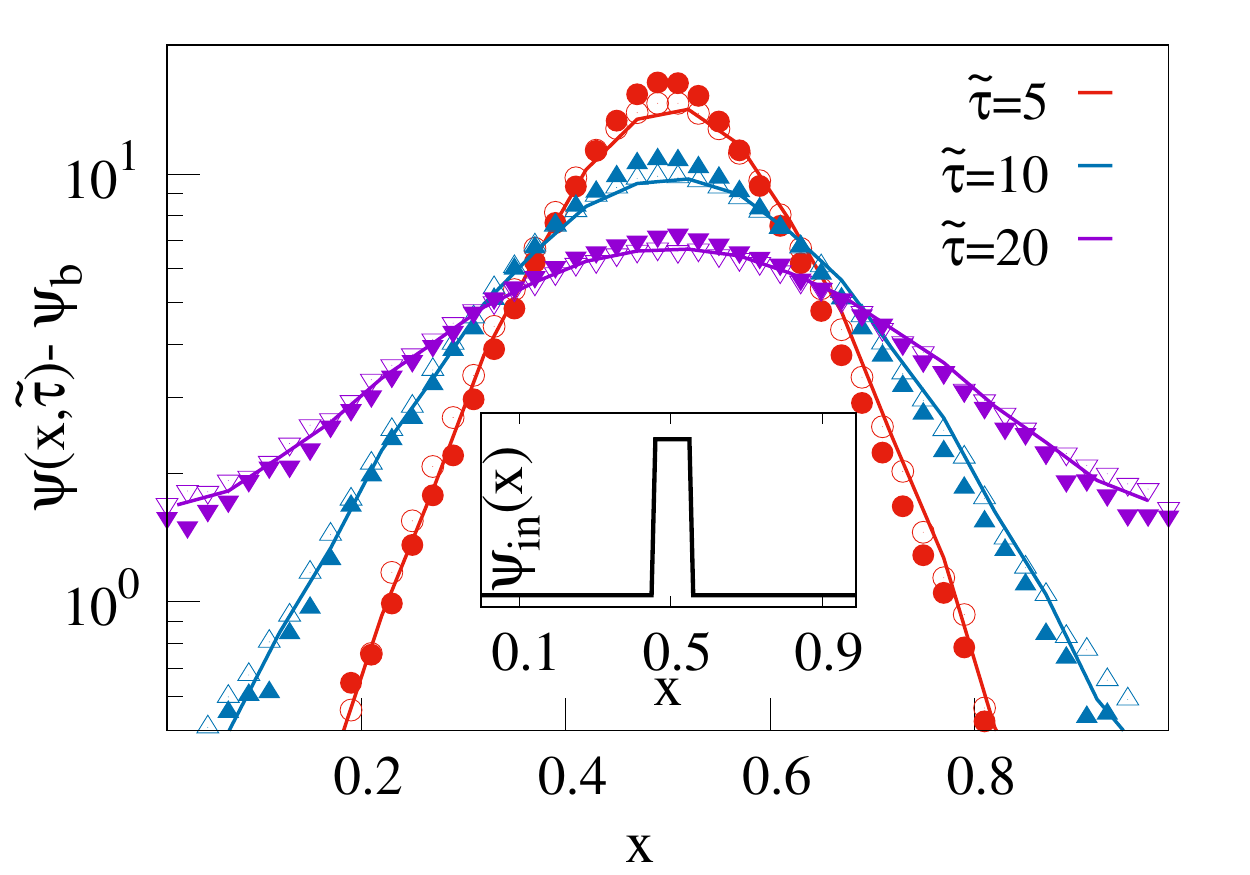}\hfill
        \includegraphics[width=0.495\textwidth]
      {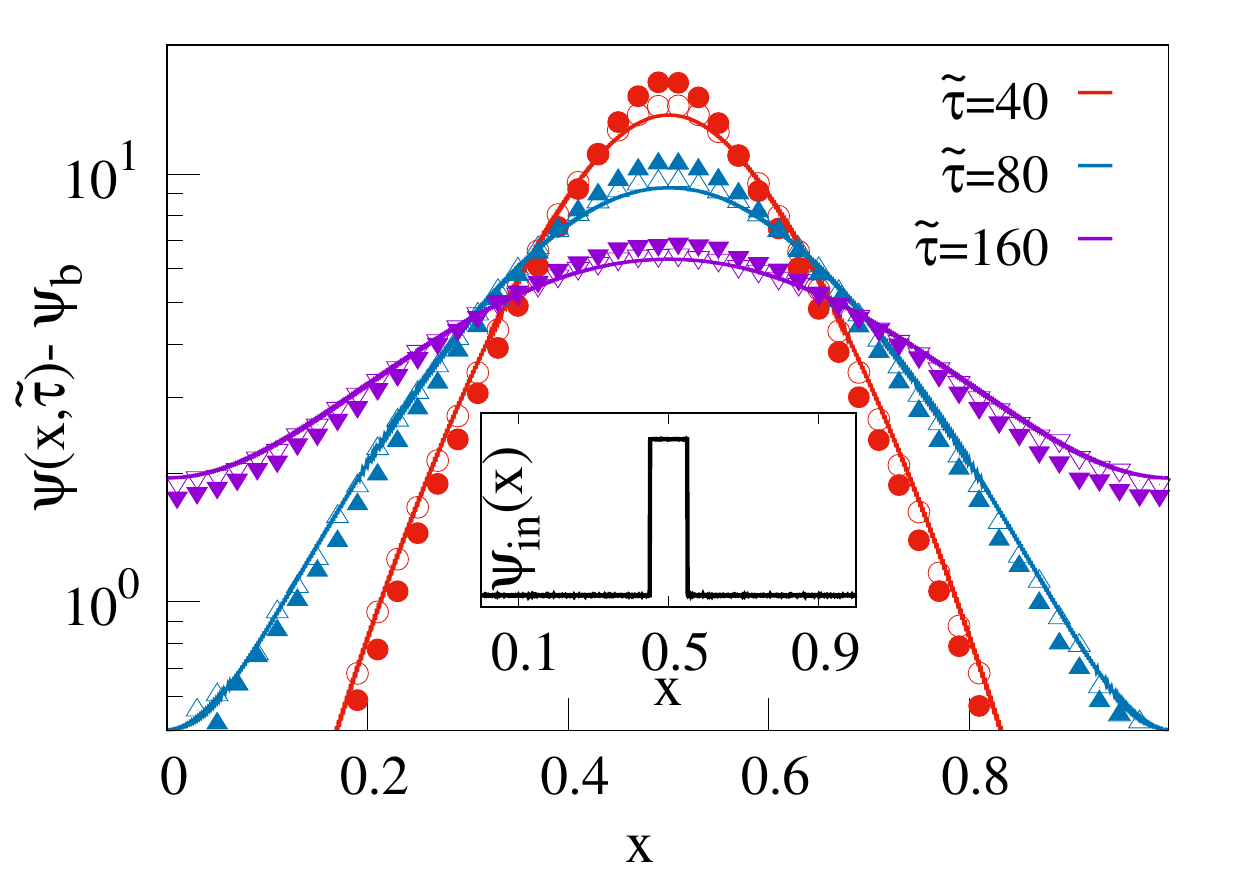}
          \caption{\textit{Verification of an emergent hydrodynamic time evolution \eqref{y_update_equation} for hardcore RTPs.--} We plot the scaled excess density field $\delta \psi(x,\tilde{\tau})= \psi(x,\tilde{\tau}) - \psi_b$, for model I (top-panel) and model II (bottom-panel), as a function of the scaled position $x$ at various rescaled time $\tilde{\tau}$ for two different $\gamma=$ $0.005$ (closed points) and $0.01$ (open points). Corresponding lines are obtained by integrating Eq.~\eqref{y_update_equation} using the scaling function $\mathcal{F}_{I}(\psi)$ obtained from simulation and $\mathcal{F}_{II}(\psi)$ from Eq.~\eqref{D_scaling_function}. We observe a good collapse between simulation points and a reasonably good agreement with the theory (lines).}
          \label{fig:scaled_density_relaxation_Dr0p01_0p005}
 \end{figure}

Remarkably, the calculated bulk-diffusion coefficients, which characterize the diffusive hydrodynamics quite well in the moderate persistence regime (see Sec.~\ref{sec-density_relaxation}), also captures quite well the anomalous behavior observed in the limit of strong persistence, i.e., $\gamma \rightarrow 0$. In the low density limit, it diverges as $1/\gamma$, while in the finite density limit, it vanishes as $ \gamma$. Furthermore, on the time scale of RTPs, the bulk-diffusion coefficient in model II, i.e., $\gamma D_{II}(\rho,\gamma)$ exhibits similar characteristics as that in model I. Consequently, the hydrodynamic description in terms of the usual density field $\rho(x,\tau)$ is not quite meaningful in the limit $\gamma \rightarrow 0$. Therefore, to obtain an appropriate hydrodynamic description, one cannot independently vary $\rho$ and $\gamma$ and, from the scaling law satisfied by $D(\rho,\gamma)$, we see that proper hydrodynamics scaling is recovered only when we rescale the density field $\psi=\rho/\gamma$ and time $\tilde{\tau}=\tau/\gamma$ (for model I) and $\tilde{\tau}=\tau/\gamma^{2}$ (for model II) and the corresponding hydrodynamic equation, in that case, is given by
 \begin{equation}\label{y_update_equation}
 \frac{\partial \psi(x,\tilde{\tau})}{\partial \tilde{\tau}}=\frac{\partial}{\partial x}\left[ \mathcal{F}(\psi) \frac{\partial \psi(x,\tilde{\tau})}{\partial x} \right],
 \end{equation}
where $\mathcal{F}(\psi)$ is the scaling function as defined in Eq.~\eqref{D_scaling}. To verify Eq.~\eqref{y_update_equation}, in Fig. (\ref{fig:scaled_density_relaxation_Dr0p01_0p005}) we plot the rescaled density field $\psi(x,\tilde{\tau})=\rho(x,\tau= \tilde{\tau} \gamma)/\gamma$ for model I  (top-panel) and $\psi(x,\tilde{\tau})=\rho(x,\tau= \tilde{\tau} \gamma^{2})/\gamma$ for model II  (bottom-panel) as a function of coarse-grained position $x=X/L$ for different $\gamma$ and different hydrodynamic times $\tau$ while keeping the rescaled hydrodynamic time $\tilde{\tau}$ fixed. The hydrodynamic equation would be verified if different curves corresponding to different $\gamma$ at the same $\tilde{\tau}$ collapse with each other and the collapsed profile at different $\tilde{\tau}$ is governed by the solution of Eq.~\eqref{y_update_equation}. In order to verify this assertion, we choose the initial scaled density profile $\psi_{in}(x)=\psi(x,\tilde{\tau}= 0)$ to be a step function with step height $\psi_1$ and width $w$ over a uniform background profile $\psi_b$, i.e.
 \begin{eqnarray} 
\psi_{in}(x) = 
\left\{
\begin{array}{ll}
\psi_b + \psi_1              & {\rm for}~ \vert x-\frac{1}{2} \vert \leq \frac{w}{2} , \\
\psi_b               & {\rm otherwise}. \\
\end{array}
\right.
\label{y_initial_verification}
\end{eqnarray}
For a fixed $\gamma$, we prepare a step profile for the density field $\rho(x,0)$ with step height $\rho_1$ and width $w$ around $x=1/2$ over a uniform background density $\rho_b$ such that the rescaled density $\psi_b=\rho_b/\gamma$ and $\psi_1=\rho_1/\gamma$ remain fixed for all $\gamma$. In Fig.~\ref{fig:scaled_density_relaxation_Dr0p01_0p005}, for a fixed $\psi_b=40$ and $\psi_1=40$, we plot the relaxation of the excess rescaled density $\psi(x,\tilde{\tau}) - \psi_b$, obtained from simulation, as a function of $x$ for different $\gamma=$ $0.01$ (open symbols) and $0.005$ (closed symbols) at various rescaled hydrodynamic time $\tilde{\tau}=$ $5$, $10$ and $20$ for model I (top-panel) and $\tilde{\tau}=$ $40$, $80$ and $160$ for model II (bottom-panel). In all of these panels, corresponding lines are obtained from numerical integration of Eq.~\eqref{y_update_equation} using $\mathcal{F}_{I}(\psi)$, obtained from simulation, for model I and $\mathcal{F}_{II}(\psi)$ for model II in Eq.~\eqref{D_scaling_function}. We observe that simulation data (points) and theory (lines) corresponding to these two different $\gamma$ collapse quite well for both these models, thus verifying the existence of an emergent hydrodynamic equation Eq.~\eqref{y_update_equation} in terms of the scaling function ${\cal F}(\psi)$.

\subsection{Anomalous transport}\label{sec:anomalous_transport}

\begin{figure}
           \centering
      \includegraphics[width=1\linewidth]{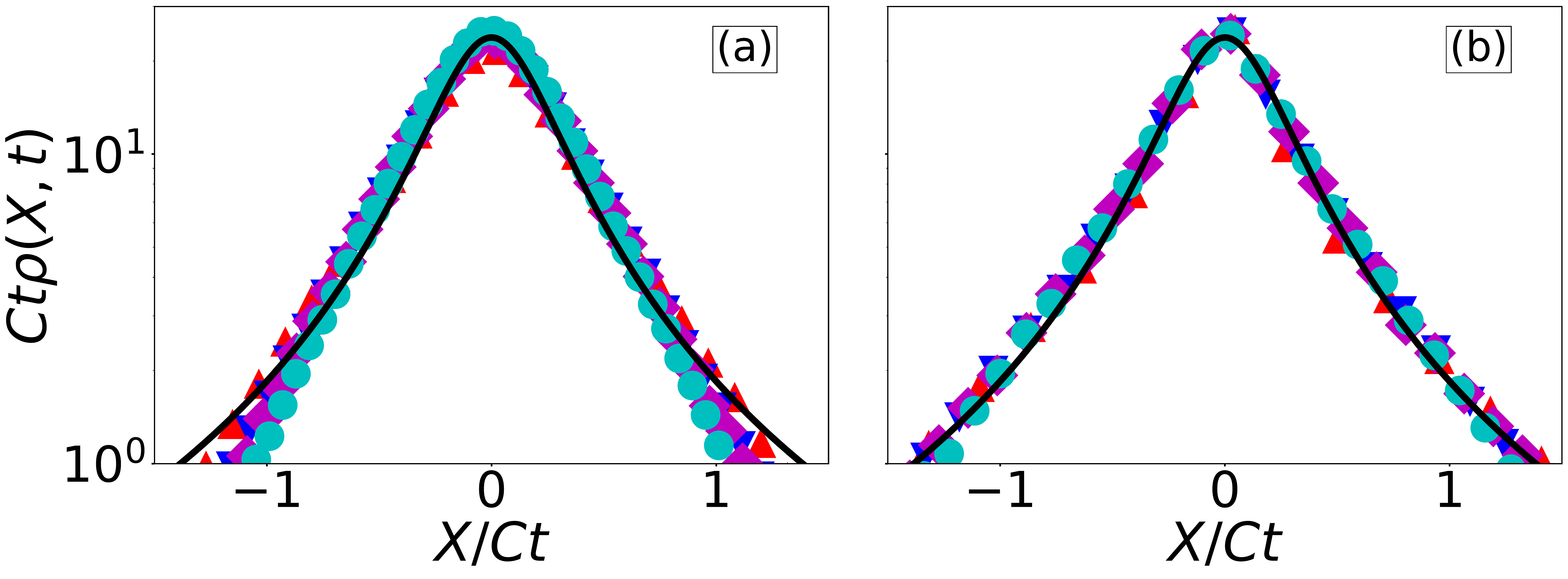} 
        
          \caption{\textit{Anomalous transport}. We plot scaled density profile $Ct \rho(X,t)$ against scaled variable $\xi=X/Ct$ at different times $t=$ $25$ (red up-triangle), $35$ (blue down-triangle), $40$ (magenta diamond), and $50$ (cyan circle) for model I (\textcolor{black}{panel a}) and $t=$ $100$ (red up-triangle), $150$ (blue down-triangle), $175$ (magenta diamond), and $200$ (cyan circle) for model II (\textcolor{black}{panel b}). \textcolor{black}{In panel (a), time $t$ is measured in the unit of $\tau_p=\gamma^{-1}$.} The spin-flipping rate $\gamma=0.05$  ($l_p=20$), \black{$N=20$} with $C=3.2$ (for model I) and $2.2$ (for model II); the black solid lines - the scaling solution $\mathcal{R}(\xi)$ in Eq.~\eqref{scaling_rho}. }
          \label{fig:ballistic_scaling_Dr0p05}
 \end{figure}

Finally, we analyze the anomalous ``early-time'' relaxation of initially localized density profiles in these two models of RTPs. We first rescale time for model I (standard RTPs) $\gamma t \rightarrow t$ (thus $D \rightarrow \gamma D$), thus putting the system on an equal footing with model II (LLG). Now, at the onset of diffusion, i.e., at {\it early but still long enough} time \black{$1 \ll t \ll  L^{2}/D$}, particles have undergone many collisions and, in that case, local density can safely be assumed to evolve through Eq.~\eqref{diffusion_equation_micro}. Importantly, depending on density and persistence length, the bulk-diffusion coefficient can have a power-law dependence on density, i.e., 
\begin{equation}\label{power_law_D}
D(\rho,\gamma) \simeq \frac{C}{\rho^{\alpha}},
\end{equation}
where the constant $C$ and $\alpha$ ($0 \leq \alpha \leq 2$) depend on the parameter regime under consideration (see Sec.~\ref{sec:D_result}). Remarkably, for the above $D(\rho,\gamma)$ and {\it delta} initial condition $\rho(X,0)=N \delta(X)$, Eq.~\eqref{diffusion_equation_micro} can be exactly solved using the following scaling ansatz,
\begin{eqnarray}
\rho(X,t)=\frac{1}{(Ct)^{\omega}}\mathcal{R}\left(\xi\right),
\end{eqnarray}
where $\xi=X/(Ct)^{\omega}$ is the scaling variable. Also, $\omega$ and $\mathcal{R}(\xi)$ are the growth exponent and the scaling function, respectively and can be calculated as discussed below. Now, plugging the above scaling solution in Eq.~\eqref{diffusion_equation_micro} and transforming $X \rightarrow \xi=X/(Ct)^{\omega}$, we obtain the following evolution equation of $\mathcal{R}(\xi)$:
\begin{equation}\label{diff_eqn_anomalous_scaled}
\omega \frac{d \left(\xi \mathcal{R}\right)}{d \xi}= - \left(Ct \right)^{1-(2-\alpha)\omega}\frac{d}{d \xi}\left[\mathcal{R}^{-\alpha}\frac{d \mathcal{R}}{d \xi}\right].
\end{equation}
However, to exhibit self-similar solution, Eq.~\eqref{diff_eqn_anomalous_scaled} should not contain $t$ explicitly, implying $1-(2-\alpha)\omega=0$, or
\begin{eqnarray}\label{growth-exponent}
 \omega= \frac{1}{2 - \alpha}.
\end{eqnarray}
Note that, Eq.~\eqref{growth-exponent} directly relates the growth exponent $\omega$ to the exponent $\alpha$, which is a parameter-dependent quantity.
We now choose a parameter regime where $\alpha=1$ and consequently $\omega=1$ (thus the dynamic exponent $z=1/\omega=1$), implying a ballistic transport, quite strikingly within a framework of a diffusion equation (\ref{diffusion_equation_micro}), albeit a nonlinear one. By using conditions $\mathcal{R}(0)=2/\xi_0^{2}$ and $\mathcal{R}^{\prime}(0)=0$, we exactly obtain,
\begin{eqnarray}\label{scaling_rho}
\mathcal{R}(\xi)=\frac{2}{\xi_0^{2}+\xi^{2}},
\end{eqnarray}
with constant $\xi_0$ fixed through conservation condition $\int_{-\infty}^{\infty} \mathcal{R}(\xi) d\xi= \rm{const.}$; in particular, for $\xi \gg 1$, $R(\xi) \sim 1/\xi^2$ signifies a power-law decay. Thus the density relaxation has anomalous spatiotemporal scaling with a non-Gaussian spread, which eventually becomes Gaussian (thus normal diffusion with $z=2$).
In simulation, we prepare the \textit{delta}-like initial profile by considering the step initial condition in Eq.~\eqref{rho_initial_verification} in the narrow width limit, i.e., $w/L \ll 1$. In particular, the simulation parameters are as following: $L=1000$, $w=20$, $\rho_0=0$, $\rho_1=1$, $\gamma=0.05$. In Fig.~\ref{fig:ballistic_scaling_Dr0p05}, we plot the scaled density profile $C t \rho(X,t)$ for model I \textcolor{black}{(panel a)} and II \textcolor{black}{(panel b)}, obtained from simulation, as a function of $X/Ct$ in the early-time regime; we have a quite good scaling collapse and a reasonably good agreement with theory (shown in solid black line); deviations at the tails are because the power-law behavior of $D(\rho,\gamma)$ is in fact cut off at very low densities.

\section{Summary}\label{sec:summary}

We study density relaxation in conventional models of {\it interacting} self-propelled particles (SPPs) on a $d$ dimensional periodic hypercubic lattice of volume $L^d$ for arbitrary density $\rho$ and tumbling rate  $\gamma =\tau_p^{-1}$. We consider two minimal models of {\it hardcore} run-and-tumble particles (RTPs): Model I is the standard version of hardcore RTPs introduced in Ref. \cite{Soto_2014}, whereas model II is a hardcore long-ranged lattice gas (LLG) - an analytically tractable variant of model I. \textcolor{black}{In model II, particles simply perform symmetric long-range hopping, with hop length chosen from a distribution $\phi(l)$, while abiding by the hardcore constraint as a hop is successful only if the empty stretch or the {\it gap} in the hopping direction is at least of length $l$; else, the particle traverses the entire stretch and sits
adjacent to its nearest occupied site [see Eq.~\eqref{hop-length} and the text above].} Note that the class of RTPs considered here are unlike those studied in Ref.~\cite{Tailleur_PRL_2018}, where the tumbling rate depends on system size and vanishes as $L \to \infty$; as a result, the large-scale evolution of the system in the latter case is exactly described by mean-field hydrodynamics, which were obtained by calculating various averages with respect to a product Bernoulli measure. On the other hand, the models considered in this paper have finite spatial correlations (which could be even long-ranged in the small tumbling-rate regime) and the nonequilibrium steady-state measure is not known.

By using models I and II, we argue that collective diffusion in RTPs is caused by subtle many-body effects arising from an intriguing interplay between interaction and persistence.
For these two systems, we characterize density relaxation through precise quantitative determination of  the density- and tumbling-rate-dependent bulk-diffusion coefficient.
To this end, we scale original position ${\bf X}$ and time $t$ as ${\bf x} = {\bf X}/L$ and time $\tau=t/L^2$,  and take the thermodynamic limit (system size $L \to \infty$ and particle number $N \to \infty$ with global density $N/L^d$ fixed).
We then show that the coarse-grained (hydrodynamic) density field $\rho({\bf x}, \tau)$ satisfies a nonlinear diffusion equation $\partial_{\tau} \rho({\bf x}, \tau) = \nabla [D(\rho,\gamma) \nabla \rho({\bf x},\tau)]$, where $D(\rho, \gamma)$ is the bulk-diffusion coefficient, which, we show, has a power-law form, i.e., $D \sim \rho^{-\alpha}$ with $0 < \alpha \le 2$ in a broad range of parameter values.
We explicitly calculate the bulk-diffusion coefficient as a function of density and tumbling rate, analytically for model II (LLG version) and  numerically in model I (standard hardcore RTPs) through an efficient Monte Carlo simulation algorithm; we also verify our analytical results for model II in simulations.
\textcolor{black}{Notably, for model II, we consider only a single-parameter (persistence length) family of step-size distribution $\phi(l)$, which, for simplicity, is taken to be exponential. However, given a single length scale in the distribution, the results should not depend on the details of $\phi(l)$. Of course, one could introduce more parameters in $\phi(l)$, which could then result in the bulk-diffusion coefficient, being a nontrivial function of density and tumbling rate and not expected to obey the scaling law in Eq.~\eqref{D_scaling}.}

Many-body correlations in these models manifest themselves through the competition between two length scales - the persistence length and a ``mean free path''\textcolor{black}{, which is a measure of the average size of an empty stretch, or the mean {\it gap}, in the direction of particle hopping.} Local structural relaxations are indeed sensitive to local densities, with low-density regions relaxing faster than high-density ones. In this way, they generate a wide range of time scales (power-law distributed), leading to dynamical heterogeneity and anomalous transport in the systems. 
In particular, we also investigate an interesting scaling regime where tumbling rate and density become small. Indeed, in the limit  $\rho \to 0$ and $\gamma \to 0$ with the ratio $\rho/\gamma$ fixed, the bulk-diffusion coefficient satisfies a scaling law $D(\rho, \gamma) = D^{(0)} {\cal F}(\rho/\gamma)$ as given in Eq. (\ref{D_scaling}). Furthermore, for model II in one dimension, we analytically calculate the scaling function ${\cal F}(\psi)$, which is in excellent agreement with simulations.
Our findings, as succinctly expressed in Eq. (\ref{D_scaling}),  are of particular significance: They are independent of dimensions and microscopic details and unveil a unique mechanism of collective transport - {\it diffusive, yet anomalous} - that has not been identified previously in the context of SPPs. However, the exact functional form of ${\cal F}(\psi)$ depends on microscopic details of the models. \textcolor{black}{It is worth mentioning here that, for any finite density and tumbling rate and in finite dimensions, the bulk-diffusion coefficient for both models I and II never vanishes and remains nonzero throughout.} 

We further stress that the tagged-particle diffusion and collective diffusion (equivalently, temporal growth
	of density perturbation as studied here) are not the same, with a particularly pronounced difference for hardcore particles in one dimension. Indeed, in the thermodynamic limit, the motion of individual hardcore RTPs, characterized by the long-time mean-squared displacement (MSD) of a tagged particle, exhibits subdiffusive growth in one dimension \cite{dolai2020, Tirthankar_cates_2022}. On the other hand, collective transport in that case, as found in the present work, is diffusive. As such, there is no contradiction here because tagged-particle transport and collective transport are, in principle, two distinct processes. Though their contrasting behavior in one dimension may appear to be somewhat counter-intuitive, it can be understood from the fact that the hardcore exclusion preserves particle ordering, resulting in confinement and thus significantly suppressed tagged-particle MSD.

It should be noted that the order of hydrodynamic limits studied in this work is important. When the large persistence-time limit is taken first, followed by the large system-size limit, the system undergoes a previously studied ``jamming" or dynamical arrest \cite{dasgupta_2020}. On the other hand, studies in Refs. \cite{Tailleur_PRL_2018,Kafri_2021} considered system-size dependent tumbling rate, which in fact differs from the conventional versions of SPPs, which are of our interest here \cite{Levis_berthier2014, Soto_2014, Fily_2012, Baskaran_2013, szamel2014self,maggi2015multidimensional}. In our study, the large system-size limit is taken first, followed by the large persistence-time limit.  Indeed, attempts to develop a rigorous hydrodynamic description of SPPs that correspond to this specific order of limit have not been successful so far. Recently, the bulk-diffusion coefficient for a dual version of hardcore RTPs  has been calculated, albeit only in the leading order of tumbling rate \cite{Dandekar_PRE_2020}, and it agrees quite well with Eq. (\ref{D_scaling}) as $\psi \rightarrow \infty$.
However, none of the earlier works actually identified the existence of a generic scaling regime for collective diffusion, which we precisely identify in our study. 
Interestingly, in a different context, a similar scaling regime for the self-diffusion coefficient in variants of SPPs has recently been observed in Ref. \cite{Debets_PRL_2021}, which nicely complements our findings. However, the precise relationship between the bulk- and self-diffusion coefficients currently remains unclear. \textcolor{black}{Furthermore, our findings imply that there is no instability in the bulk-diffusion coefficient for the class of RTPs studied here. In this scenario, it will be rather  interesting to explicitly calculate the bulk-diffusion coefficient as a funtion of density and tumbling rate to see if other classes of conventional SPPs exhibit any diffusive instability. We believe our proposed mechanism of density relaxations could provide useful insights into collective transport in SPPs and will open up new research avenues in the future.}

 \begin{acknowledgments}
We thank Subhadip Chakraborti, Arghya Das, Rahul Dandekar and R. Rajesh for discussions. P.P. gratefully acknowledges the Science and Engineering Research Board (SERB), India, under Grant No. MTR/2019/000386, for financial support. T.C. acknowledges a research fellowship [Grant No. 09/575 (0124)/2019-EMRI] from the Council of Scientific and Industrial Research (CSIR), India.
\end{acknowledgments}

  \begin{widetext}
  \section*{Appendix}
In the appendix section, we provide additional calculation and derivation details, as well as simulation results used in the main text.

\subsection{ Verification of the scaling form of gap distribution $P(g)$}\label{sec:app2}

In this section, we numerically verify the scaling law satisfied by the typical gap size $g_*$ in the regime of low density and strong persistence used in the main text in Eq.~\eqref{g_*}, i.e.,
\begin{equation}\label{g_*_app}
g_*=\frac{1}{\rho} \mathcal{G}(\psi).
\end{equation}
Note that, using the assumed exponential form of the gap distribution, i.e.,
\begin{eqnarray}\label{pg_app}
P(g) \simeq N_* e^{-g/g_*}
\end{eqnarray}
and the above scaling form of $g_*$ in the following conservation equation
\begin{eqnarray}
\langle g \rangle = \sum_{g=1}^{\infty} g P(g)= 1/\rho,
\end{eqnarray}
which is exact in the low density limit, we straightforwardly obtain the proportionality constant
\begin{equation}\label{norm}
N_*=\frac{1}{\rho} \frac{[\exp(\rho/\mathcal{G}(\psi))-1]^{2}}{\exp(\rho/\mathcal{G}(\psi))}.
\end{equation}
Now, inserting $g_*$ and $N_*$, as shown in Eqs.~\eqref{g_*_app} and \eqref{norm}, in the expression of $P(g)$ in Eq.~\eqref{pg_app}, we obtain the following scaling law for $P(g)$,
 \begin{equation}\label{P(g)_scaling}
 -\ln\left(\frac{P(g)}{\rho} \right) = \frac{\rho g}{\mathcal{G}(\psi)} + 2\ln\left( \mathcal{G}(\psi) \right).
 \end{equation}
Clearly, the above scaling law is a direct consequence of the assumed scaling form of $g_*$ in Eq.~\eqref{g_*_app}. Hence, verifying Eq.~\eqref{P(g)_scaling} yields the existence of the proposed scaling law of $g *$ right away. In order to do so,  we compute the steady-state $P(g)$ for model I (left panel) and model II (right panel) in $1D$ at various tumbling rates $\gamma$ (various persistent length for model II) and densities $\rho$, while keeping the ratio $\psi=\rho v/\gamma$ constant. The scaling law would be verified if data points corresponding to different $\rho$ and $\gamma$ at a fixed $\psi$ collapse with each other.

\begin{figure}
           \centering
      \includegraphics[width=0.495\textwidth]{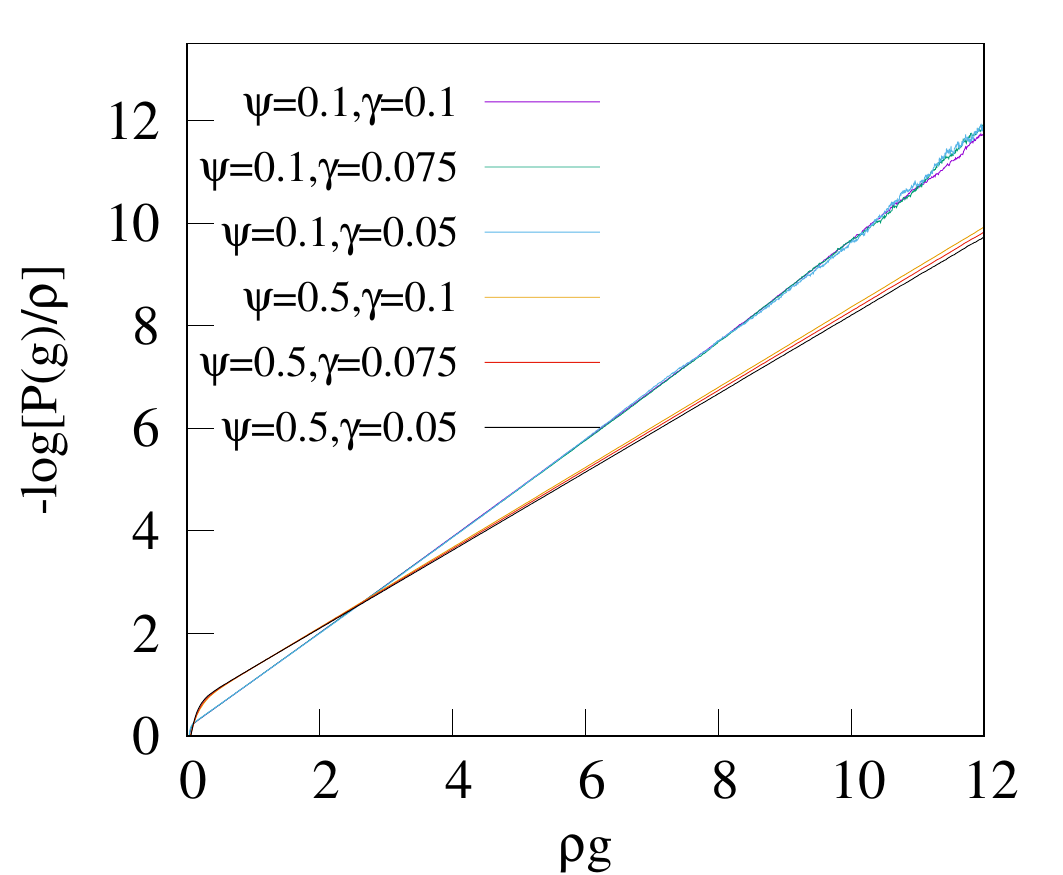} \hfill
      \includegraphics[width=0.495\textwidth]{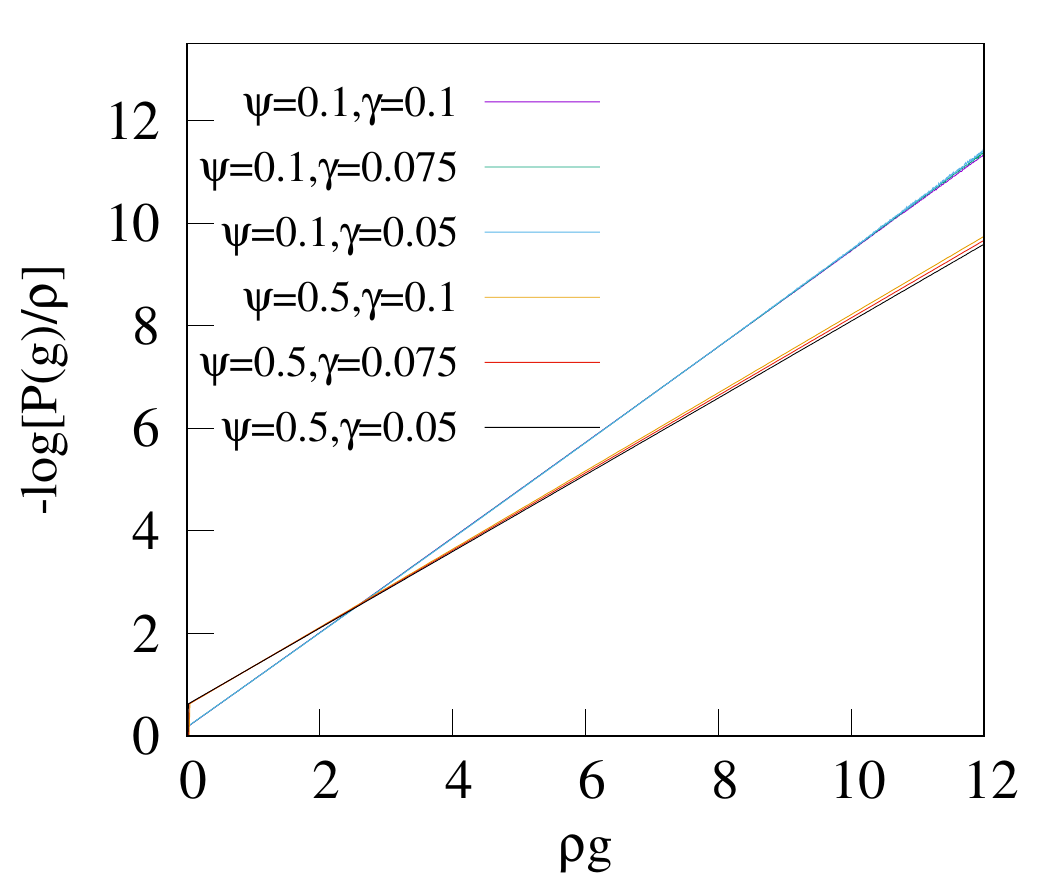} 
          \caption{\textit{Verification of Eq.~\eqref{P(g)_scaling} }. We plot $-\ln\left( P(g)/\rho\right)$ as a function of $\rho g$ for model I (left-panel) and model II (right-panel) in $1D$, for various combinations of $\rho$ and $\gamma$ such that the scaling variable $\psi$ remains fixed at $0.1$ and $0.5$. In both these panels, for $\psi = 0.1$, we have used $\gamma=$ $0.1$ (magenta line), $0.05$ (green line) and $0.01$ (sky blue line); while the same for $\psi=0.5$ are shown in orange line, red line and black line, respectively. }
 \label{fig:P(g)_scaling}
 \end{figure}
To check this, in Fig.~\ref{fig:P(g)_scaling}, we plot $-\ln (P(g)/\rho)$ as a function of the scaled gap size $\rho g$ for model I (left) and model II (right) at two different $\psi= 0.1$ and $0.5$. For each $\psi$, we have used three different tumbling rates $\gamma=$ $0.05$, $0.075$, and $0.1$ and the corresponding density $\rho$ is chosen from the relation $\rho= \gamma \psi/v$ (where $v=1$). Data belonging to different parameter spaces $(\rho, \gamma)$ corresponding to the same $\psi$ collapse quite well, as shown in the figure. This observation immediately validates the scaling form of $g_*$ for both models I and II.

\subsection{Derivation of scaling function $\mathcal{F}_{II}(\psi)$ of the bulk-diffusion coefficient in model II}\label{sec:app3}

In this section, we derive the analytic expressions for the bulk-diffusion coefficient $D_{II}(\rho,\gamma)$ and the corresponding scaling function $\mathcal{F}_{II}(\psi)$, presented in the main text in Eq.~\eqref{D_scaling_function_G(psi)}. Using the gap distribution function given in Eq.~\eqref{pg_app} in the calculated expression of the bulk-diffusion coefficient $D_{II}(\rho, \gamma)$, shown in the main text in Eq~\eqref{bulk-diffusivity-LLG}, we obtain,
\begin{eqnarray}\label{D_1}
D_{II}(\rho,\gamma)=-\frac{1}{2d} \frac{\partial}{\partial \rho}\Big[ N_* \rho \sum_{l=1}^{\infty} \phi(l) \Big(\sum_{g=1}^{\infty} g e^{-g/g_*} + (l-1)\sum_{g=l}^{\infty} (g-l)e^{-g/g_*} \Big) \Big]. 
\end{eqnarray}
We now note the following identities:
\begin{eqnarray}
\sum_{g=1}^{\infty} g e^{-g/g_*} =  \frac{\langle g \rangle}{ N_*}, \\
\sum_{g=l}^{\infty} (g-l) e^{-g/g_*} =  e^{-l/g_*} \frac{\langle g \rangle}{ N_*},\\
\langle g \rangle =\frac{1}{\rho} -1.
\end{eqnarray}
Here $\langle g \rangle$ is the mean gap length at density $\rho$. Now, using these identities and $\phi(l)=B e^{-l/l_p}$ in Eq.~\eqref{D_1}, we obtain
 
\begin{equation}\label{D_2}
 D_{II}(\rho,\gamma) = -\frac{B}{2d} \frac{\partial}{\partial \rho} \left[(1-\rho)\Bigg\{\sum_{l=1}^{\infty} e^{-l/l_p} + \sum_{l=1}^{\infty} (l-1) e^{-l/\xi}\Bigg\}\right],
 \end{equation}
where we define $\xi=1/\left(1/l_p+1/g_*\right)$. It is now easy to perform the summations appearing in Eq.~\eqref{D_2}, which are given by,
\begin{eqnarray}\label{D_3}
\sum_{l=1}^{\infty} e^{-l/l_p}=\frac{1}{e^{1/l_p}-1}, \\
\sum_{l=1}^{\infty}(l-1)e^{-l/\xi} = \frac{1}{(e^{1/\xi}-1)^{2}}.
\label{D_4}
\end{eqnarray}
Finally, replacing Eqs.~\eqref{D_3} and \eqref{D_4} in Eq.~\eqref{D_2}, we obtain the following simplified form:
\begin{eqnarray}\label{D_4}
D_{II}(\rho,\gamma) = -\frac{B}{2d} \frac{\partial}{\partial \rho} \left[(1-\rho)\Bigg\{\frac{1}{e^{1/l_p}-1} +  \frac{1}{(e^{1/\xi}-1)^{2}} \Bigg\}\right].
\end{eqnarray}
Notably, the above form of $D_{II}(\rho,\gamma)$ is used in the main text in Eq.~\eqref{bulk-diffusivity-LLG_most_general} and is valid for arbitrary values of $\rho$ and $\gamma$. However, in the subsequent analysis, we consider the following three special cases:
\begin{itemize}
\item Case I, $\rho$ arbitrary, $\gamma \rightarrow \infty:$ In this case, we have $B=1$ and the typical gap size $g_*$ is given by,
\begin{eqnarray}
g_* = -\frac{1}{\log (1-\rho)}.
\end{eqnarray}
Using this particular $g_*$ and $\gamma \rightarrow \infty$ or $l_p \rightarrow 0$ in Eq.~\eqref{D_4}, it is straightforward to show that the second term in the curly bracket does not contribute, and the resulting expression of $D_{II}(\rho, \gamma)$ is given by,  
\begin{equation}
D_{II}(\rho,\gamma) = e^{-1/l_p} D_{SSEP},
\end{equation}
where $D_{SSEP}=1/2d$ is the bulk-diffusion coefficient for SSEP in $d$ dimension.

\item Case II, $\rho$ finite and large, $\gamma \rightarrow 0:$ In this case, we have $B=1/l_p$ and the typical gap size, as calculated in Ref.~\cite{subhadip_PRE_2021}, is given by
\begin{equation}
g_*=\sqrt{l_p \langle g \rangle},
\end{equation}
which immediately implies,
\begin{equation}
\frac{\partial g_{*}}{\partial \rho}=-\frac{1}{2\rho^{2}}\sqrt{\frac{l_p}{\langle g \rangle}}.
\end{equation}
Moreover, using the $g_*$ mentioned above, we obtain
\begin{eqnarray}
\xi=\frac{l_p}{1+\sqrt{\frac{l_p}{\langle g \rangle}}}.
\end{eqnarray}

Notably, in this regime, $l_p$, $g_{*}$ and $\xi$ are infinitely large. Using all of the above relations in Eq.~\eqref{D_4} and after some algebraic manipulations, we finally obtain a simplified expression of $D(\rho,\gamma)$ in this regime, which is given by,
\begin{eqnarray}
D_{II}(\rho,\gamma)=\frac{1}{2\rho^{2}}.
\end{eqnarray} 

\item Case III, $\rho \rightarrow 0$, $\gamma \rightarrow 0:$ In this particular regime of interest, the typical gap size $g_*$ obeys the following scaling law,
\begin{eqnarray}
g_*=\frac{1}{\rho}\mathcal{G}(\psi),
\end{eqnarray}
where $\psi=\rho/\gamma$. Using the above expression of $g_*$, we can easily caluclate
\begin{equation}
 \frac{\partial g_*}{\partial \rho}=-\frac{g_*}{\rho} \left( 1- \psi \frac{\mathcal{G}'(\psi)}{\mathcal{G}(\psi)} \right).
 \end{equation}
 Note that, since $\rho \rightarrow 0$, we can easily write $1-\rho \simeq 1$, which simplifies Eq.~\eqref{D_4} substantially. Finally, substituting all of the results in Eq.~\eqref{D_4}, some algebraic manipulations leads to the following expression of the bulk-diffusion coefficient:
 \begin{equation}\label{bulk_diff_LH_final}
 D_{II}(\rho,\gamma)=\frac{ e^{1/\xi}}{l_p d(e^{1/\xi}-1)^{3}}\frac{1}{\mathcal{G}(\psi)}\left(1 - \psi \frac{\mathcal{G}'(\psi)}{\mathcal{G}(\psi)} \right).
 \end{equation}
\textit{Asymptotic expansion of $D_{II}(\rho,\gamma)$:}  We further expand the bulk-diffusion coefficient $D_{II}(\rho,\gamma)$, obtained in Eq.~\eqref{bulk_diff_LH_final}, in the limit of $\gamma\rightarrow0$ and $\rho \rightarrow 0$, such that the dimensionless quantity $\psi=\rho v/\gamma$ remains finite. Let us first make the replacement,
\begin{eqnarray}
\frac{1}{\xi} = n = \frac{1}{l_p} + \frac{\rho}{\mathcal{G}(\psi)} = \frac{1}{l_p} \left( 1 + \frac{\psi}{\mathcal{G}(\psi)} \right).
\end{eqnarray}
Note that this particular limit of interest automatically implies $n \rightarrow 0$. We now expand the following expression as
\begin{eqnarray}\label{n_approx}
\frac{e^n}{l_p(e^n -1)^{3}} &=& \frac{\left(1 + n + \frac{n^{2}}{2} \dots \right)}{l_p n^{3}\left(1+\frac{n}{2}+\frac{n^{2}}{6} \dots\right)^{3}} \nonumber \\
			&\simeq& \frac{1}{l_p} \left(\frac{1}{n^{3}}-\frac{1}{2n^{2}} - \frac{3}{2n} \dots \right).
\end{eqnarray}
Since we are working in the regime $n \rightarrow 0$, the leading order contribution to Eq.~\eqref{n_approx} emerges from $1/n^{3}$ term only and consequently the bulk-diffusion coefficient can be written as

\begin{eqnarray}
D_{II}(\rho,\gamma)=\frac{1}{l_pdn^{3}\mathcal{G}(\psi)}\left(1 - \psi \frac{\mathcal{G}'(\psi)}{\mathcal{G}(\psi)} \right) + \mathcal{O}\left(\frac{\gamma}{n^{2}}\right), \nonumber \\
   \simeq \frac{l_p^{2}\mathcal{G}^{2}(\psi)}{d\left(\mathcal{G}(\psi)+\psi \right)^{3}}\left(1 - \psi \frac{\mathcal{G}'(\psi)}{\mathcal{G}(\psi)} \right) + \mathcal{O}\left(l_p\right). 
\end{eqnarray}
From the above equation, it is quite evident that $D_{II}(\rho,\gamma)$ satisfies a scaling law and the corresponding scaling function can be written as
\begin{equation}
\mathcal{F}_{II}(\rho,\gamma) = \frac{1}{d}\left(\frac{\gamma}{v}\right)^{2} D_{II}(\rho,\gamma) \simeq\frac{\mathcal{G}^{2}(\psi)}{\left(\mathcal{G}(\psi)+\psi \right)^{3}}\left(1 - \psi \frac{G'(\psi)}{\mathcal{G}(\psi)} \right),
\end{equation} 
which we have used in the main text in Eq.~\eqref{D_scaling_function_G(psi)}.
 
\end{itemize}

 \end{widetext}
\bibliography{rtp_diffusion_new}

\begin{thebibliography}{69}%
\makeatletter
\providecommand \@ifxundefined [1]{%
 \@ifx{#1\undefined}
}%
\providecommand \@ifnum [1]{%
 \ifnum #1\expandafter \@firstoftwo
 \else \expandafter \@secondoftwo
 \fi
}%
\providecommand \@ifx [1]{%
 \ifx #1\expandafter \@firstoftwo
 \else \expandafter \@secondoftwo
 \fi
}%
\providecommand \natexlab [1]{#1}%
\providecommand \enquote  [1]{``#1''}%
\providecommand \bibnamefont  [1]{#1}%
\providecommand \bibfnamefont [1]{#1}%
\providecommand \citenamefont [1]{#1}%
\providecommand \href@noop [0]{\@secondoftwo}%
\providecommand \href [0]{\begingroup \@sanitize@url \@href}%
\providecommand \@href[1]{\@@startlink{#1}\@@href}%
\providecommand \@@href[1]{\endgroup#1\@@endlink}%
\providecommand \@sanitize@url [0]{\catcode `\\12\catcode `\$12\catcode
  `\&12\catcode `\#12\catcode `\^12\catcode `\_12\catcode `\%12\relax}%
\providecommand \@@startlink[1]{}%
\providecommand \@@endlink[0]{}%
\providecommand \url  [0]{\begingroup\@sanitize@url \@url }%
\providecommand \@url [1]{\endgroup\@href {#1}{\urlprefix }}%
\providecommand \urlprefix  [0]{URL }%
\providecommand \Eprint [0]{\href }%
\providecommand \doibase [0]{http://dx.doi.org/}%
\providecommand \selectlanguage [0]{\@gobble}%
\providecommand \bibinfo  [0]{\@secondoftwo}%
\providecommand \bibfield  [0]{\@secondoftwo}%
\providecommand \translation [1]{[#1]}%
\providecommand \BibitemOpen [0]{}%
\providecommand \bibitemStop [0]{}%
\providecommand \bibitemNoStop [0]{.\EOS\space}%
\providecommand \EOS [0]{\spacefactor3000\relax}%
\providecommand \BibitemShut  [1]{\csname bibitem#1\endcsname}%
\let\auto@bib@innerbib\@empty
%</preamble>
\bibitem [{\citenamefont {Marchetti}\ \emph {et~al.}(2013)\citenamefont
  {Marchetti}, \citenamefont {Joanny}, \citenamefont {Ramaswamy}, \citenamefont
  {Liverpool}, \citenamefont {Prost}, \citenamefont {Rao},\ and\ \citenamefont
  {Simha}}]{Ramaswamy2013}%
  \BibitemOpen
  \bibfield  {author} {\bibinfo {author} {\bibfnamefont {M.~C.}\ \bibnamefont
  {Marchetti}}, \bibinfo {author} {\bibfnamefont {J.~F.}\ \bibnamefont
  {Joanny}}, \bibinfo {author} {\bibfnamefont {S.}~\bibnamefont {Ramaswamy}},
  \bibinfo {author} {\bibfnamefont {T.~B.}\ \bibnamefont {Liverpool}}, \bibinfo
  {author} {\bibfnamefont {J.}~\bibnamefont {Prost}}, \bibinfo {author}
  {\bibfnamefont {M.}~\bibnamefont {Rao}}, \ and\ \bibinfo {author}
  {\bibfnamefont {R.~A.}\ \bibnamefont {Simha}},\ }\href {\doibase
  10.1103/RevModPhys.85.1143} {\bibfield  {journal} {\bibinfo  {journal} {Rev.
  Mod. Phys.}\ }\textbf {\bibinfo {volume} {85}},\ \bibinfo {pages} {1143}
  (\bibinfo {year} {2013})}\BibitemShut {NoStop}%
\bibitem [{\citenamefont {Wu}\ and\ \citenamefont
  {Libchaber}(2000)}]{Libchaber_2000}%
  \BibitemOpen
  \bibfield  {author} {\bibinfo {author} {\bibfnamefont {X.-L.}\ \bibnamefont
  {Wu}}\ and\ \bibinfo {author} {\bibfnamefont {A.}~\bibnamefont {Libchaber}},\
  }\href {\doibase 10.1103/PhysRevLett.84.3017} {\bibfield  {journal} {\bibinfo
   {journal} {Phys. Rev. Lett.}\ }\textbf {\bibinfo {volume} {84}},\ \bibinfo
  {pages} {3017} (\bibinfo {year} {2000})}\BibitemShut {NoStop}%
\bibitem [{\citenamefont {Wang}\ \emph {et~al.}(2009)\citenamefont {Wang},
  \citenamefont {Anthony}, \citenamefont {Bae},\ and\ \citenamefont
  {Granick}}]{granick_pnas_2009}%
  \BibitemOpen
  \bibfield  {author} {\bibinfo {author} {\bibfnamefont {B.}~\bibnamefont
  {Wang}}, \bibinfo {author} {\bibfnamefont {S.~M.}\ \bibnamefont {Anthony}},
  \bibinfo {author} {\bibfnamefont {S.~C.}\ \bibnamefont {Bae}}, \ and\
  \bibinfo {author} {\bibfnamefont {S.}~\bibnamefont {Granick}},\ }\href
  {\doibase 10.1073/pnas.0903554106} {\bibfield  {journal} {\bibinfo  {journal}
  {Proceedings of the National Academy of Sciences}\ }\textbf {\bibinfo
  {volume} {106}},\ \bibinfo {pages} {15160} (\bibinfo {year} {2009})},\
  \Eprint
  {http://arxiv.org/abs/https://www.pnas.org/doi/pdf/10.1073/pnas.0903554106}
  {https://www.pnas.org/doi/pdf/10.1073/pnas.0903554106} \BibitemShut {NoStop}%
\bibitem [{\citenamefont {Leptos}\ \emph {et~al.}(2009)\citenamefont {Leptos},
  \citenamefont {Guasto}, \citenamefont {Gollub}, \citenamefont {Pesci},\ and\
  \citenamefont {Goldstein}}]{goldstein_prl_2009}%
  \BibitemOpen
  \bibfield  {author} {\bibinfo {author} {\bibfnamefont {K.~C.}\ \bibnamefont
  {Leptos}}, \bibinfo {author} {\bibfnamefont {J.~S.}\ \bibnamefont {Guasto}},
  \bibinfo {author} {\bibfnamefont {J.~P.}\ \bibnamefont {Gollub}}, \bibinfo
  {author} {\bibfnamefont {A.~I.}\ \bibnamefont {Pesci}}, \ and\ \bibinfo
  {author} {\bibfnamefont {R.~E.}\ \bibnamefont {Goldstein}},\ }\href {\doibase
  10.1103/PhysRevLett.103.198103} {\bibfield  {journal} {\bibinfo  {journal}
  {Phys. Rev. Lett.}\ }\textbf {\bibinfo {volume} {103}},\ \bibinfo {pages}
  {198103} (\bibinfo {year} {2009})}\BibitemShut {NoStop}%
\bibitem [{\citenamefont {Kudrolli}(2010)}]{Kudroli_2010}%
  \BibitemOpen
  \bibfield  {author} {\bibinfo {author} {\bibfnamefont {A.}~\bibnamefont
  {Kudrolli}},\ }\href {\doibase 10.1103/PhysRevLett.104.088001} {\bibfield
  {journal} {\bibinfo  {journal} {Phys. Rev. Lett.}\ }\textbf {\bibinfo
  {volume} {104}},\ \bibinfo {pages} {088001} (\bibinfo {year}
  {2010})}\BibitemShut {NoStop}%
\bibitem [{\citenamefont {Wang}\ \emph {et~al.}(2012)\citenamefont {Wang},
  \citenamefont {Kuo}, \citenamefont {Bae},\ and\ \citenamefont
  {Granick}}]{granick_nature}%
  \BibitemOpen
  \bibfield  {author} {\bibinfo {author} {\bibfnamefont {B.}~\bibnamefont
  {Wang}}, \bibinfo {author} {\bibfnamefont {J.}~\bibnamefont {Kuo}}, \bibinfo
  {author} {\bibfnamefont {S.~C.}\ \bibnamefont {Bae}}, \ and\ \bibinfo
  {author} {\bibfnamefont {S.}~\bibnamefont {Granick}},\ }\href {\doibase
  10.1038/nmat3308} {\bibfield  {journal} {\bibinfo  {journal} {Nature
  Materials}\ }\textbf {\bibinfo {volume} {11}},\ \bibinfo {pages} {481}
  (\bibinfo {year} {2012})}\BibitemShut {NoStop}%
\bibitem [{\citenamefont {Ariel}\ \emph {et~al.}(2015)\citenamefont {Ariel},
  \citenamefont {Rabani}, \citenamefont {Benisty}, \citenamefont {Partridge},
  \citenamefont {Harshey},\ and\ \citenamefont {Be'Er}}]{ariel2015swarming}%
  \BibitemOpen
  \bibfield  {author} {\bibinfo {author} {\bibfnamefont {G.}~\bibnamefont
  {Ariel}}, \bibinfo {author} {\bibfnamefont {A.}~\bibnamefont {Rabani}},
  \bibinfo {author} {\bibfnamefont {S.}~\bibnamefont {Benisty}}, \bibinfo
  {author} {\bibfnamefont {J.~D.}\ \bibnamefont {Partridge}}, \bibinfo {author}
  {\bibfnamefont {R.~M.}\ \bibnamefont {Harshey}}, \ and\ \bibinfo {author}
  {\bibfnamefont {A.}~\bibnamefont {Be'Er}},\ }\href@noop {} {\bibfield
  {journal} {\bibinfo  {journal} {Nature communications}\ }\textbf {\bibinfo
  {volume} {6}},\ \bibinfo {pages} {1} (\bibinfo {year} {2015})}\BibitemShut
  {NoStop}%
\bibitem [{\citenamefont {L\'opez}\ \emph {et~al.}(2015)\citenamefont
  {L\'opez}, \citenamefont {Gachelin}, \citenamefont {Douarche}, \citenamefont
  {Auradou},\ and\ \citenamefont {Cl\'ement}}]{Clement_PRL_2015}%
  \BibitemOpen
  \bibfield  {author} {\bibinfo {author} {\bibfnamefont {H.~M.}\ \bibnamefont
  {L\'opez}}, \bibinfo {author} {\bibfnamefont {J.}~\bibnamefont {Gachelin}},
  \bibinfo {author} {\bibfnamefont {C.}~\bibnamefont {Douarche}}, \bibinfo
  {author} {\bibfnamefont {H.}~\bibnamefont {Auradou}}, \ and\ \bibinfo
  {author} {\bibfnamefont {E.}~\bibnamefont {Cl\'ement}},\ }\href {\doibase
  10.1103/PhysRevLett.115.028301} {\bibfield  {journal} {\bibinfo  {journal}
  {Phys. Rev. Lett.}\ }\textbf {\bibinfo {volume} {115}},\ \bibinfo {pages}
  {028301} (\bibinfo {year} {2015})}\BibitemShut {NoStop}%
\bibitem [{\citenamefont {Cavagna}\ \emph {et~al.}(2017)\citenamefont
  {Cavagna}, \citenamefont {Conti}, \citenamefont {Creato}, \citenamefont
  {Del~Castello}, \citenamefont {Giardina}, \citenamefont {Grigera},
  \citenamefont {Melillo}, \citenamefont {Parisi},\ and\ \citenamefont
  {Viale}}]{Cavagna_2017}%
  \BibitemOpen
  \bibfield  {author} {\bibinfo {author} {\bibfnamefont {A.}~\bibnamefont
  {Cavagna}}, \bibinfo {author} {\bibfnamefont {D.}~\bibnamefont {Conti}},
  \bibinfo {author} {\bibfnamefont {C.}~\bibnamefont {Creato}}, \bibinfo
  {author} {\bibfnamefont {L.}~\bibnamefont {Del~Castello}}, \bibinfo {author}
  {\bibfnamefont {I.}~\bibnamefont {Giardina}}, \bibinfo {author}
  {\bibfnamefont {T.~S.}\ \bibnamefont {Grigera}}, \bibinfo {author}
  {\bibfnamefont {S.}~\bibnamefont {Melillo}}, \bibinfo {author} {\bibfnamefont
  {L.}~\bibnamefont {Parisi}}, \ and\ \bibinfo {author} {\bibfnamefont
  {M.}~\bibnamefont {Viale}},\ }\href@noop {} {\bibfield  {journal} {\bibinfo
  {journal} {Nature Physics}\ }\textbf {\bibinfo {volume} {13}},\ \bibinfo
  {pages} {914} (\bibinfo {year} {2017})}\BibitemShut {NoStop}%
\bibitem [{\citenamefont {Sokolov}\ \emph {et~al.}(2018)\citenamefont
  {Sokolov}, \citenamefont {Rubio}, \citenamefont {Brady},\ and\ \citenamefont
  {Aranson}}]{Sokolov_2018}%
  \BibitemOpen
  \bibfield  {author} {\bibinfo {author} {\bibfnamefont {A.}~\bibnamefont
  {Sokolov}}, \bibinfo {author} {\bibfnamefont {L.~D.}\ \bibnamefont {Rubio}},
  \bibinfo {author} {\bibfnamefont {J.~F.}\ \bibnamefont {Brady}}, \ and\
  \bibinfo {author} {\bibfnamefont {I.~S.}\ \bibnamefont {Aranson}},\
  }\href@noop {} {\bibfield  {journal} {\bibinfo  {journal} {Nature
  communications}\ }\textbf {\bibinfo {volume} {9}},\ \bibinfo {pages} {1}
  (\bibinfo {year} {2018})}\BibitemShut {NoStop}%
\bibitem [{\citenamefont {Cherstvy}\ \emph {et~al.}(2018)\citenamefont
  {Cherstvy}, \citenamefont {Nagel}, \citenamefont {Beta},\ and\ \citenamefont
  {Metzler}}]{Metzler_2018}%
  \BibitemOpen
  \bibfield  {author} {\bibinfo {author} {\bibfnamefont {A.~G.}\ \bibnamefont
  {Cherstvy}}, \bibinfo {author} {\bibfnamefont {O.}~\bibnamefont {Nagel}},
  \bibinfo {author} {\bibfnamefont {C.}~\bibnamefont {Beta}}, \ and\ \bibinfo
  {author} {\bibfnamefont {R.}~\bibnamefont {Metzler}},\ }\href@noop {}
  {\bibfield  {journal} {\bibinfo  {journal} {Physical Chemistry Chemical
  Physics}\ }\textbf {\bibinfo {volume} {20}},\ \bibinfo {pages} {23034}
  (\bibinfo {year} {2018})}\BibitemShut {NoStop}%
\bibitem [{\citenamefont {Lagarde}\ \emph {et~al.}(2020)\citenamefont
  {Lagarde}, \citenamefont {Dag{\`e}s}, \citenamefont {Nemoto}, \citenamefont
  {D{\'e}mery}, \citenamefont {Bartolo},\ and\ \citenamefont
  {Gibaud}}]{Takahiro_2020}%
  \BibitemOpen
  \bibfield  {author} {\bibinfo {author} {\bibfnamefont {A.}~\bibnamefont
  {Lagarde}}, \bibinfo {author} {\bibfnamefont {N.}~\bibnamefont {Dag{\`e}s}},
  \bibinfo {author} {\bibfnamefont {T.}~\bibnamefont {Nemoto}}, \bibinfo
  {author} {\bibfnamefont {V.}~\bibnamefont {D{\'e}mery}}, \bibinfo {author}
  {\bibfnamefont {D.}~\bibnamefont {Bartolo}}, \ and\ \bibinfo {author}
  {\bibfnamefont {T.}~\bibnamefont {Gibaud}},\ }\href@noop {} {\bibfield
  {journal} {\bibinfo  {journal} {Soft Matter}\ }\textbf {\bibinfo {volume}
  {16}},\ \bibinfo {pages} {7503} (\bibinfo {year} {2020})}\BibitemShut
  {NoStop}%
\bibitem [{\citenamefont {Underhill}\ \emph {et~al.}(2008)\citenamefont
  {Underhill}, \citenamefont {Hernandez-Ortiz},\ and\ \citenamefont
  {Graham}}]{Graham_PRL_2008}%
  \BibitemOpen
  \bibfield  {author} {\bibinfo {author} {\bibfnamefont {P.~T.}\ \bibnamefont
  {Underhill}}, \bibinfo {author} {\bibfnamefont {J.~P.}\ \bibnamefont
  {Hernandez-Ortiz}}, \ and\ \bibinfo {author} {\bibfnamefont {M.~D.}\
  \bibnamefont {Graham}},\ }\href {\doibase 10.1103/PhysRevLett.100.248101}
  {\bibfield  {journal} {\bibinfo  {journal} {Phys. Rev. Lett.}\ }\textbf
  {\bibinfo {volume} {100}},\ \bibinfo {pages} {248101} (\bibinfo {year}
  {2008})}\BibitemShut {NoStop}%
\bibitem [{\citenamefont {Liao}\ \emph {et~al.}(2019)\citenamefont {Liao},
  \citenamefont {Han}, \citenamefont {Fruchart}, \citenamefont {Vitelli},\ and\
  \citenamefont {Vaikuntanathan}}]{Suriyanarayanan_2019}%
  \BibitemOpen
  \bibfield  {author} {\bibinfo {author} {\bibfnamefont {Z.}~\bibnamefont
  {Liao}}, \bibinfo {author} {\bibfnamefont {M.}~\bibnamefont {Han}}, \bibinfo
  {author} {\bibfnamefont {M.}~\bibnamefont {Fruchart}}, \bibinfo {author}
  {\bibfnamefont {V.}~\bibnamefont {Vitelli}}, \ and\ \bibinfo {author}
  {\bibfnamefont {S.}~\bibnamefont {Vaikuntanathan}},\ }\href@noop {}
  {\bibfield  {journal} {\bibinfo  {journal} {The Journal of chemical physics}\
  }\textbf {\bibinfo {volume} {151}},\ \bibinfo {pages} {194108} (\bibinfo
  {year} {2019})}\BibitemShut {NoStop}%
\bibitem [{\citenamefont {Breoni}\ \emph {et~al.}(2020)\citenamefont {Breoni},
  \citenamefont {Schmiedeberg},\ and\ \citenamefont {L\"owen}}]{Lowen_2020}%
  \BibitemOpen
  \bibfield  {author} {\bibinfo {author} {\bibfnamefont {D.}~\bibnamefont
  {Breoni}}, \bibinfo {author} {\bibfnamefont {M.}~\bibnamefont
  {Schmiedeberg}}, \ and\ \bibinfo {author} {\bibfnamefont {H.}~\bibnamefont
  {L\"owen}},\ }\href {\doibase 10.1103/PhysRevE.102.062604} {\bibfield
  {journal} {\bibinfo  {journal} {Phys. Rev. E}\ }\textbf {\bibinfo {volume}
  {102}},\ \bibinfo {pages} {062604} (\bibinfo {year} {2020})}\BibitemShut
  {NoStop}%
\bibitem [{\citenamefont {Hatwalne}\ \emph {et~al.}(2004)\citenamefont
  {Hatwalne}, \citenamefont {Ramaswamy}, \citenamefont {Rao},\ and\
  \citenamefont {Simha}}]{Hatwalne_PRL_2004}%
  \BibitemOpen
  \bibfield  {author} {\bibinfo {author} {\bibfnamefont {Y.}~\bibnamefont
  {Hatwalne}}, \bibinfo {author} {\bibfnamefont {S.}~\bibnamefont {Ramaswamy}},
  \bibinfo {author} {\bibfnamefont {M.}~\bibnamefont {Rao}}, \ and\ \bibinfo
  {author} {\bibfnamefont {R.~A.}\ \bibnamefont {Simha}},\ }\href {\doibase
  10.1103/PhysRevLett.92.118101} {\bibfield  {journal} {\bibinfo  {journal}
  {Phys. Rev. Lett.}\ }\textbf {\bibinfo {volume} {92}},\ \bibinfo {pages}
  {118101} (\bibinfo {year} {2004})}\BibitemShut {NoStop}%
\bibitem [{\citenamefont {Belan}\ and\ \citenamefont
  {Kardar}(2019)}]{Kardar_2019}%
  \BibitemOpen
  \bibfield  {author} {\bibinfo {author} {\bibfnamefont {S.}~\bibnamefont
  {Belan}}\ and\ \bibinfo {author} {\bibfnamefont {M.}~\bibnamefont {Kardar}},\
  }\href@noop {} {\bibfield  {journal} {\bibinfo  {journal} {The Journal of
  Chemical Physics}\ }\textbf {\bibinfo {volume} {150}},\ \bibinfo {pages}
  {064907} (\bibinfo {year} {2019})}\BibitemShut {NoStop}%
\bibitem [{\citenamefont {Dulaney}\ and\ \citenamefont
  {Brady}(2020)}]{Brady_PRE_2020}%
  \BibitemOpen
  \bibfield  {author} {\bibinfo {author} {\bibfnamefont {A.~R.}\ \bibnamefont
  {Dulaney}}\ and\ \bibinfo {author} {\bibfnamefont {J.~F.}\ \bibnamefont
  {Brady}},\ }\href {\doibase 10.1103/PhysRevE.101.052609} {\bibfield
  {journal} {\bibinfo  {journal} {Phys. Rev. E}\ }\textbf {\bibinfo {volume}
  {101}},\ \bibinfo {pages} {052609} (\bibinfo {year} {2020})}\BibitemShut
  {NoStop}%
\bibitem [{\citenamefont {Le~Doussal}\ \emph {et~al.}(2020)\citenamefont
  {Le~Doussal}, \citenamefont {Majumdar},\ and\ \citenamefont
  {Schehr}}]{majumdar_epl_2020}%
  \BibitemOpen
  \bibfield  {author} {\bibinfo {author} {\bibfnamefont {P.}~\bibnamefont
  {Le~Doussal}}, \bibinfo {author} {\bibfnamefont {S.~N.}\ \bibnamefont
  {Majumdar}}, \ and\ \bibinfo {author} {\bibfnamefont {G.}~\bibnamefont
  {Schehr}},\ }\href@noop {} {\bibfield  {journal} {\bibinfo  {journal} {EPL
  (Europhysics Letters)}\ }\textbf {\bibinfo {volume} {130}},\ \bibinfo {pages}
  {40002} (\bibinfo {year} {2020})}\BibitemShut {NoStop}%
\bibitem [{\citenamefont {Takatori}\ \emph {et~al.}(2016)\citenamefont
  {Takatori}, \citenamefont {De~Dier}, \citenamefont {Vermant},\ and\
  \citenamefont {Brady}}]{takatori2016}%
  \BibitemOpen
  \bibfield  {author} {\bibinfo {author} {\bibfnamefont {S.~C.}\ \bibnamefont
  {Takatori}}, \bibinfo {author} {\bibfnamefont {R.}~\bibnamefont {De~Dier}},
  \bibinfo {author} {\bibfnamefont {J.}~\bibnamefont {Vermant}}, \ and\
  \bibinfo {author} {\bibfnamefont {J.~F.}\ \bibnamefont {Brady}},\ }\href@noop
  {} {\bibfield  {journal} {\bibinfo  {journal} {Nature communications}\
  }\textbf {\bibinfo {volume} {7}},\ \bibinfo {pages} {1} (\bibinfo {year}
  {2016})}\BibitemShut {NoStop}%
\bibitem [{\citenamefont {Dolai}\ \emph {et~al.}(2020)\citenamefont {Dolai},
  \citenamefont {Das}, \citenamefont {Kundu}, \citenamefont {Dasgupta},
  \citenamefont {Dhar},\ and\ \citenamefont {Kumar}}]{dolai2020}%
  \BibitemOpen
  \bibfield  {author} {\bibinfo {author} {\bibfnamefont {P.}~\bibnamefont
  {Dolai}}, \bibinfo {author} {\bibfnamefont {A.}~\bibnamefont {Das}}, \bibinfo
  {author} {\bibfnamefont {A.}~\bibnamefont {Kundu}}, \bibinfo {author}
  {\bibfnamefont {C.}~\bibnamefont {Dasgupta}}, \bibinfo {author}
  {\bibfnamefont {A.}~\bibnamefont {Dhar}}, \ and\ \bibinfo {author}
  {\bibfnamefont {K.~V.}\ \bibnamefont {Kumar}},\ }\href@noop {} {\bibfield
  {journal} {\bibinfo  {journal} {Soft Matter}\ }\textbf {\bibinfo {volume}
  {16}},\ \bibinfo {pages} {7077} (\bibinfo {year} {2020})}\BibitemShut
  {NoStop}%
\bibitem [{\citenamefont {Peruani}\ \emph {et~al.}(2012)\citenamefont
  {Peruani}, \citenamefont {Starru\ss{}}, \citenamefont {Jakovljevic},
  \citenamefont {S\o{}gaard-Andersen}, \citenamefont {Deutsch},\ and\
  \citenamefont {B\"ar}}]{Bar_2012}%
  \BibitemOpen
  \bibfield  {author} {\bibinfo {author} {\bibfnamefont {F.}~\bibnamefont
  {Peruani}}, \bibinfo {author} {\bibfnamefont {J.}~\bibnamefont
  {Starru\ss{}}}, \bibinfo {author} {\bibfnamefont {V.}~\bibnamefont
  {Jakovljevic}}, \bibinfo {author} {\bibfnamefont {L.}~\bibnamefont
  {S\o{}gaard-Andersen}}, \bibinfo {author} {\bibfnamefont {A.}~\bibnamefont
  {Deutsch}}, \ and\ \bibinfo {author} {\bibfnamefont {M.}~\bibnamefont
  {B\"ar}},\ }\href {\doibase 10.1103/PhysRevLett.108.098102} {\bibfield
  {journal} {\bibinfo  {journal} {Phys. Rev. Lett.}\ }\textbf {\bibinfo
  {volume} {108}},\ \bibinfo {pages} {098102} (\bibinfo {year}
  {2012})}\BibitemShut {NoStop}%
\bibitem [{\citenamefont {Slowman}\ \emph {et~al.}(2016)\citenamefont
  {Slowman}, \citenamefont {Evans},\ and\ \citenamefont
  {Blythe}}]{Blythe_PRL_2016}%
  \BibitemOpen
  \bibfield  {author} {\bibinfo {author} {\bibfnamefont {A.~B.}\ \bibnamefont
  {Slowman}}, \bibinfo {author} {\bibfnamefont {M.~R.}\ \bibnamefont {Evans}},
  \ and\ \bibinfo {author} {\bibfnamefont {R.~A.}\ \bibnamefont {Blythe}},\
  }\href {\doibase 10.1103/PhysRevLett.116.218101} {\bibfield  {journal}
  {\bibinfo  {journal} {Phys. Rev. Lett.}\ }\textbf {\bibinfo {volume} {116}},\
  \bibinfo {pages} {218101} (\bibinfo {year} {2016})}\BibitemShut {NoStop}%
\bibitem [{\citenamefont {Kourbane-Houssene}\ \emph {et~al.}(2018)\citenamefont
  {Kourbane-Houssene}, \citenamefont {Erignoux}, \citenamefont {Bodineau},\
  and\ \citenamefont {Tailleur}}]{Tailleur_PRL_2018}%
  \BibitemOpen
  \bibfield  {author} {\bibinfo {author} {\bibfnamefont {M.}~\bibnamefont
  {Kourbane-Houssene}}, \bibinfo {author} {\bibfnamefont {C.}~\bibnamefont
  {Erignoux}}, \bibinfo {author} {\bibfnamefont {T.}~\bibnamefont {Bodineau}},
  \ and\ \bibinfo {author} {\bibfnamefont {J.}~\bibnamefont {Tailleur}},\
  }\href {\doibase 10.1103/PhysRevLett.120.268003} {\bibfield  {journal}
  {\bibinfo  {journal} {Phys. Rev. Lett.}\ }\textbf {\bibinfo {volume} {120}},\
  \bibinfo {pages} {268003} (\bibinfo {year} {2018})}\BibitemShut {NoStop}%
\bibitem [{\citenamefont {Dal~Cengio}\ \emph {et~al.}(2019)\citenamefont
  {Dal~Cengio}, \citenamefont {Levis},\ and\ \citenamefont
  {Pagonabarraga}}]{pagonabarraga_2019}%
  \BibitemOpen
  \bibfield  {author} {\bibinfo {author} {\bibfnamefont {S.}~\bibnamefont
  {Dal~Cengio}}, \bibinfo {author} {\bibfnamefont {D.}~\bibnamefont {Levis}}, \
  and\ \bibinfo {author} {\bibfnamefont {I.}~\bibnamefont {Pagonabarraga}},\
  }\href {\doibase 10.1103/PhysRevLett.123.238003} {\bibfield  {journal}
  {\bibinfo  {journal} {Phys. Rev. Lett.}\ }\textbf {\bibinfo {volume} {123}},\
  \bibinfo {pages} {238003} (\bibinfo {year} {2019})}\BibitemShut {NoStop}%
\bibitem [{\citenamefont {Metson}\ \emph {et~al.}(2020)\citenamefont {Metson},
  \citenamefont {Evans},\ and\ \citenamefont {Blythe}}]{Blythe_2020}%
  \BibitemOpen
  \bibfield  {author} {\bibinfo {author} {\bibfnamefont {M.~J.}\ \bibnamefont
  {Metson}}, \bibinfo {author} {\bibfnamefont {M.~R.}\ \bibnamefont {Evans}}, \
  and\ \bibinfo {author} {\bibfnamefont {R.~A.}\ \bibnamefont {Blythe}},\
  }\href@noop {} {\bibfield  {journal} {\bibinfo  {journal} {Journal of
  Statistical Mechanics: Theory and Experiment}\ }\textbf {\bibinfo {volume}
  {2020}},\ \bibinfo {pages} {103207} (\bibinfo {year} {2020})}\BibitemShut
  {NoStop}%
\bibitem [{\citenamefont {Shi}\ \emph {et~al.}(2020)\citenamefont {Shi},
  \citenamefont {Fausti}, \citenamefont {Chat\'e}, \citenamefont {Nardini},\
  and\ \citenamefont {Solon}}]{Chate_2020}%
  \BibitemOpen
  \bibfield  {author} {\bibinfo {author} {\bibfnamefont {X.-q.}\ \bibnamefont
  {Shi}}, \bibinfo {author} {\bibfnamefont {G.}~\bibnamefont {Fausti}},
  \bibinfo {author} {\bibfnamefont {H.}~\bibnamefont {Chat\'e}}, \bibinfo
  {author} {\bibfnamefont {C.}~\bibnamefont {Nardini}}, \ and\ \bibinfo
  {author} {\bibfnamefont {A.}~\bibnamefont {Solon}},\ }\href {\doibase
  10.1103/PhysRevLett.125.168001} {\bibfield  {journal} {\bibinfo  {journal}
  {Phys. Rev. Lett.}\ }\textbf {\bibinfo {volume} {125}},\ \bibinfo {pages}
  {168001} (\bibinfo {year} {2020})}\BibitemShut {NoStop}%
\bibitem [{\citenamefont {Agranov}\ \emph {et~al.}(2021)\citenamefont
  {Agranov}, \citenamefont {Ro}, \citenamefont {Kafri},\ and\ \citenamefont
  {Lecomte}}]{Kafri_2021}%
  \BibitemOpen
  \bibfield  {author} {\bibinfo {author} {\bibfnamefont {T.}~\bibnamefont
  {Agranov}}, \bibinfo {author} {\bibfnamefont {S.}~\bibnamefont {Ro}},
  \bibinfo {author} {\bibfnamefont {Y.}~\bibnamefont {Kafri}}, \ and\ \bibinfo
  {author} {\bibfnamefont {V.}~\bibnamefont {Lecomte}},\ }\href@noop {}
  {\bibfield  {journal} {\bibinfo  {journal} {Journal of Statistical Mechanics:
  Theory and Experiment}\ }\textbf {\bibinfo {volume} {2021}},\ \bibinfo
  {pages} {083208} (\bibinfo {year} {2021})}\BibitemShut {NoStop}%
\bibitem [{\citenamefont {Chepizhko}\ and\ \citenamefont
  {Peruani}(2013)}]{Peruani_PRL_2013}%
  \BibitemOpen
  \bibfield  {author} {\bibinfo {author} {\bibfnamefont {O.}~\bibnamefont
  {Chepizhko}}\ and\ \bibinfo {author} {\bibfnamefont {F.}~\bibnamefont
  {Peruani}},\ }\href {\doibase 10.1103/PhysRevLett.111.160604} {\bibfield
  {journal} {\bibinfo  {journal} {Phys. Rev. Lett.}\ }\textbf {\bibinfo
  {volume} {111}},\ \bibinfo {pages} {160604} (\bibinfo {year}
  {2013})}\BibitemShut {NoStop}%
\bibitem [{\citenamefont {Levis}\ and\ \citenamefont
  {Berthier}(2014)}]{Levis_berthier2014}%
  \BibitemOpen
  \bibfield  {author} {\bibinfo {author} {\bibfnamefont {D.}~\bibnamefont
  {Levis}}\ and\ \bibinfo {author} {\bibfnamefont {L.}~\bibnamefont
  {Berthier}},\ }\href {\doibase 10.1103/PhysRevE.89.062301} {\bibfield
  {journal} {\bibinfo  {journal} {Phys. Rev. E}\ }\textbf {\bibinfo {volume}
  {89}},\ \bibinfo {pages} {062301} (\bibinfo {year} {2014})}\BibitemShut
  {NoStop}%
\bibitem [{\citenamefont {Liluashvili}\ \emph {et~al.}(2017)\citenamefont
  {Liluashvili}, \citenamefont {\'Onody},\ and\ \citenamefont
  {Voigtmann}}]{Voigtmann_2017_PRE}%
  \BibitemOpen
  \bibfield  {author} {\bibinfo {author} {\bibfnamefont {A.}~\bibnamefont
  {Liluashvili}}, \bibinfo {author} {\bibfnamefont {J.}~\bibnamefont
  {\'Onody}}, \ and\ \bibinfo {author} {\bibfnamefont {T.}~\bibnamefont
  {Voigtmann}},\ }\href {\doibase 10.1103/PhysRevE.96.062608} {\bibfield
  {journal} {\bibinfo  {journal} {Phys. Rev. E}\ }\textbf {\bibinfo {volume}
  {96}},\ \bibinfo {pages} {062608} (\bibinfo {year} {2017})}\BibitemShut
  {NoStop}%
\bibitem [{\citenamefont {Bertrand}\ \emph {et~al.}(2018)\citenamefont
  {Bertrand}, \citenamefont {Zhao}, \citenamefont {B\'enichou}, \citenamefont
  {Tailleur},\ and\ \citenamefont {Voituriez}}]{Benichou_2018}%
  \BibitemOpen
  \bibfield  {author} {\bibinfo {author} {\bibfnamefont {T.}~\bibnamefont
  {Bertrand}}, \bibinfo {author} {\bibfnamefont {Y.}~\bibnamefont {Zhao}},
  \bibinfo {author} {\bibfnamefont {O.}~\bibnamefont {B\'enichou}}, \bibinfo
  {author} {\bibfnamefont {J.}~\bibnamefont {Tailleur}}, \ and\ \bibinfo
  {author} {\bibfnamefont {R.}~\bibnamefont {Voituriez}},\ }\href {\doibase
  10.1103/PhysRevLett.120.198103} {\bibfield  {journal} {\bibinfo  {journal}
  {Phys. Rev. Lett.}\ }\textbf {\bibinfo {volume} {120}},\ \bibinfo {pages}
  {198103} (\bibinfo {year} {2018})}\BibitemShut {NoStop}%
\bibitem [{\citenamefont {Put}\ \emph {et~al.}(2019)\citenamefont {Put},
  \citenamefont {Berx},\ and\ \citenamefont {Vanderzande}}]{Vanderzande_2019}%
  \BibitemOpen
  \bibfield  {author} {\bibinfo {author} {\bibfnamefont {S.}~\bibnamefont
  {Put}}, \bibinfo {author} {\bibfnamefont {J.}~\bibnamefont {Berx}}, \ and\
  \bibinfo {author} {\bibfnamefont {C.}~\bibnamefont {Vanderzande}},\
  }\href@noop {} {\bibfield  {journal} {\bibinfo  {journal} {Journal of
  Statistical Mechanics: Theory and Experiment}\ }\textbf {\bibinfo {volume}
  {2019}},\ \bibinfo {pages} {123205} (\bibinfo {year} {2019})}\BibitemShut
  {NoStop}%
\bibitem [{\citenamefont {Singh}\ and\ \citenamefont
  {Kundu}(2021)}]{Anupam_2021}%
  \BibitemOpen
  \bibfield  {author} {\bibinfo {author} {\bibfnamefont {P.}~\bibnamefont
  {Singh}}\ and\ \bibinfo {author} {\bibfnamefont {A.}~\bibnamefont {Kundu}},\
  }\href@noop {} {\bibfield  {journal} {\bibinfo  {journal} {Journal of Physics
  A: Mathematical and Theoretical}\ }\textbf {\bibinfo {volume} {54}},\
  \bibinfo {pages} {305001} (\bibinfo {year} {2021})}\BibitemShut {NoStop}%
\bibitem [{\citenamefont {Debets}\ \emph {et~al.}(2021)\citenamefont {Debets},
  \citenamefont {de~Wit},\ and\ \citenamefont {Janssen}}]{Debets_PRL_2021}%
  \BibitemOpen
  \bibfield  {author} {\bibinfo {author} {\bibfnamefont {V.~E.}\ \bibnamefont
  {Debets}}, \bibinfo {author} {\bibfnamefont {X.~M.}\ \bibnamefont {de~Wit}},
  \ and\ \bibinfo {author} {\bibfnamefont {L.~M.~C.}\ \bibnamefont {Janssen}},\
  }\href {\doibase 10.1103/PhysRevLett.127.278002} {\bibfield  {journal}
  {\bibinfo  {journal} {Phys. Rev. Lett.}\ }\textbf {\bibinfo {volume} {127}},\
  \bibinfo {pages} {278002} (\bibinfo {year} {2021})}\BibitemShut {NoStop}%
\bibitem [{\citenamefont {Rizkallah}\ \emph {et~al.}(2022)\citenamefont
  {Rizkallah}, \citenamefont {Sarracino}, \citenamefont {B\'enichou},\ and\
  \citenamefont {Illien}}]{Benichou_2022}%
  \BibitemOpen
  \bibfield  {author} {\bibinfo {author} {\bibfnamefont {P.}~\bibnamefont
  {Rizkallah}}, \bibinfo {author} {\bibfnamefont {A.}~\bibnamefont
  {Sarracino}}, \bibinfo {author} {\bibfnamefont {O.}~\bibnamefont
  {B\'enichou}}, \ and\ \bibinfo {author} {\bibfnamefont {P.}~\bibnamefont
  {Illien}},\ }\href {\doibase 10.1103/PhysRevLett.128.038001} {\bibfield
  {journal} {\bibinfo  {journal} {Phys. Rev. Lett.}\ }\textbf {\bibinfo
  {volume} {128}},\ \bibinfo {pages} {038001} (\bibinfo {year}
  {2022})}\BibitemShut {NoStop}%
\bibitem [{\citenamefont {Granek}\ \emph {et~al.}(2022)\citenamefont {Granek},
  \citenamefont {Kafri},\ and\ \citenamefont {Tailleur}}]{Tailleur_2022}%
  \BibitemOpen
  \bibfield  {author} {\bibinfo {author} {\bibfnamefont {O.}~\bibnamefont
  {Granek}}, \bibinfo {author} {\bibfnamefont {Y.}~\bibnamefont {Kafri}}, \
  and\ \bibinfo {author} {\bibfnamefont {J.}~\bibnamefont {Tailleur}},\
  }\href@noop {} {\bibfield  {journal} {\bibinfo  {journal} {Physical Review
  Letters}\ }\textbf {\bibinfo {volume} {129}},\ \bibinfo {pages} {038001}
  (\bibinfo {year} {2022})}\BibitemShut {NoStop}%
\bibitem [{\citenamefont {Kurzthaler}\ \emph {et~al.}(2018)\citenamefont
  {Kurzthaler}, \citenamefont {Devailly}, \citenamefont {Arlt}, \citenamefont
  {Franosch}, \citenamefont {Poon}, \citenamefont {Martinez},\ and\
  \citenamefont {Brown}}]{Kurzthaler_PRL_2018}%
  \BibitemOpen
  \bibfield  {author} {\bibinfo {author} {\bibfnamefont {C.}~\bibnamefont
  {Kurzthaler}}, \bibinfo {author} {\bibfnamefont {C.}~\bibnamefont
  {Devailly}}, \bibinfo {author} {\bibfnamefont {J.}~\bibnamefont {Arlt}},
  \bibinfo {author} {\bibfnamefont {T.}~\bibnamefont {Franosch}}, \bibinfo
  {author} {\bibfnamefont {W.~C.~K.}\ \bibnamefont {Poon}}, \bibinfo {author}
  {\bibfnamefont {V.~A.}\ \bibnamefont {Martinez}}, \ and\ \bibinfo {author}
  {\bibfnamefont {A.~T.}\ \bibnamefont {Brown}},\ }\href {\doibase
  10.1103/PhysRevLett.121.078001} {\bibfield  {journal} {\bibinfo  {journal}
  {Phys. Rev. Lett.}\ }\textbf {\bibinfo {volume} {121}},\ \bibinfo {pages}
  {078001} (\bibinfo {year} {2018})}\BibitemShut {NoStop}%
\bibitem [{\citenamefont {Irani}\ \emph {et~al.}(2022)\citenamefont {Irani},
  \citenamefont {Mokhtari},\ and\ \citenamefont {Zippelius}}]{Irani_PRL_2022}%
  \BibitemOpen
  \bibfield  {author} {\bibinfo {author} {\bibfnamefont {E.}~\bibnamefont
  {Irani}}, \bibinfo {author} {\bibfnamefont {Z.}~\bibnamefont {Mokhtari}}, \
  and\ \bibinfo {author} {\bibfnamefont {A.}~\bibnamefont {Zippelius}},\ }\href
  {\doibase 10.1103/PhysRevLett.128.144501} {\bibfield  {journal} {\bibinfo
  {journal} {Phys. Rev. Lett.}\ }\textbf {\bibinfo {volume} {128}},\ \bibinfo
  {pages} {144501} (\bibinfo {year} {2022})}\BibitemShut {NoStop}%
\bibitem [{\citenamefont {Arnoulx~de Pirey}\ \emph {et~al.}(2019)\citenamefont
  {Arnoulx~de Pirey}, \citenamefont {Lozano},\ and\ \citenamefont {van
  Wijland}}]{Wijland_2019_PRL}%
  \BibitemOpen
  \bibfield  {author} {\bibinfo {author} {\bibfnamefont {T.}~\bibnamefont
  {Arnoulx~de Pirey}}, \bibinfo {author} {\bibfnamefont {G.}~\bibnamefont
  {Lozano}}, \ and\ \bibinfo {author} {\bibfnamefont {F.}~\bibnamefont {van
  Wijland}},\ }\href {\doibase 10.1103/PhysRevLett.123.260602} {\bibfield
  {journal} {\bibinfo  {journal} {Phys. Rev. Lett.}\ }\textbf {\bibinfo
  {volume} {123}},\ \bibinfo {pages} {260602} (\bibinfo {year}
  {2019})}\BibitemShut {NoStop}%
\bibitem [{\citenamefont {Takatori}\ \emph {et~al.}(2014)\citenamefont
  {Takatori}, \citenamefont {Yan},\ and\ \citenamefont
  {Brady}}]{Brady_PRL2014}%
  \BibitemOpen
  \bibfield  {author} {\bibinfo {author} {\bibfnamefont {S.~C.}\ \bibnamefont
  {Takatori}}, \bibinfo {author} {\bibfnamefont {W.}~\bibnamefont {Yan}}, \
  and\ \bibinfo {author} {\bibfnamefont {J.~F.}\ \bibnamefont {Brady}},\ }\href
  {\doibase 10.1103/PhysRevLett.113.028103} {\bibfield  {journal} {\bibinfo
  {journal} {Phys. Rev. Lett.}\ }\textbf {\bibinfo {volume} {113}},\ \bibinfo
  {pages} {028103} (\bibinfo {year} {2014})}\BibitemShut {NoStop}%
\bibitem [{\citenamefont {Solon}\ \emph {et~al.}(2015)\citenamefont {Solon},
  \citenamefont {Stenhammar}, \citenamefont {Wittkowski}, \citenamefont
  {Kardar}, \citenamefont {Kafri}, \citenamefont {Cates},\ and\ \citenamefont
  {Tailleur}}]{Solon_2015_PRL}%
  \BibitemOpen
  \bibfield  {author} {\bibinfo {author} {\bibfnamefont {A.~P.}\ \bibnamefont
  {Solon}}, \bibinfo {author} {\bibfnamefont {J.}~\bibnamefont {Stenhammar}},
  \bibinfo {author} {\bibfnamefont {R.}~\bibnamefont {Wittkowski}}, \bibinfo
  {author} {\bibfnamefont {M.}~\bibnamefont {Kardar}}, \bibinfo {author}
  {\bibfnamefont {Y.}~\bibnamefont {Kafri}}, \bibinfo {author} {\bibfnamefont
  {M.~E.}\ \bibnamefont {Cates}}, \ and\ \bibinfo {author} {\bibfnamefont
  {J.}~\bibnamefont {Tailleur}},\ }\href {\doibase
  10.1103/PhysRevLett.114.198301} {\bibfield  {journal} {\bibinfo  {journal}
  {Phys. Rev. Lett.}\ }\textbf {\bibinfo {volume} {114}},\ \bibinfo {pages}
  {198301} (\bibinfo {year} {2015})}\BibitemShut {NoStop}%
\bibitem [{\citenamefont {Omar}\ \emph {et~al.}(2023)\citenamefont {Omar},
  \citenamefont {Row}, \citenamefont {Mallory},\ and\ \citenamefont
  {Brady}}]{Brady_2023_PNAS}%
  \BibitemOpen
  \bibfield  {author} {\bibinfo {author} {\bibfnamefont {A.~K.}\ \bibnamefont
  {Omar}}, \bibinfo {author} {\bibfnamefont {H.}~\bibnamefont {Row}}, \bibinfo
  {author} {\bibfnamefont {S.~A.}\ \bibnamefont {Mallory}}, \ and\ \bibinfo
  {author} {\bibfnamefont {J.~F.}\ \bibnamefont {Brady}},\ }\href@noop {}
  {\bibfield  {journal} {\bibinfo  {journal} {Proceedings of the National
  Academy of Sciences}\ }\textbf {\bibinfo {volume} {120}},\ \bibinfo {pages}
  {e2219900120} (\bibinfo {year} {2023})}\BibitemShut {NoStop}%
\bibitem [{\citenamefont {Speck}(2021)}]{Speck_2021_PRE}%
  \BibitemOpen
  \bibfield  {author} {\bibinfo {author} {\bibfnamefont {T.}~\bibnamefont
  {Speck}},\ }\href@noop {} {\bibfield  {journal} {\bibinfo  {journal}
  {Physical Review E}\ }\textbf {\bibinfo {volume} {103}},\ \bibinfo {pages}
  {012607} (\bibinfo {year} {2021})}\BibitemShut {NoStop}%
\bibitem [{\citenamefont {Chakraborti}\ \emph {et~al.}(2016)\citenamefont
  {Chakraborti}, \citenamefont {Mishra},\ and\ \citenamefont
  {Pradhan}}]{Subhadip_PRE_2016}%
  \BibitemOpen
  \bibfield  {author} {\bibinfo {author} {\bibfnamefont {S.}~\bibnamefont
  {Chakraborti}}, \bibinfo {author} {\bibfnamefont {S.}~\bibnamefont {Mishra}},
  \ and\ \bibinfo {author} {\bibfnamefont {P.}~\bibnamefont {Pradhan}},\ }\href
  {\doibase 10.1103/PhysRevE.93.052606} {\bibfield  {journal} {\bibinfo
  {journal} {Phys. Rev. E}\ }\textbf {\bibinfo {volume} {93}},\ \bibinfo
  {pages} {052606} (\bibinfo {year} {2016})}\BibitemShut {NoStop}%
\bibitem [{\citenamefont {Chakraborty}\ and\ \citenamefont
  {Pradhan}(2023)}]{Tanmoy-condmat2023}%
  \BibitemOpen
  \bibfield  {author} {\bibinfo {author} {\bibfnamefont {T.}~\bibnamefont
  {Chakraborty}}\ and\ \bibinfo {author} {\bibfnamefont {P.}~\bibnamefont
  {Pradhan}},\ }\href@noop {} {\  (\bibinfo {year} {2023})},\ \Eprint
  {http://arxiv.org/abs/2309.02896} {arXiv:2309.02896 [cond-mat.stat-mech]}
  \BibitemShut {NoStop}%
\bibitem [{\citenamefont {Bodineau}\ and\ \citenamefont
  {Derrida}(2004)}]{Derrida_2004}%
  \BibitemOpen
  \bibfield  {author} {\bibinfo {author} {\bibfnamefont {T.}~\bibnamefont
  {Bodineau}}\ and\ \bibinfo {author} {\bibfnamefont {B.}~\bibnamefont
  {Derrida}},\ }\href@noop {} {\bibfield  {journal} {\bibinfo  {journal}
  {Physical review letters}\ }\textbf {\bibinfo {volume} {92}},\ \bibinfo
  {pages} {180601} (\bibinfo {year} {2004})}\BibitemShut {NoStop}%
\bibitem [{\citenamefont {Bertini}\ \emph {et~al.}(2001)\citenamefont
  {Bertini}, \citenamefont {De~Sole}, \citenamefont {Gabrielli}, \citenamefont
  {Jona-Lasinio},\ and\ \citenamefont {Landim}}]{Bertini_2001}%
  \BibitemOpen
  \bibfield  {author} {\bibinfo {author} {\bibfnamefont {L.}~\bibnamefont
  {Bertini}}, \bibinfo {author} {\bibfnamefont {A.}~\bibnamefont {De~Sole}},
  \bibinfo {author} {\bibfnamefont {D.}~\bibnamefont {Gabrielli}}, \bibinfo
  {author} {\bibfnamefont {G.}~\bibnamefont {Jona-Lasinio}}, \ and\ \bibinfo
  {author} {\bibfnamefont {C.}~\bibnamefont {Landim}},\ }\href@noop {}
  {\bibfield  {journal} {\bibinfo  {journal} {Physical Review Letters}\
  }\textbf {\bibinfo {volume} {87}},\ \bibinfo {pages} {040601} (\bibinfo
  {year} {2001})}\BibitemShut {NoStop}%
\bibitem [{\citenamefont {Kipnis}\ and\ \citenamefont
  {Landim}(1998)}]{Landim_1998}%
  \BibitemOpen
  \bibfield  {author} {\bibinfo {author} {\bibfnamefont {C.}~\bibnamefont
  {Kipnis}}\ and\ \bibinfo {author} {\bibfnamefont {C.}~\bibnamefont
  {Landim}},\ }\href@noop {} {\emph {\bibinfo {title} {Scaling limits of
  interacting particle systems}}},\ Vol.\ \bibinfo {volume} {320}\ (\bibinfo
  {publisher} {Springer Science \& Business Media},\ \bibinfo {year}
  {1998})\BibitemShut {NoStop}%
\bibitem [{\citenamefont {Spohn}(1991)}]{Spohn_2012}%
  \BibitemOpen
  \bibfield  {author} {\bibinfo {author} {\bibfnamefont {H.}~\bibnamefont
  {Spohn}},\ }\href@noop {} {\emph {\bibinfo {title} {Large scale dynamics of
  interacting particles}}}\ (\bibinfo  {publisher} {Springer Berlin
  Heidelberg},\ \bibinfo {year} {1991})\BibitemShut {NoStop}%
\bibitem [{\citenamefont {Arita}\ \emph {et~al.}(2014)\citenamefont {Arita},
  \citenamefont {Krapivsky},\ and\ \citenamefont {Mallick}}]{Krapivsky_2014}%
  \BibitemOpen
  \bibfield  {author} {\bibinfo {author} {\bibfnamefont {C.}~\bibnamefont
  {Arita}}, \bibinfo {author} {\bibfnamefont {P.~L.}\ \bibnamefont
  {Krapivsky}}, \ and\ \bibinfo {author} {\bibfnamefont {K.}~\bibnamefont
  {Mallick}},\ }\href {\doibase 10.1103/PhysRevE.90.052108} {\bibfield
  {journal} {\bibinfo  {journal} {Phys. Rev. E}\ }\textbf {\bibinfo {volume}
  {90}},\ \bibinfo {pages} {052108} (\bibinfo {year} {2014})}\BibitemShut
  {NoStop}%
\bibitem [{\citenamefont {Carlson}\ \emph {et~al.}(1993)\citenamefont
  {Carlson}, \citenamefont {Grannan},\ and\ \citenamefont
  {Swindle}}]{Carlson_1993}%
  \BibitemOpen
  \bibfield  {author} {\bibinfo {author} {\bibfnamefont {J.~M.}\ \bibnamefont
  {Carlson}}, \bibinfo {author} {\bibfnamefont {E.~R.}\ \bibnamefont
  {Grannan}}, \ and\ \bibinfo {author} {\bibfnamefont {G.~H.}\ \bibnamefont
  {Swindle}},\ }\href {\doibase 10.1103/PhysRevE.47.93} {\bibfield  {journal}
  {\bibinfo  {journal} {Phys. Rev. E}\ }\textbf {\bibinfo {volume} {47}},\
  \bibinfo {pages} {93} (\bibinfo {year} {1993})}\BibitemShut {NoStop}%
\bibitem [{\citenamefont {Malakar}\ \emph {et~al.}(2018)\citenamefont
  {Malakar}, \citenamefont {Jemseena}, \citenamefont {Kundu}, \citenamefont
  {Kumar}, \citenamefont {Sabhapandit}, \citenamefont {Majumdar}, \citenamefont
  {Redner},\ and\ \citenamefont {Dhar}}]{Malakar}%
  \BibitemOpen
  \bibfield  {author} {\bibinfo {author} {\bibfnamefont {K.}~\bibnamefont
  {Malakar}}, \bibinfo {author} {\bibfnamefont {V.}~\bibnamefont {Jemseena}},
  \bibinfo {author} {\bibfnamefont {A.}~\bibnamefont {Kundu}}, \bibinfo
  {author} {\bibfnamefont {K.~V.}\ \bibnamefont {Kumar}}, \bibinfo {author}
  {\bibfnamefont {S.}~\bibnamefont {Sabhapandit}}, \bibinfo {author}
  {\bibfnamefont {S.~N.}\ \bibnamefont {Majumdar}}, \bibinfo {author}
  {\bibfnamefont {S.}~\bibnamefont {Redner}}, \ and\ \bibinfo {author}
  {\bibfnamefont {A.}~\bibnamefont {Dhar}},\ }\href@noop {} {\bibfield
  {journal} {\bibinfo  {journal} {Journal of Statistical Mechanics: Theory and
  Experiment}\ }\textbf {\bibinfo {volume} {2018}},\ \bibinfo {pages} {043215}
  (\bibinfo {year} {2018})}\BibitemShut {NoStop}%
\bibitem [{\citenamefont {Slowman}\ \emph {et~al.}(2017)\citenamefont
  {Slowman}, \citenamefont {Evans},\ and\ \citenamefont
  {Blythe}}]{Blythe_2017}%
  \BibitemOpen
  \bibfield  {author} {\bibinfo {author} {\bibfnamefont {A.}~\bibnamefont
  {Slowman}}, \bibinfo {author} {\bibfnamefont {M.}~\bibnamefont {Evans}}, \
  and\ \bibinfo {author} {\bibfnamefont {R.}~\bibnamefont {Blythe}},\
  }\href@noop {} {\bibfield  {journal} {\bibinfo  {journal} {Journal of Physics
  A: Mathematical and Theoretical}\ }\textbf {\bibinfo {volume} {50}},\
  \bibinfo {pages} {375601} (\bibinfo {year} {2017})}\BibitemShut {NoStop}%
\bibitem [{\citenamefont {Das}\ \emph {et~al.}(2020)\citenamefont {Das},
  \citenamefont {Dhar},\ and\ \citenamefont {Kundu}}]{Das_2020}%
  \BibitemOpen
  \bibfield  {author} {\bibinfo {author} {\bibfnamefont {A.}~\bibnamefont
  {Das}}, \bibinfo {author} {\bibfnamefont {A.}~\bibnamefont {Dhar}}, \ and\
  \bibinfo {author} {\bibfnamefont {A.}~\bibnamefont {Kundu}},\ }\href@noop {}
  {\bibfield  {journal} {\bibinfo  {journal} {Journal of Physics A:
  Mathematical and Theoretical}\ }\textbf {\bibinfo {volume} {53}},\ \bibinfo
  {pages} {345003} (\bibinfo {year} {2020})}\BibitemShut {NoStop}%
\bibitem [{\citenamefont {Fily}\ and\ \citenamefont
  {Marchetti}(2012)}]{Fily_2012}%
  \BibitemOpen
  \bibfield  {author} {\bibinfo {author} {\bibfnamefont {Y.}~\bibnamefont
  {Fily}}\ and\ \bibinfo {author} {\bibfnamefont {M.~C.}\ \bibnamefont
  {Marchetti}},\ }\href {\doibase 10.1103/PhysRevLett.108.235702} {\bibfield
  {journal} {\bibinfo  {journal} {Phys. Rev. Lett.}\ }\textbf {\bibinfo
  {volume} {108}},\ \bibinfo {pages} {235702} (\bibinfo {year}
  {2012})}\BibitemShut {NoStop}%
\bibitem [{\citenamefont {Redner}\ \emph {et~al.}(2013)\citenamefont {Redner},
  \citenamefont {Hagan},\ and\ \citenamefont {Baskaran}}]{Baskaran_2013}%
  \BibitemOpen
  \bibfield  {author} {\bibinfo {author} {\bibfnamefont {G.~S.}\ \bibnamefont
  {Redner}}, \bibinfo {author} {\bibfnamefont {M.~F.}\ \bibnamefont {Hagan}}, \
  and\ \bibinfo {author} {\bibfnamefont {A.}~\bibnamefont {Baskaran}},\ }\href
  {\doibase 10.1103/PhysRevLett.110.055701} {\bibfield  {journal} {\bibinfo
  {journal} {Phys. Rev. Lett.}\ }\textbf {\bibinfo {volume} {110}},\ \bibinfo
  {pages} {055701} (\bibinfo {year} {2013})}\BibitemShut {NoStop}%
\bibitem [{\citenamefont {Bialk{\'e}}\ \emph {et~al.}(2013)\citenamefont
  {Bialk{\'e}}, \citenamefont {L{\"o}wen},\ and\ \citenamefont
  {Speck}}]{Speck_2013_EPL}%
  \BibitemOpen
  \bibfield  {author} {\bibinfo {author} {\bibfnamefont {J.}~\bibnamefont
  {Bialk{\'e}}}, \bibinfo {author} {\bibfnamefont {H.}~\bibnamefont
  {L{\"o}wen}}, \ and\ \bibinfo {author} {\bibfnamefont {T.}~\bibnamefont
  {Speck}},\ }\href@noop {} {\bibfield  {journal} {\bibinfo  {journal}
  {Europhysics Letters}\ }\textbf {\bibinfo {volume} {103}},\ \bibinfo {pages}
  {30008} (\bibinfo {year} {2013})}\BibitemShut {NoStop}%
\bibitem [{\citenamefont {Soto}\ and\ \citenamefont
  {Golestanian}(2014)}]{Soto_2014}%
  \BibitemOpen
  \bibfield  {author} {\bibinfo {author} {\bibfnamefont {R.}~\bibnamefont
  {Soto}}\ and\ \bibinfo {author} {\bibfnamefont {R.}~\bibnamefont
  {Golestanian}},\ }\href {\doibase 10.1103/PhysRevE.89.012706} {\bibfield
  {journal} {\bibinfo  {journal} {Phys. Rev. E}\ }\textbf {\bibinfo {volume}
  {89}},\ \bibinfo {pages} {012706} (\bibinfo {year} {2014})}\BibitemShut
  {NoStop}%
\bibitem [{\citenamefont {Schnitzer}(1993)}]{Schnitzer93}%
  \BibitemOpen
  \bibfield  {author} {\bibinfo {author} {\bibfnamefont {M.~J.}\ \bibnamefont
  {Schnitzer}},\ }\href {\doibase 10.1103/PhysRevE.48.2553} {\bibfield
  {journal} {\bibinfo  {journal} {Phys. Rev. E}\ }\textbf {\bibinfo {volume}
  {48}},\ \bibinfo {pages} {2553} (\bibinfo {year} {1993})}\BibitemShut
  {NoStop}%
\bibitem [{\citenamefont {Weiss}(2002)}]{WEISS2002_PRW}%
  \BibitemOpen
  \bibfield  {author} {\bibinfo {author} {\bibfnamefont {G.~H.}\ \bibnamefont
  {Weiss}},\ }\href {\doibase https://doi.org/10.1016/S0378-4371(02)00805-1}
  {\bibfield  {journal} {\bibinfo  {journal} {Physica A: Statistical Mechanics
  and its Applications}\ }\textbf {\bibinfo {volume} {311}},\ \bibinfo {pages}
  {381} (\bibinfo {year} {2002})}\BibitemShut {NoStop}%
\bibitem [{\citenamefont {Chakraborty}\ \emph {et~al.}(2020)\citenamefont
  {Chakraborty}, \citenamefont {Chakraborti}, \citenamefont {Das},\ and\
  \citenamefont {Pradhan}}]{Tanmoy_PRE_2020}%
  \BibitemOpen
  \bibfield  {author} {\bibinfo {author} {\bibfnamefont {T.}~\bibnamefont
  {Chakraborty}}, \bibinfo {author} {\bibfnamefont {S.}~\bibnamefont
  {Chakraborti}}, \bibinfo {author} {\bibfnamefont {A.}~\bibnamefont {Das}}, \
  and\ \bibinfo {author} {\bibfnamefont {P.}~\bibnamefont {Pradhan}},\ }\href
  {\doibase 10.1103/PhysRevE.101.052611} {\bibfield  {journal} {\bibinfo
  {journal} {Phys. Rev. E}\ }\textbf {\bibinfo {volume} {101}},\ \bibinfo
  {pages} {052611} (\bibinfo {year} {2020})}\BibitemShut {NoStop}%
\bibitem [{\citenamefont {Chakraborti}\ \emph {et~al.}(2021)\citenamefont
  {Chakraborti}, \citenamefont {Chakraborty}, \citenamefont {Das},
  \citenamefont {Dandekar},\ and\ \citenamefont {Pradhan}}]{subhadip_PRE_2021}%
  \BibitemOpen
  \bibfield  {author} {\bibinfo {author} {\bibfnamefont {S.}~\bibnamefont
  {Chakraborti}}, \bibinfo {author} {\bibfnamefont {T.}~\bibnamefont
  {Chakraborty}}, \bibinfo {author} {\bibfnamefont {A.}~\bibnamefont {Das}},
  \bibinfo {author} {\bibfnamefont {R.}~\bibnamefont {Dandekar}}, \ and\
  \bibinfo {author} {\bibfnamefont {P.}~\bibnamefont {Pradhan}},\ }\href
  {\doibase 10.1103/PhysRevE.103.042133} {\bibfield  {journal} {\bibinfo
  {journal} {Phys. Rev. E}\ }\textbf {\bibinfo {volume} {103}},\ \bibinfo
  {pages} {042133} (\bibinfo {year} {2021})}\BibitemShut {NoStop}%
\bibitem [{\citenamefont {Katz}\ \emph {et~al.}(1984)\citenamefont {Katz},
  \citenamefont {Lebowitz},\ and\ \citenamefont
  {Spohn}}]{katz1984nonequilibrium}%
  \BibitemOpen
  \bibfield  {author} {\bibinfo {author} {\bibfnamefont {S.}~\bibnamefont
  {Katz}}, \bibinfo {author} {\bibfnamefont {J.~L.}\ \bibnamefont {Lebowitz}},
  \ and\ \bibinfo {author} {\bibfnamefont {H.}~\bibnamefont {Spohn}},\
  }\href@noop {} {\bibfield  {journal} {\bibinfo  {journal} {Journal of
  statistical physics}\ }\textbf {\bibinfo {volume} {34}},\ \bibinfo {pages}
  {497} (\bibinfo {year} {1984})}\BibitemShut {NoStop}%
\bibitem [{\citenamefont {Banerjee}\ \emph {et~al.}(2022)\citenamefont
  {Banerjee}, \citenamefont {Jack},\ and\ \citenamefont
  {Cates}}]{Tirthankar_cates_2022}%
  \BibitemOpen
  \bibfield  {author} {\bibinfo {author} {\bibfnamefont {T.}~\bibnamefont
  {Banerjee}}, \bibinfo {author} {\bibfnamefont {R.~L.}\ \bibnamefont {Jack}},
  \ and\ \bibinfo {author} {\bibfnamefont {M.~E.}\ \bibnamefont {Cates}},\
  }\href@noop {} {\bibfield  {journal} {\bibinfo  {journal} {Journal of
  Statistical Mechanics: Theory and Experiment}\ }\textbf {\bibinfo {volume}
  {2022}},\ \bibinfo {pages} {013209} (\bibinfo {year} {2022})}\BibitemShut
  {NoStop}%
\bibitem [{\citenamefont {Mandal}\ \emph {et~al.}(2020)\citenamefont {Mandal},
  \citenamefont {Bhuyan}, \citenamefont {Chaudhuri}, \citenamefont {Dasgupta},\
  and\ \citenamefont {Rao}}]{dasgupta_2020}%
  \BibitemOpen
  \bibfield  {author} {\bibinfo {author} {\bibfnamefont {R.}~\bibnamefont
  {Mandal}}, \bibinfo {author} {\bibfnamefont {P.~J.}\ \bibnamefont {Bhuyan}},
  \bibinfo {author} {\bibfnamefont {P.}~\bibnamefont {Chaudhuri}}, \bibinfo
  {author} {\bibfnamefont {C.}~\bibnamefont {Dasgupta}}, \ and\ \bibinfo
  {author} {\bibfnamefont {M.}~\bibnamefont {Rao}},\ }\href@noop {} {\bibfield
  {journal} {\bibinfo  {journal} {Nature communications}\ }\textbf {\bibinfo
  {volume} {11}},\ \bibinfo {pages} {2581} (\bibinfo {year}
  {2020})}\BibitemShut {NoStop}%
\bibitem [{\citenamefont {Szamel}(2014)}]{szamel2014self}%
  \BibitemOpen
  \bibfield  {author} {\bibinfo {author} {\bibfnamefont {G.}~\bibnamefont
  {Szamel}},\ }\href@noop {} {\bibfield  {journal} {\bibinfo  {journal}
  {Physical Review E}\ }\textbf {\bibinfo {volume} {90}},\ \bibinfo {pages}
  {012111} (\bibinfo {year} {2014})}\BibitemShut {NoStop}%
\bibitem [{\citenamefont {Maggi}\ \emph {et~al.}(2015)\citenamefont {Maggi},
  \citenamefont {Marconi}, \citenamefont {Gnan},\ and\ \citenamefont
  {Di~Leonardo}}]{maggi2015multidimensional}%
  \BibitemOpen
  \bibfield  {author} {\bibinfo {author} {\bibfnamefont {C.}~\bibnamefont
  {Maggi}}, \bibinfo {author} {\bibfnamefont {U.~M.~B.}\ \bibnamefont
  {Marconi}}, \bibinfo {author} {\bibfnamefont {N.}~\bibnamefont {Gnan}}, \
  and\ \bibinfo {author} {\bibfnamefont {R.}~\bibnamefont {Di~Leonardo}},\
  }\href@noop {} {\bibfield  {journal} {\bibinfo  {journal} {Scientific
  reports}\ }\textbf {\bibinfo {volume} {5}},\ \bibinfo {pages} {1} (\bibinfo
  {year} {2015})}\BibitemShut {NoStop}%
\bibitem [{\citenamefont {Dandekar}\ \emph {et~al.}(2020)\citenamefont
  {Dandekar}, \citenamefont {Chakraborti},\ and\ \citenamefont
  {Rajesh}}]{Dandekar_PRE_2020}%
  \BibitemOpen
  \bibfield  {author} {\bibinfo {author} {\bibfnamefont {R.}~\bibnamefont
  {Dandekar}}, \bibinfo {author} {\bibfnamefont {S.}~\bibnamefont
  {Chakraborti}}, \ and\ \bibinfo {author} {\bibfnamefont {R.}~\bibnamefont
  {Rajesh}},\ }\href {\doibase 10.1103/PhysRevE.102.062111} {\bibfield
  {journal} {\bibinfo  {journal} {Phys. Rev. E}\ }\textbf {\bibinfo {volume}
  {102}},\ \bibinfo {pages} {062111} (\bibinfo {year} {2020})}\BibitemShut
  {NoStop}%
\end{thebibliography}%

\end{document}